\documentclass[a4paper]{article}
\pdfoutput=1
\usepackage{jhepmod}

\usepackage[utf8]{inputenc}

\usepackage[table ]{ xcolor}

\usepackage{enumitem}

\usepackage{multirow}

\usepackage{amsmath}

\usepackage{tikz}
\usetikzlibrary{shapes,arrows}
\tikzstyle{startstop} = [rectangle,rounded corners, minimum width=3cm,minimum height=1cm,text centered, draw=black,fill=red!30]
\tikzstyle{io} = [trapezium, trapezium left angle = 70,trapezium right angle=110,minimum width=3cm,minimum height=1cm,text centered,draw=black,fill=blue!30]
\tikzstyle{process} = [rectangle,minimum width=3cm,minimum height=1cm,text centered,text width =3cm,draw=black,fill=orange!30]
\tikzstyle{decision} = [diamond,minimum width=3cm,minimum height=1cm,shape aspect=3,inner sep = 0.4pt,text centered,draw=black,fill=green!30]
\tikzstyle{arrow} = [thick,->,>=stealth]
\tikzstyle{shadow}=[preaction={fill=black,opacity=.5,transform canvas={xshift=0.5mm,yshift=-0.5mm},shading=radial,shading angle=20},fill=red]

\tikzstyle{ellipse}=[draw, rectangle, minimum width=2.8em, rounded corners=6pt,line width=0.5pt]
\tikzstyle{pxsbx}=[trapezium, trapezium left angle=75, trapezium right angle=105, minimum width=3em, text centered, draw = black, fill=white,line width=0.5pt] 
\tikzstyle{lingxing}=[draw,diamond,shape aspect=3,inner sep = 0.4pt,thick,font=\itshape,line width=0.5pt]

\usepackage{amssymb}
\usepackage{graphicx,color}
\usepackage{appendix}

\usepackage{amsmath,amsthm}

\usepackage{mathrsfs}

\usepackage{bm}

\def\beq{\begin{equation}}
\def\eeq{\end{equation}}
\newcommand{\bea}{\begin{eqnarray}}
\newcommand{\eea}{\end{eqnarray}}
\def\bi{\begin{itemize}}
\def\ei{\end{itemize}}
\def\ba{\begin{array}}
\usepackage{verbatim}
\def\ea{\end{array}}
\def\bfig{\begin{figure}}
\def\efig{\end{figure}}

\def\R{\mathbb{R}}

\newtheorem{theorem}{Theorem}[section]
\newtheorem{definition}{Definition}[section]

\newtheorem{lemma}[theorem]{Lemma}

\newcommand{\EH}{{\rm EH}}

\def\be{\begin{eqnarray}}
\def\ee{\end{eqnarray}}

\newcommand{\cb}{\mathcal B}

\newcommand{\ck}{\mathcal K}
\newcommand{\cl}{\mathcal L}

\newcommand{\calp}{\mathcal P}

\newcommand{\cs}{\mathcal S}

\newcommand{\sa}{\mathscr{A}}

\renewcommand{\a}{\alpha}
\renewcommand{\b}{\beta}

\newcommand{\sig}{\sigma}

\renewcommand{\l}{\lambda}

\renewcommand{\o}{\omega}
\renewcommand{\O}{\Omega}

\newcommand{\rmd}{\mathrm d}

\newcommand{\lt}{\left}
\newcommand{\rt}{\right}

\newcommand{\ds}{\mathrm{dust}}
\newcommand{\tot}{\mathrm{tot}}
\newcommand{\geo}{\mathrm{geo}}

\title{Symmetry charges on reduced phase space and BMS algebra}

\author[1,2]{Muxin Han}

\author[3,4,5]{\ Zichang Huang}  

\author[1]{\ Hongwei Tan}

\affiliation[1]{Department of Physics, Florida Atlantic University, 777 Glades Road, Boca Raton, FL 33431-0991, USA}

\affiliation[2]{Institut f\"ur Quantengravitation, Universit\"at Erlangen-N\"urnberg, Staudtstr. 7/B2, 91058 Erlangen, Germany}

\affiliation[3]{Department of Physics, Center for Field Theory and Particle Physics,and Institute for Nano- electronic devices and Quantum computing, Fudan University, Shanghai 200433, China}
\affiliation[4]{State Key Laboratory of Surface Physics, Fudan University, Shanghai 200433, China}
\affiliation[5]{College of Science, University of Shanghai for Science and Technology, Shanghai 200093, P. R. China}

\emailAdd{hanm(At)fau.edu}
\emailAdd{htan2018(AT)fau.edu}

\abstract{This paper studies the reduced phase space formulation (relational formalism) of gravity coupling to the Brown-Kucha\v r dust for asymptotic flat spacetimes. A set of boundary conditions for the asymptotic flatness are formulated for Dirac observables on the reduced phase space. The physical Hamiltonian generates the time translation of the dust clock. We compute the boundary term of the physical Hamiltonian, which is identical to the ADM mass. We construct a set of the symmetry charges on the reduced phase space, which are conserved by the physical Hamiltonian evolution. The symmetry charges generate transformations preserving the asymptotically flat boundary condition. Under the reduced-phase-space Poisson bracket, the symmetry charges form an infinite dimensional Lie algebra $\mathscr{A}_G$ after adding a central charge. A suitable quotient of $\mathscr{A}_G$ closely relates to the BMS algebra at spatial infinity by Henneaux and Troessaert.  }


\begin{document}

\maketitle

\section{Introduction}

The canonical formulation of gravity is a theory of constraints \cite{wald2010general,thiemann2008modern}. 
The Hamiltonian of 4-dimensional gravity consists of four constraints: one Hamiltonian constraint and three components of diffeomorphism constraint. 
All hypersurface-deformations generated by these constraints are gauge transformations. Constructing gauge invariant quantities is important for understanding the physics of gravitational field at both classical and quantum levels \cite{mukhanov2005physical,dodelson2003modern,dirac2001lectures}.

An interesting approach of contructing gauge invariant observables is the relational formalism \cite{Rovelli_1991,Rovelli_1991_quan_ref_sys,Rovelli_1991_Time,Rovelli_1991_QM, rovelli2004quantum, carlo1994physical, kucha1995lie, dittrich2007partial,Dittrich_2006,tambornino2012relational, Ashtekar2011LQC, Domagala2010quantized,giesel2010manifestly,Giesel_2010, domenech2018hamiltonian, hwang2017gauge,tomikawa2020gauge,de2020gauge, yuan2020scalar,noh2004second}. 
The idea is to couple gravity with certain matter fields, e.g., pressure-less dust fields or scalar fields. In this work, we consider the Brown-Kucha\v r (BK) dust as the matter and refer to this formalism as BK formalism \cite{giesel2010manifestly, brown1995dust}. In this formalism, BK dust fields play the role of rulers and clocks that provide a physical reference frame at every spacetime point, and they relate to the gauge fixing scheme that allows us to construct a reduced phase space of the gravity-dust system. The gauge invariant Dirac observables defined on the reduced phase space are canonical fields evaluated in the physical reference frame made by the dust fields. Importantly, when formulating the dynamics on the reduced phase space in terms of Dirac observables, the gravity-dust system in the BK formalism becomes free of any constraint, since the Hamiltonian and diffeomorphism constraints are resolved on the reduced phase space. As a result, the formalism provides a physical Hamiltonian $\mathbf{H}_{\text {phys }}:=\int \mathrm{d}^{3} \sigma\, H_{\text{phys}}$ generates the physical time evolution of the Dirac observables. $\vec{\sigma}$ and $\tau$ are the spatial and time coordinates in the physical reference frame. There are infinitely many conserved charges $H_{\rm phys}(\vec{\sig})$ and $C_j(\vec{\sig})$ descended from the hypersurface deformation. $H_{\rm phys}(\vec{\sig})$ and $C_j(\vec{\sig})$ relates to the time translation and spatial diffeomorphism in the space of $(\tau,\vec{\sigma})$.

The purpose of this work is to apply the relational formalism to spacetimes with boundary at infinity. In particular we focus on the situation of asymptotically flat spacetimes. We propose a set of asymptotic boundary conditions for the asymptotically flatness in the BK formalism. These boundary conditions, discussed in Section \ref{sec:ABC}, are formulated for Dirac observables on the reduced phase space, and thus they implement the asymptotically flatness in the gauge-invariant manner. In addition, the compatibility with the asymptotically flatness requires the BK dust density to have a fall-off behavior of $O(r^{-4}\log r)$ as the radius $r\to\infty$.
 
As a technical aspect, the BK dust turns out to either be incompatible to the usual fall-off behavior and the parity condition of the canonical fields in the literature,  or it rules out many interesting boundary charges (e.g. in \cite{regge_1974_role,hennADM,beig1987poincare}). As the resolution, some additional $r^{-n}\log(r)$-terms are added to the fall-off behavior, and the parity condition is relaxed and replaced by other suitable boundary conditions. In this case, a counter-term has to be added to the symplectic form to remove the divergence. These treatments are partially inspired by \cite{mann2006holographic,compere2011relaxing}.

A key development is to define construct the boundary-preserving symmetry charge on the reduced phase space:
\be
G\lt(\xi,\vec{\xi}\rt)=\int \rmd^3\sig \lt(\xi H_{\rm phys}+\xi^jC_j\rt)+\cb\lt(\xi,\vec{\xi}\rt).
\ee
The smearing fields $\xi,\vec{\xi}$ satisfy certain boundary conditions relating to a subset of the traditional Poincaré transformations and the supertranslations. Unlike the usual situations in general relativity, the bulk terms of the symmetry charge do not vanish on-shell, and they generate physical symmetries in the bulk of the dust space (the space of $\vec{\sigma}$). The symmetry charge contains the boundary term $\cb(\xi,\vec{\xi})$, which we call the boundary charge. $\cb(\xi,\vec{\xi})$ is determined by the compatibility to the asymptotically flat boundary condition and making the variation $\delta G(\xi,\vec{\xi})$ well-defined on the reduced phase space. The boundary charge $\cb(\xi,\vec{\xi})$ is computed explicitly in Section \ref{sec:bounchg}.

We show that the boundary-preserving symmetry charge $G(\xi,\vec{\xi})$ enjoys the following important properties (the 1st and 2nd properties are what we mean by $G(\xi,\vec{\xi})$ being symmetry and boundary-perserving):
\begin{itemize}
\item It commutes with the total physical Hamiltonian ${\bf H}_{\rm phys}+{\bf H}_{\rm bdy}$, where ${\bf H}_{\rm bdy}$ is the boundary term relating to the ADM mass;
\item The Hamiltonian flow generated by $G(\xi,\vec{\xi})$ preserves the boundary conditions;

\item With the Poisson bracket on the reduced phase space, the set of charges $G(\xi,\vec{\xi})$ for all $\xi,\vec{\xi}$ form an infinite-dimensional Lie algebra up to a central charge $\mathscr{C}$, namely
\be
\lt\{G\lt(\xi_1, \vec{\xi}_1\rt),G\lt(\xi_2, \vec{\xi}_2\rt)\rt\}
    =G\lt(\hat{\xi}, \hat{\vec{\xi}}\rt)
    +\mathscr{C}\lt(\hat{\xi}, \hat{\vec{\xi}}\rt).
\ee
where $\hat{\xi}=\xi_1^iD_i\xi_2-\xi_2^iD_i\xi_1$ and $ \hat{\xi}^j=[\vec{\xi}_1,\vec{\xi}_2]^j$. In other words, the boundary-preserving symmetry charge $G$ and the central charge $\mathscr{C}$ form a closed Lie algebra. The derivation of the algebra is given in Section \ref{sec:ABSG}.
\end{itemize}

We denote by $\sa_{\rm G}$ the algebra of the boundary-preserving symmetry charges $G(\xi,\vec{\xi})$, and we would like to compare $\sa_{\rm G}$ to the original BMS algebra, $\sa_{\rm BMS}$ \cite{bondi1962gravitational, sachs1962gravitational, barnich2010aspects,barnich2011bms}. 
The original BMS charges are defined as boundary charges that generate symmetry transformations on the null infinity of an asymptotically flat spacetime. These symmetry transformations include the traditional Poincar\'e transformations and the supertranslations. The supertranslation charges form an infinitesimal Abelian subalgebra of $\sa_{\rm BMS}$. In contrast, the charges $G(\xi,\vec{\xi})$ defined in this paper incorporate both bulk terms and boundary terms and are Dirac observables. The bulk terms in the charges do not correspond to any constraint and thus non-vanishing. Indeed, $\sa_{\rm G}$ contains an ideal $\tilde{\sa}_{\rm G}$ of the symmetries in the bulk of the dust space that are not extended to the asymptotic boundary. The quotient Lie algebra $\hat\sa_{\rm G}=\sa_{\rm G}/\tilde{\sa}_{\rm G}$ is comparable to $\sa_{\rm BMS}$.


As mentioned earlier, the supertranslations themselves form an Abelian subalgebra within the original BMS algebra. Correspondingly, the supertranslations in $\hat\sa_{\rm G}$ along the physical time and radial directions form an Abelian subalgebra when the central charge vanishes. If we impose appropriate parity conditions on the parameters of the supertranslations, the central extension vanishes, then the original BMS subalgebra of supertranslations can be embedded as an subalgebra in $\hat\sa_{\rm G}$. In our work, we do not generally require the parameters of the supertranslations to satisfy parity conditions, and we introduce two additional degrees of freedom contributed by the angular components of the supertranslations in the spatial direction. Therefore, $\hat\sa_{\rm G}$ represents a generalization of $\sa_{\rm BMS}$ as far as the supertranslations are concerned.

The asymptotic spatial rotations in $\hat\sa_{\rm G}$ form an SO(3) algebra thus is isomorphic to the corresponding subalgebra in BMS. The commutator between the supertranslations in the time direction and in the spatial directions involves the central charge of the algebra. In this work, we only consider the isometric spatial rotations generated by the killing vectors on $S^2$. It would be interesting to generalize the analysis to the conformal transformations on $S^2$. But we leave this aspect as a future research.

The symmetry algebra $\hat\sa_{\rm G}$ does not take account of the asymptotic boost transformation, since it is difficult to make the boost transformation preserving the boundary condition. The detailed discussion is given in Section \ref{sec: boost}. 

$\hat\sa_{\rm G}$ closely relates to the BMS group at spatial infinity proposed in \cite{hennADM}. In particular, certain restriction of $\xi,\vec{\xi}$ selects a subalgebra of $\hat\sa_{\rm G}$ that recovers the BMS algebra at spatial infinity in \cite{hennADM} with vanishing boost generator, as shown in Section \ref{subsec:efa}.

The relational formalism enables the gravity-dust system to be formulated on the reduced phase space and free of constraints, and the dynamics are expressed similarly to common Hamiltonian dynamical systems. Intriguingly, the symmetry of the gravity-dust system can be formulated similarly to symmetries in typical Hamiltonian dynamical systems. This is illustrated in this study for asymptotically flat spacetimes: All symmetry charges $G(\xi,\vec{\xi})$ are obtained as phase space functions that Poisson commute with the physical Hamiltonian. The symmetry charge, when expressed in terms of Dirac observables, includes both the bulk and boundary contributions. We are concentrating on asymptotically flat spacetimes at this stage because our aim is to compare this with the familiar result on the BMS algebra \cite{bondi1962gravitational, sachs1962gravitational, barnich2010aspects,barnich2011bms}. However, the same symmetry analysis can be carried out for other boundary conditions, or even spacetimes without a boundary. 

The relational formalism is an approach for resolving the problem of time in classical and quantum gravity. On the reduced phase space, the physical Hamiltonian generates the physical time translation, whose quantization makes sense the unitarity of quantum gravity \cite{Domagala2010quantized,giesel2010algebraic,Han:2020chr}. In contrast, in the usual formulation of gravity, the time translation and unitarity are meaningful only on the boundary of the spacetime, and this is the idea behind the holographic duality. It's worth exploring the relation between the physical time evolution formulated on the reduce phase space and the time evolution of the holographic boundary dynamics. This work may present an initial effort towards understanding this relation, by comparing $\sa_{\rm G}$ to the BMS algebra on the boundary.

This paper is organized as follows: Section \ref{sec:ralaform} reviews the relational formalism. Section \ref{sec:ABC} introduces the asymptotic boundary conditions and defines the finite symplectic structure with a counter term. Section \ref{sec:bounchg} constructs the boundary-preserving symmetry charges. Section \ref{sec:ABSG} compute the Poisson bracket between a pair of the boundary-preserving symmetry charges. Section \ref{sec: boost} demonstrates that the generator of the boost term does not satisfy the definition of the boundary-preserving generator. Section \ref{conclude} summarizes the results and discusses a few future perspectives.
 
\section{A Review of the Brown-Kuchař Formalism}\label{sec:ralaform}

\subsection{Lagrangian Formalism}\label{subsec: bklag}

The BK formalism is a realization of relational formalism with the BK dust field consisting of four scalars $T,\ S^{j=1,2,3}$. The total action in the BK formalism is 
\begin{equation}\label{eq:totalaction}
    S=S_\EH+S_\ds.
\end{equation}
$S_\EH$ is the Einstein-Hilbert action\footnote{In our convention, the constant $\kappa={16\pi G}$ is set to be $1$.}.
The BK dust action $S_\ds$ depends on the dust fields $T, S^j$ and Lagrangian multipliers $\rho,W_j$ \cite{brown1995dust}:
\begin{equation}\label{eq:dustAction}
S_{\text {dust }}=-\frac{1}{2} \int_{M} \mathrm{~d}^{4} x \sqrt{|\operatorname{det}(g)|} \rho\left[g^{\mu \nu} U_{\mu} U_{\nu}+1\right],
\end{equation}
with 
\begin{equation}\label{eq:4-velo}
U_{\mu}=-\partial_{\mu} T+W_{j} \partial_{\mu} S^{j}.
\end{equation}

The equation of motion $\frac{\delta S}{\delta \rho}=0$ yields 
\begin{equation}\label{4velo}
    g^{\mu \nu} U_{\mu} U_{\nu}=-1.
\end{equation}
Thus, $U^{\mu}$ is the 4-velocity of the dust. 
Another equation of motion $\frac{\delta S}{\delta g_{\mu\nu}}=0$ relates the Einstein tensor to the energy-stress tensor of the dust field: 
\begin{equation}
            T_{\mu \nu}^{\ds}=\rho U_{\mu} U_{\nu},
\end{equation}
which indicates that the BK dust is a pressure-less perfect fluid. $\rho$ is interpreted as the dust density, and
\begin{equation}
\rho= \begin{cases}>0& \text { physical dust } \\ 
<0 & \text { phantom dust }\end{cases}.
\end{equation}
Note that the case of phantom dust can still fulfill the energy condition when coupling to additional matter fields \cite{giesel2010manifestly}. Our following discussion applies to both cases of physical and phantom dusts.

Other equations of motion $\frac{\delta S}{\delta T}=0$, $\frac{\delta S}{\delta S^j}=0$, and $\frac{\delta S}{\delta W_j}=0$ provide
\be\label{eom}
U^\nu\nabla_\nu U_\mu=0,\qquad
U^\nu\nabla_\nu T=1,\qquad 
U^\nu\nabla_\nu S_j=0.
\ee
The first two equations in \eqref{eom} indicate that the integral line of $U^\mu$ is a time-like geodesic and $T$ is its line-length parameter. The last one indicates that $S_j$ are constants along the integral line of $U^\mu$. 
The dust field naturally introduces a reference frame to the spacetime. On this spacetime,  the dust fields $T$ and $\vec{S}$ can be viewed as providing a coordinate system where $T$ and $\vec{S}$ are the time and space coordinates respectively.

\subsection{Hamiltonian analysis}\label{subsec: bkhamil}
Following the ADM formalism, a spacetime is decomposed into $\Sigma\times \mathbb{R}$, where $\Sigma$ is the $3$-dimensional spatial slice. We denote the $3$-metric on $\Sigma$ by $q_{ab}$ and its conjugate momentum by $p^{ab}$\footnote{The Latin letters $a,b$ are the indices of on $\Sigma$. }. The Hamiltonian analysis provides four first-class constraints
\be
 c^\tot=c^\geo+c^\ds,\label{hamilcon}\\
 c^\tot_a=c^\geo_a+c^\ds_a, \label{diffcon}
\ee
with
\be
 c^{\geo}&=&\frac{1}{\sqrt{\operatorname{det}(q)}}\left[q_{a c} q_{b d}-\frac{1}{2} q_{a b} q_{c d}\right] p^{a b} p^{c d}-\sqrt{\operatorname{det}(q)} {}^{(3)}R_{0},\\
c^{\ds}&=&\frac{1}{2}\left[\frac{P^{2} / \rho}{\sqrt{\operatorname{det}(q)}}+\sqrt{\operatorname{det}(q)} \rho\left(q^{a b} U_{a} U_{b}+1\right)\right],\\
 c_{a}^{\geo}&=&-2 q_{a c} D_{b} p^{b c},\\
c_{a}^{\ds}&=P&\left[T_{, a}-W_{j} S_{, a}^{j}\right],\label{dusdiff}
\ee
where $P$ and $P_j$ are conjugate momenta of $T$ and $S^j$ respectively, and ${}^{(3)}R_{0}$ is the Ricci scalar of $q_{ab}$. 
There are eight second-class constraints.
Four of them are
\be
\quad W_{j}&=&-P_{j} / P,\label{solforwj}\\
\rho^{2}&=& \frac{P^{2}}{\operatorname{det}(q)}\left(q^{a b} U_{a} U_{b}+1\right)^{-1}.\label{consofrhop} 
\ee
Eq. \eqref{consofrhop} gives
\begin{equation}
   \rho =\epsilon\frac{P}{\sqrt{\operatorname{det}(q)}}\left(q^{a b} U_{a} U_{b}+1\right)^{-1/2},\,\epsilon=\pm 1.
\end{equation}
The future-pointing timelike $U^\mu$ fixes $\epsilon = 1$ \cite{giesel2010manifestly}. 
By \eqref{hamilcon}, $ c^\tot=0$ yields
\begin{equation}\label{relacgeorho}
    c^\geo=-\sqrt{\operatorname{det}(q)}\rho\left(q^{a b} U_{a} U_{b}+1\right).
\end{equation}
With \eqref{eq:4-velo}, \eqref{dusdiff}, and $c^{tot}_a=0$, we get
\begin{equation}\label{velo}
    U_a= \frac{c^\geo_a}{P}.
\end{equation}
These formulae are useful in analysing the asymptotic behaviors of the dust density $\rho$, which we will see in the next section. Other second-class constraints are less important for our discussion, so we refer the reader to \cite{giesel2010manifestly} for further details.  

In this section, we do not introduce the boundary condition and boundary terms to the constraints or Dirac observables. We will introduce the asymptotically flat boundary condition directly at the level of reduced phase space, as to be discussed in Section \ref{sec:ABC}.

\subsection{Dirac Observables and the Bulk symmetry charges}\label{subsec:bkrelat}
BK formalism defines the gauge invariant quantities -- Dirac observables through the deparametrisation process \cite{giesel2010manifestly}. Firstly, the first-class constraints \eqref{hamilcon} and $\eqref{diffcon}$ are solved by
\be
P=\operatorname{sgn}(P)\, h,\qquad 
P_j=-h_j,\label{relaph}
\ee
with
\be
h=\sqrt{\lt(c^\geo\rt)^2-q^{a b} c^\geo_a c^\geo_b},\qquad 
h_j=S_j^a\left(-h T_{, a}+c^\geo_a\right).
\ee 
and $S_j^a$ is the inverse matrix of $\partial_a S^j$, and $h$ does not depends on $\lt(T, P\rt)$. 

Secondly, the smeared constraint is 
\begin{equation}
K_\beta \equiv \int_{\Sigma} \mathrm{d}^3 x\left[\beta(\vec{x}) c^{\mathrm{tot}}(\vec{x})+\beta^j(\vec{x}) c_j^{\mathrm{tot}}(\vec{x})\right]. 
\end{equation}
The Dirac observable from a generic phase space function $f$ is constructed by the exponential map generated by $K_\beta$
\begin{equation}
O_f[\tau, \sigma] \equiv\left[\sum_{n=0}^{\infty} \frac{1}{n !}\left\{f, K_\beta\right\}_{(n)}\right]_{\substack{\beta \rightarrow \tau-T \\ \beta^j \rightarrow \sigma^j-S^j}},
\end{equation}
Here $\left\{f, K_\beta\right\}_{(n)}$  means doing $n$ times of the Poisson bracket between $f$ and $K_\beta$. For example: $\left\{f, K_\beta\right\}_{(0)}=f$, and $\left\{f, K_\beta\right\}_{(2)}=\left\{f,\left\{f, K_\beta\right\}\right\}$. $\tau,\sigma^j\in\R$ are coordinates in the reference frame defined by the dust fields. One can prove that $O_f[\tau, \sigma] $ is gauge invariant on the constraint surface, namely, it is a (weakly) Dirac observable \cite{thiemann2008modern}. $O_f$ is a field defined on the dust space $\cs$ \footnote{ Here the $\cs$ is the space-like hypersurface whose coordinates are given by $\sigma^j$. We have a dust space $\cs\lt[\tau\rt]$ at each instance of the physical time $\tau$.}. In this way, we construct the Dirac observables of $q_{ab}$ and $p^{ab}$ and denote them by $g_{ij}$ and $\pi^{ij}$ respectively\footnote{The indices of the Dirac observables are the Greek letter since the observables are defined on the dust space. }. The symplectic structure on the reduced phase space is then given by 
\begin{equation}
    {\O}:=\int_\cs \rmd^3\sigma\, \delta g_{ij}\wedge \delta\pi^{ij}.
\end{equation}
We have ignored the the possible modification of $\O$ in the case that $\cs$ has a boundary, we will come back to the symplectic structure in the next section. 

$c^{\geo}$ and $c^{\geo}_a$ can be promoted to the Dirac observables $C$ and $C_j$ by substituting $\lt(q_{ab}, p^{ab}\rt)$ with $\lt(g_{ij}, \pi^{ij}\rt)$ in their expressions:
\be
C&=&\frac{1}{\sqrt{\operatorname{det}(g)}}\left[g_{i k} g_{j l}-\frac{1}{2} g_{ij} g_{kl}\right] \pi^{ij} \pi^{kl}-\sqrt{\operatorname{det}(g)} {}^{(3)}R,\label{eq:C_in_Dirac}\\
C_j&=&-2 g_{j k} D_{i} \pi^{i k},
\ee
with ${}^{(3)}R$ is the Ricci scalar of $g_{ij}$.
The Dirac observable of $h$ is given by
\begin{equation}\label{hamdensity}
    H_{\text{phys}}=\sqrt{C^2-g^{ij}C_iC_j}.
\end{equation}
Ignoring again the situation that $\cs$ has a boundary, the integration of $H_{\rm phys}$ over $\cs$ gives the physical Hamiltonian
\begin{equation}\label{genH}
    \mathbf{H}_{\text {phys }}:=\int_{\mathcal{S}} \mathrm{d}^{3} \sigma H_{\text{phys}
    }.
\end{equation}
Given a generic Dirac observable $F$ on the reduced phase space, its physical time evolution with respect to $\tau$ is generated by ${\bf H}_{\rm phys}$
\begin{equation}\label{eq:phy_time_dire}
    \frac{\mathrm{d} F}{\mathrm{d}\tau}:=\lt\{F, \mathbf{H}_{\text {phys }}\rt\}.
\end{equation}
The variation of the ${\rm H_{phy}}$ is given by
\begin{equation}\label{variaofph}
    \delta \mathbf{H}_{\text {phys }}=\int_{\mathcal{S}} \mathrm{d}^{3} \sigma \left(N \delta C+N^i \delta C_i+\frac{1}{2} H_{\text{phys}} N^i N^j \delta g_{i j}\right).
\end{equation}
In analogy to the lapse function and the shift vector field in general relativity, $N$ and $N^{i}$ are called the dynamical lapse function and the dynamical shift vector field respectively.
The expressions of $N^i$ and $N$ are
\begin{equation}\label{dyns}
    N^{i}=-\frac{g^{i j} C_{j}}{H_{\text{phys}}},
\end{equation}
 \begin{equation}\label{dynl}
N=C/H_{\text{phys}}= \begin{cases}-\sqrt{1+g^{jk}N_jN_k} & \text { physical dust } 
\\ \sqrt{1+g^{jk}N_jN_k} & \text { phantom dust }\end{cases}.
\end{equation}
A difference between the general relativity and the BK formalism is that both $N$ and $N^j$ depend on the canonical variables. 
When ignoring the boundary of $\cs$, we have infinitely many conserved charges $H_{\rm phys}(\vec{\sig})$ and $C_j(\vec{\sig})$
\be
\lt\{C_j(\vec{\sig}), \mathbf{H}_{\text {phys }}\rt\}=0,\qquad
\lt\{H_{\text{phys}}(\vec{\sig}), \mathbf{H}_{\text {phys }}\rt\}=0.
\ee
The Hamiltonian density $H_{\rm phys}(\vec{\sig})$ and the dust-space diffeomorphism $C_j(\vec{\sig})$ generate symmetries that are analogs of the hypersurface deformations in canonical general relativity, although here they are physical symmetries on the reduced phase space rather than gauge symmetries \cite{brown1995dust}. The modification of these symmetries charges in presence of boundary will be discussed in section \ref{sec:bounchg}.

\section{Asymptotic Boundary Conditions}\label{sec:ABC}

Starting from this section, we analyze the situation that the dust space $\cs$ has an asymptotic boundary at infinity. We consider the asymptotically flat boundary condition, and accordingly, we modify the symmetry charges $H_{\text{phys}}$ and $C_j$ by adding the corresponding boundary charges. Subsection \ref{sec:phasym} introduces the boundary conditions and the symplectic structure. Subsection \ref{sec:varphyham} shows that the boundary charge of the physical Hamiltonian corresponds to the ADM mass.

\subsection{The Phase Space and the symplectic structure}\label{sec:phasym}

In any asymptotically flat spacetime, there exits an asymptotic Cartesian coordinate $X^\a$ , such that the 4-metric has the following asymptotic behavior at $r_0\to\infty$:
\be
g_{\a\b}=\eta_{\a\b}+\frac{f_{\a\b}\lt(X^0,\vec{X}/r_0\rt)}{r_0}+o\lt(r_0^{-1}\rt),\label{asympflat}
\ee
where $r_0=\sqrt{{X}^i{X}_i}$, and $f_{\a\b}\lt(X^0,\vec{X}/r_0\rt)$ is a smooth tensor field on $S^2$ at $X^0$ \footnote{In our notations, $o\lt(r^a\rt) $ means decaying faster than $r^a$ while $O\lt(r^a\rt)$ means decaying as fast as $r^a$ for some constant $a$.}.
To study the asymptotically flat spacetime in the reduced phase space, we propose that the material reference frame given by the dust is asymptotically Cartesian. Namely, we identify $\tau=X^0$, $\vec{\sigma}=\vec{X}$, and we define $r=\sqrt{\sigma^i\sigma_i}$. Then Eq.\eqref{asympflat} leads to the following asymptotic behaviors of $q_{ij}$ and $\pi^{ij}$ on the dust space $\cs$:
\be\label{convefalof}
g_{ij}=\delta_{ij}+\frac{\bar{h}_{ij }}{r}+o\lt(r^{-1}\rt),\quad \pi^{ij}=\frac{\bar{\pi}^{ij}}{r^2}+o\lt(r^{-1}\rt).
\ee

A naively approach might be to adapt the standard procedure \cite{regge_1974_role,hennADM,beig1987poincare} in our context, by expanding
\be
g_{ij}&=&\delta_{ij}+\frac{\bar{h}_{ij }}{r}+\frac{h^{(2)}_{ij}}{r^2}+{o}\left(r^{-2}\right),\label{asympq000} \\
\pi^{ij}&=&\frac{\bar{\pi}^{ij}}{r^2}+\frac{\pi^{(2)ij}}{r^3}
+\frac{\pi^{(3)ij}}{r^4}
+{o}\left(r^{-4}\right),\label{asympp000}
\ee
and imposing parity conditions to $\bar{h}_{ij}$ and $\bar{\pi}^{ij}$ \footnote{For a function $f\lt(\vec{\sigma}\rt)$ on $S^2$, $f\lt(\vec{x}\rt)$ has odd parity if $f\lt(-\vec{\sigma}\rt)=-f\lt(\vec{\sigma}\rt)$; $f\lt(\vec{\sigma}\rt)$ has even parity if $f\lt(-\vec{\sigma}\rt)=f\lt(\vec{\sigma}\rt)$.}. The reason of the parity condition is to make the following symplectic form finite as $R\to\infty$
\begin{equation}\label{defomgp}
    \O_0:=\int_{\cs_R} \rmd^3\sigma\, \delta g_{ij}\wedge \delta\pi^{ij},
\end{equation}
where $\cs_R$ is the dust space with radial cut-off $R$. $\O_0$ diverges logarithmically as $R\to\infty$ without the parity conditions.

However, as to be clarified in a moment, the above naive approach fails here because it is incompatible with either physical dust or phantom dust (see the remark at the end of this subsection), or it rules out many interesting boundary charges . We make two modifications: (1) We consider more general boundary condition by relaxing the parity conditions, similar to the approach in \cite{mann2006holographic,compere2011relaxing}. Then a counter term has to be added to $\O_0$ to cancel the divergent terms in \eqref{defomgp}, as we see in a moment. (2) we consider the following more general fall-off conditions including logarithmic terms 
\be
g_{ij}&=&\delta_{ij}+\frac{\bar{h}_{ij }}{r}+\frac{
\log r}{r^2}h^{(\log)}_{ij}+\frac{h^{(2)}_{ij}}{r^2}+{o}\left(r^{-2}\right),\label{asympq} \\
\pi^{ij}&=&\frac{\bar{\pi}^{ij}}{r^2}+\frac{
\log r}{r^3}\pi^{(\log)ij}+\frac{\pi^{(2)ij}}{r^3}
+\frac{\log r}{r^4}\pi^{(ll)ij}
+\frac{\pi^{(3)ij}}{r^4}
+{o}\left(r^{-4}\right).\label{asympp}
\ee
Here, $\bar{h}_{ij }$, $\bar{\pi}^{ij }$, ${h}^{\rm (log)}_{ij }$, ${\pi}^{\mathrm{(log)}ij }$, ${h}^{\rm (2)}_{ij }$, ${\pi}^{\mathrm{(2)}ij }$, $\pi^{(ll)ij}$ and $\pi^{(3)ij}$ are called the boundary fields, since they are functions on $S^2$, which is the boundary of $\cs$. In order to analyze the higher order asymptotic behavior of $C_j$, we expand $\pi^{ij}$ to $O\lt(r^{-4}\rt)$. 
The motivation of adding $\log r$-terms in \eqref{asympq} and \eqref{asympp} is to include more general solutions of $C_j=0$ (see \cite{compere2011relaxing} and appendix B of \cite{beig1987poincare}).
We next transform \eqref{asympq} and \eqref{asympp} to spherical coordinates $\lt\{r,\sigma^A\rt\}\,\lt(A=1, 2\rt)$. Here $\sigma^A$ are the angular coordinates, and they are the functions of the traditional spherical coordinates $\theta$, $\varphi$. They have odd parity $\sigma^A\lt(\pi-\theta,\varphi+\pi\rt)=-\sigma^A\lt(\pi,\varphi\rt)$. 
\be
g_{r r}&=&1+\frac{1}{r} \bar{h}_{r r}+\frac{\log  r}{r^2} h^{(\log)}_{r r}+\frac{1}{r^{2}} h_{r r}^{(2)}+o\left(r^{-2}\right),\label{asymsphere1}\\
g_{r A}&=&\bar{h}_{rA}+\frac{\log  r}{r} h_{r A}^{(\log)}+\frac{1}{r} h_{r A}^{(2)}+o\left(r^{-1}\right),\\
g_{A B}&=&r^{2} \bar{\gamma}_{A B}+r \bar{h}_{A B}+\log (r) h_{A B}^{(\log)}+h_{A B}^{(2)}+o\lt(r^0\rt),\label{asymsphere3}\\
\pi^{r r}&=&\bar{\pi}^{r r}+\frac{\log r}{r} \pi^{(\log)r r}+\frac{1}{r} \pi^{(2) r r}+\frac{\log r}{r^2}\pi^{(ll)rr}
+\frac{\pi^{(3)rr}}{r^2}
+{o}\left(r^{-2}\right),\\
\pi^{r A}&=&\frac{1}{r} \bar{\pi}^{r A}+\frac{\log  r}{r^2} \pi^{(\log)r A}+\frac{1}{r^{2}} {\pi^{(2)rA}}+\frac{\log r}{r^3}\pi^{(ll)rA}
+\frac{\pi^{(3)rA}}{r^3}
+{o}\left(r^{-3}\right),\\
\pi^{A B}&=&\frac{1}{r^{2}} \bar{\pi}^{A B}+\frac{\log  r}{r^3} \pi^{(\log)A B}+\frac{1}{r^{3}} \pi^{(2) A B}+\frac{\log r}{r^4}\pi^{(ll)AB}
+\frac{\pi^{(3)AB}}{r^4}
+{o}\left(r^{-4}\right).\label{asymsphere6}
\ee
Here $A$ and $B$ are indices of the angular coordinates on $S^2$, and $\bar{\gamma}_{AB}$ is the metric on a unit sphere. We introduce some short-hand notations for the following discussion 
\begin{equation}
\begin{aligned}
    &\bar{\lambda}=\frac{1}{2}\bar h_{r r}, \quad \tilde{\bar{k}}_{A B}=\frac{1}{2} \bar{h}_{A B}+\bar{\lambda}\bar{\gamma}_{A B},\quad  \bar{\lambda}_{A}=\bar{h}_{rA},\\
&\bar{p}=2\left(\bar{\pi}^{r r}-\bar{\pi}_{A}^{A}\right), \quad \pi_{(k)}^{A B}=2 \bar{\pi}^{A B}.
\end{aligned}
\end{equation}

We need to impose some additional boundary conditions to $g_{ij}$ and $\pi^{ij}$. Firstly, there is a parity condition that is imposed to $\pi^{(\log)r A}$ and is compatible to the dust: 
\begin{equation}\label{paroflogpi}
    \pi^{(\log)r A}\lt(-\vec{\sigma}\rt) = \pi^{(\log)r A}\lt(\vec{\sigma}\rt).
\end{equation}
This parity condition results in that the ADM angular momentum becomes finite (it will be shown in section \ref{sec:bounchg}).
Equations \eqref{asymsphere1}-\eqref{asymsphere6} yield
\begin{equation}\label{crexp}
    C_r=\frac{C_{r}^{(1)}}{r}+\frac{\log(r)}{r^2}C^{(\log)}_{r}+\frac{ C^{(2)}_r}{r^2}
+\frac{\log r}{r^3}C^{(ll)}_r
+O\lt(r^{-3}\rt),
\end{equation}
\begin{equation}\label{cAexp}
    C_A=C_{A}^{(0)}+\frac{\log(r)}{r}C^{(\log)}_{A}+\frac{ C^{(1)}_A}{r}
+\frac{\log r}{r^2}C^{(ll)}_A
+O\lt(r^{-2}\rt),
\end{equation}
\begin{equation}\label{cexp}
    C=\frac{C^{(1)}}{r}+\frac{\log(r)}{r^2}C^{(\log)}+O\lt(r^{-2}\rt),
\end{equation}
where
\be
C_r^{(1)}&=&2\lt(\bar{\pi}^A_{\ A}-\partial_A\bar{\pi}^{rA}\rt),\label{Cj-exp}\\
C_r^{(\log)}&=&2\lt(\pi^{(\log)rr}-\partial_A\pi^{(\log)rA}+\pi^{(\log)}{}^A_{\ A}\rt),\\
C_A^{(0)}&=&-2\lt(\bar{\pi}^{r}_{\ A}+\bar{D}_B\bar{\pi}^{B}_{\ A}\rt),\\
C_A^{(\log)}&=&-2\bar{D}_B\pi^{(\log)}{}^B_{\ A},\\
 C^{(1)}&=&-2 \sqrt{\bar{\gamma}}\left(\bar{D}_{A} \bar{D}_{B} \tilde{\bar{k}}^{A B}-\bar{D}_{A} \bar{D}^{A} \tilde{\bar{k}}+\bar{D}_A\bar{\lambda}^A\right).\label{c1}
\ee
The explicit  expression of $C^{(\log)}$ is complicate and unnecessary for the following discussions. We will provide further comments on this term in a moment.
 
Generically, the physical Hamiltonian density $H_{\rm{phy}}$ decays as $ O(r^{-1})$ by \eqref{crexp}-\eqref{cexp} and \eqref{hamdensity}. Then the bulk physical Hamiltonian diverges logarithmically,
\begin{equation}\label{newsym}
    {\bf H}_{\rm phys}=\lim_{R\to\infty}\int_{\cs_R} \rmd^3\mathscr{V}\, H_{\text {phys }}=\lim_{R\to\infty}\lt|C^{(1)}\rt|\log (R) +{\rm finite}.
\end{equation}
This integral is written in the sphereical coordinate. The coordinate volume is 
\be
\rmd^3\mathscr{V}=\rmd r\rmd\sigma^1\rmd\sigma^2. 
\ee
To remove the logarithmic divergence, we introduce the boundary condition
\begin{equation}\label{asyc1}
    C^{(1)} = 0.
\end{equation}
The leading order contribution to $H_{\rm phys}$ is $C^{(\log)}$. The parity condition \eqref{paroflogpi} has no impact on $C^{(\log)}$ based on power counting. It is convenient to analyze this in asymptotic Cartesian coordinates, where the contribution of $C^{(\log)}$ is in $O\lt(\frac{\log r}{r^4}\rt)$. On the other hand, with \eqref{eq:C_in_Dirac} and \eqref{asympp}, one finds that the leading order of the contribution from $\pi^{\lt(\log\rt)ij}$ in $C$ is at $O\lt(\frac{\log r}{r^5}\rt)$. As the result, regardless what parity condition assigned to $\pi^{\lt(\log\rt)ij}$, it does not influence $C^{(\log)}$.

Given a functional $f$ of the fields $g_{ij},\,\pi^{ij}$, the boundary term of $\delta f$ is a differential if  
\be
\delta f=A_{ij}\delta\pi^{ij}+B^{ij}\delta g_{ij}-\delta \mathcal{B}_f|_{r\to\infty},
\ee
for $\mathcal{B}_f|_{r\to\infty}$ depending on only boundary fields. In this case, the variation of $f+\cb_f$ is well-defined, i.e. $f+\cb_f$ is a differentiable functional. If $f$ relates to a symmetry, we call $f+\cb_f$ the symmetry charge and $\cb_f$ the boundary charge.
If we only impose \eqref{asympq} and \eqref{asympp}, the boundary term of $\delta \mathbf{H}_{\text {phys }}$ is not a differential thus cannot be cancelled by adding boundary charges. A resolution is to impose some additional boundary conditions:
\begin{equation}\label{intial}
\begin{aligned}
&C_{A}^{(0)}=C^{(\log)}_A=C_{A}^{(1)}=C_{r}^{(1)}=C^{(\log)}_{r}=C_{r}^{(2)}=0,\\
&C^{(\log)} \neq 0.
\end{aligned}
\end{equation}
In Subsection \ref{sec:varphyham}, we will show that these boundary conditions ensure the non-differential boundary terms in $\delta \mathbf{H}_{\text {phys }}$ decaying fast enough and vanishing in the spatial infinity.

Inspired by \cite{mann2006holographic,compere2011relaxing}, the symplectic form $\O$ becomes finite at $R\to\infty$ after adding a counter term: 
\be\label{defofsym}
\O=\lim_{R\to\infty}\int_{\cs_R} \rmd^3\sig \, \delta g_{ij}\wedge \delta\pi^{ij}-\log(R)\oint_{S^2} \rmd^2 S\, \delta \bar{h}_{ij}\wedge \delta \bar{\pi}^{ij},
\ee
with $\rmd^2 S=\rmd\sigma^1\rmd\sigma^2$, and the $\delta$ indicates the variation of fields satisfying the fall-off conditions \eqref{asympq} and \eqref{asympp} and boundary conditions \eqref{paroflogpi}, \eqref{asyc1}, and \eqref{intial}. The counter term of \eqref{defofsym} may relate to the corner term in the formalism of \cite{Harlow:2019yfa}.
Note that $\O$ does not depend on the variations of the boundary fields $\bar{\pi}^{ij},\bar{h}_{ij}, {\pi}^{({\rm log})ij}, h^{\rm (log)}_{ij}$, etc, in the expansions \eqref{asympq} and \eqref{asympp}. This can be seen by inserting the expansions of $g_{ij},\pi^{ij}$ in $\O$. Indeed, in Cartesian coordinate, the boundary contribution of \eqref{defofsym} is $\frac{\log(R)}{R}\oint_{S^2}\rmd^2 S\,\lt(\delta\bar{\pi}^{ij}\wedge\delta\bar{h}^{(\log)}_{ij}+\delta\pi^{(\log)ij}\wedge\delta\bar{h}_{ij}\rt)+O\lt(R^{-1
}\rt)$, and it vanishes as $R\to\infty$.

We define the reduced phase space $\calp$ to be the space of fields $g_{ij},\pi^{ij}$ satisfying the fall-off conditions \eqref{asympq} and \eqref{asympp} and boundary conditions \eqref{paroflogpi}, \eqref{asyc1}, and \eqref{intial}. $\calp$ is equipped with the symplectic form $\O$. For any differentiable function $f$ on $\calp$, its variation on $\calp$ is given by
\be
\delta f =\int_{\cs}\rmd^3\sigma \lt[\frac{\delta f}{\delta g_{ij}(\sigma)}\delta g_{ij}(\sig)+\frac{\delta f}{\delta \pi^{ij}(\sigma)}\delta \pi^{ij}(\sig)\rt],
\ee
where the variations $\delta g_{ij}$ and $\delta \pi^{ij}$ preserve the fall-off conditions. If two differentiable functions $f,f'$ have the corresponding Hamiltonian vector fields, their Poisson bracket is given by
\be
\{f,f'\}:=\int_{\cs}\rmd^3\sigma \lt[\frac{\delta f}{\delta g_{ij}(\sigma)}\frac{\delta f'}{\delta \pi^{ij}(\sig)}-\frac{\delta f}{\delta \pi^{ij}(\sigma)}\frac{\delta f'}{\delta g_{ij}(\sig)}\rt].\label{pbk}
\ee  
In the above two formulae and formulae in the following discussion, we adopt the following notation
\be
\int_\cs\cdots =\lim_{R\to\infty}\int_{\cs_R}\cdots\ . 
\ee

There are two remarks about the boundary conditions:
\begin{itemize}
    \item
    By \eqref{crexp}-\eqref{cexp}, and \eqref{intial}, we have $C=O\lt(\frac{\log r}{r^4}\rt)$ and $\frac{C_j}{C}=O\lt(r^{-1}\rt)$ in the Cartesian coordinates. Then  \eqref{hamdensity}, \eqref{relaph} and \eqref{velo} give $U_{j}=\frac{C_j}{P}=O\lt(r^{-1}\rt)$. 
    Recall \eqref{relacgeorho}, we have $C=-\sqrt{\operatorname{det}(g)}\rho\left(g^{ij} U_{i} U_{j}+1\right)$ in the dust space, and $g^{ij}U_iU_j=O(r^{-2})$. Thus,
    $\rho=O\lt(\frac{\log r}{r^4}\rt)$. Ref. \cite{andersson2019asymptotic} analyses the spacetime coupling to a perfect fluid with certain boundary conditions. The asymptotic flatness conditions would be violated if the energy density $\rho$ decays as $O(r^{-2})$. In our case, asymptotic flatness conditions, however, are preserved due to the faster fall-off behavior of $\rho$.

    \item As mentioned above, naively adding parity conditions to the expansions \eqref{asympq000} and \eqref{asympp000} does not work in the BK formalism.
    In the cases without introducing logarithmic terms to the fall-off conditions \eqref{convefalof}, we need to impose $C^{(2)}\neq 0$ as the boundary condition. The explicit expression of $C^{(2)}$ is in appendix B of \cite{hennADM}, where $C^{(2)}$ contains the contributions of the canonical variables with odd parity. It turns out that the sign of $C^{(2)}$ can flip under an antipodal map. Since $C\propto \sqrt{g}\rho$, $\rho$ can be smaller than $0$ in some regions and be greater than $0$ in other regions, whereas the BK formalism requires $\mathrm{sgn}(\rho)$ to be constant. Introducing logarithmic terms to the fall-off conditions with appropriate parity conditions of $h^{(\log)}_{ij}$ removes this problem.
    In Cartesian coordinates, the leading order of $C$ is $\frac{\log(r)}{r^4}C^{(\log)}$ with $C^{(1)}=0$. 
    The terms contribute to $\frac{\log(r)}{r^4}C^{(\log)}$ are the second derivative of $\frac{\log(r)}{r^{2}}h^{(\log)}_{ij}$. Henceforth, introducing even parity to $h^{(\log)}_{ij}$ ensures $C^{(\log)}$ has even parity. Its sign does not flip under the antipodal map.
    Nevertheless, relaxing the parity conditions give us more general results. 
    We will concentrate on the relaxed parity conditions cases in our discussions throughout this paper.

\end{itemize}

 \subsection{The Boundary Term of the Physical Hamiltonian}\label{sec:varphyham}
 In this subsection, we show that the boundary term that arises from the variation $\delta \mathbf{H}_{\text {phys }}$ is a differential under our boundary conditions. Thus, we can add a boundary Hamiltonian $\mathbf{H}_{\text {bdy }}$ whose variation cancels the boundary term.  $\mathbf{H}_{\text {bdy }}$ turns out to be the ADM mass. 

Recall \eqref{variaofph},  $\int_{\cs} \rmd^3\mathscr{V}\,g^{ij}N_i\delta C_j$ and $\int_{\cs} \rmd^3\mathscr{V}\,N\delta C$ may give boundary terms after integration by parts. Firstly,
\begin{equation}\label{variaofNDC}
\begin{aligned}
&\int_{\cs}\rmd^3\mathscr{V}\, g^{ij}N_i\delta C_j\\
=&\int_{\cs} \rmd^3\mathscr{V}\,
\lt(
\delta\pi^{ij}\mathcal{L}_{\vec{N}}g_{ij}
-\delta g_{ij}\mathcal{L}_{\vec{N}}\pi^{ij}
\rt)
-\lim_{R\to\infty}\oint_{S^2} \rmd^2 S\, \lt(2N^j\delta 
\lt(\pi^{rk}g_{jk}\rt)-N^r\pi^{ij}\delta g_{ij}\rt).
\end{aligned}
\end{equation}
The boundary conditions turns out to make the boundary terms in \eqref{variaofNDC} vanish at $R\to\infty$. Indeed, recall \eqref{dyns}, we have
\begin{equation}\label{exp_of_N_i}
     N_i=-\frac{C_i}{H_{\text {phys }}}.
\end{equation}
By \eqref{asymsphere1}-\eqref{asymsphere6}, we find that the boundary term in \eqref{variaofNDC} vanishes at $R\to \infty$ once $C_j$ decays faster than $H_{\text {phys}}$ in Cartesian coordinates. 
Eq. \eqref{convefalof} indicates $C_j$ decays as fast as $H_{\text {phys}}$ without additional conditions and their leading terms are $O\lt(r^{-3}\rt)$.
The approach in Ref. \cite{giesel2010manifestly} is to add an $\epsilon$-regulator to the leading order of $C_j$ such that $C_j$ decays as $C_j=O\lt(r^{-3-\epsilon}\rt)$. Here $\epsilon$ is a positive constant. Then 
\begin{equation}\label{eq:asy_of_ori_Nj}
    N_j=O\lt(r^{-\epsilon}\rt).
\end{equation}
with \eqref{eq:asy_of_ori_Nj}, the boundary term of \eqref{variaofNDC} vanishes. In this paper, we take a different approach by imposing \eqref{asyc1}-\eqref{intial}. Firstly, in spherical coordinates \eqref{exp_of_N_i} reads
\begin{equation}
\begin{aligned}
 N_r=-\frac{C_r}{H_{\text {phys }}},\qquad
  N_A=-\frac{C_A}{H_{\text {phys }}}.
    \end{aligned}
    \end{equation}
By \eqref{dyns},  \eqref{crexp}-\eqref{cAexp}, \eqref{asyc1}, and \eqref{intial},  we find the following asymptotic behaviors of $N_r, N_A$:
\begin{equation}
    N_r=O(r^{-1}),\qquad  N_A=O(1).
\end{equation}
Or for $N^r$ and $N^A$,
\begin{equation}\label{asydysh}
    N^r=O(r^{-1}),\qquad  N^A=O(r^{-2}).
\end{equation}
Then we analyze the boundary terms in \eqref{variaofNDC} by using \eqref{asymsphere1}-\eqref{asymsphere6}, and we obtain
\begin{equation}
\begin{aligned}
\oint_{S^2}\rmd^2 S\, N^j\delta 
\lt(\pi^{rk}g_{jk}\rt)
=&\oint_{S^2}\rmd^2 S\,
\lt[ 
    N^r\delta 
\lt(\pi^{rr}g_{rr}+\pi^{rD}g_{rD}\rt)
+N^C\delta 
(\pi^{rr}g_{Cr}+\pi^{rD}g_{CD})\rt]\\
=&O\lt(r^{-1}\rt),
\end{aligned}
\end{equation}
and
\begin{equation}
    \begin{aligned}
    \oint_{S^2}\rmd^2 S\,N^r\pi^{jk}\delta g_{jk}
=&\oint_{S^2}\rmd^2 S\,N^r\lt(\pi^{rr}\delta g_{rr}
+2\pi^{rA}\delta g_{rA}
+\pi^{AB}\delta g_{AB}\rt)\\
=&O\lt(r^{-2}\rt).
    \end{aligned}
\end{equation}
When taking the limit $R\to\infty$, both terms vanish:
\begin{equation}
    \lim_{R\to\infty}\oint_{S^2} \rmd^2 S\, \lt(2N^j\delta 
\lt(\pi^{rk}g_{jk}\rt)-N^r\pi^{ij}\delta g_{ij}\rt)=0.
\end{equation}

On the other hand,
\begin{equation}
   \int_{\cs} \rmd^3\mathscr{V}\,N\delta C
    =\int_{\cs} \rmd^3\mathscr{V}\,\lt(\mathscr{A}_{ij}\delta\pi^{ij}
    -\mathscr{B}^{ij}\delta g_{ij}\rt)
    +\mathcal{K}_N.
\end{equation}
The explicit expressions of $\mathscr{A}_{ij}$ and $\mathscr{B}^{ij}$ can be found in \cite{giesel2010manifestly}:
\begin{equation}\label{eq:del_N_C}
\begin{aligned}
  \mathscr{A}_{ij}=&2 N g^{-\frac{1}{2}}\left(\pi_{i j}-\frac{1}{2} g_{i j} \pi\right),\\
    \mathscr{B}^{ij}=&-N g^{\frac{1}{2}}\left(R^{i j}-\frac{1}{2} g^{i j} R\right)+\frac{1}{2} N g^{-\frac{1}{2}}g^{ij}\left(\pi_{m n} \pi^{m n}-\frac{1}{2} \pi^2\right)\\
    &-2 N g^{-\frac{1}{2}}\left(\pi^{i m} \pi_m^j-\frac{1}{2} \pi^{i j} \pi\right)+g^{\frac{1}{2}}\left(D^iD^jN-g^{i j} D^mD_mN\right).\\
    \end{aligned}
    \end{equation}
Here $N$ is the dynamical lapse function, rather than the Lagrange multiplier in the original ADM formalism. The boundary term $\mathcal{K}_N$ reads (see appendix \ref{bunofc} for more details):
\begin{equation}\label{NdCboun}
\begin{aligned}
\mathcal{K}_N
=&\lim_{R\to\infty}\oint_{S^2}\rmd^2 S\,\sqrt{\gamma}\lt(-\frac{1}{\lambda} \gamma^{AB}\lt(D_{r} N\rt) \delta \gamma_{AB}
+\frac{\lambda^A}{\lambda} \gamma^{BC}\lt(D_{A} N\rt) \delta  \gamma_{BC}\rt)\\
&-2\lim_{R\to\infty}\oint_{S^2}\rmd^2 S\,\lt(N\delta K
+N\delta\lt(\gamma_{AC}\rt)\gamma^{AB}  K^{C}_B\rt)
\\
&+\lim_{R\to\infty}\oint_{S^2} \rmd^2 S\,\sqrt{\gamma}\frac{N}{\lambda}
\delta
\lt[
\lt(
    -\frac{\lambda^{A}}{\lambda}\left(\partial_{A} \lambda+K_{A C} \lambda^{C}\right)+D_{A} \lambda^{A}
\rt)
-D_{A} \delta\lambda^{A}
\rt]\\
&-\lim_{R\to\infty}\oint_{S^2} \rmd^2 S\,\sqrt{\gamma}N\frac{\lambda^A}{\lambda}\lt(
    \frac{1}{2}\gamma^{BC}D_A\delta \gamma_{BC}-\delta \lt(\frac{\lambda^{B}}{\lambda} K_{B A}\rt)
\rt)\\
&+\lim_{R\to\infty}\oint_{S^2}\rmd^2 S\,\sqrt{\gamma}N\frac{\lambda^A}{\lambda}
\delta
\lt(
    \frac{1}{\lambda}\left(\partial_{A} \lambda+K_{A B} \lambda^{B}\right)
\rt)\\
&+\lim_{R\to\infty}\oint_{S^2}\rmd^2 S\,\lambda\sqrt{\gamma}\lt(-N\frac{\lambda^A\lambda^B}{\lambda^5}\delta K_{AB}
+NK_{AB}\frac{\lambda^A\lambda^B}{\lambda^4}\frac{\delta\lambda}{\lambda^2}\rt)\\
&+\lim_{R\to\infty}\oint_{S^2}\rmd^2 S\,\sqrt{\gamma}\lt(
- ND^{A} \lt(\frac{1}{\lambda}\rt)\delta \lambda_A
-\frac{\lambda^A}{\lambda}  \gamma^{BC}\lt(D_{C} N\rt) \delta \gamma_{AB}\rt).
\end{aligned}
\end{equation}
Recall \eqref{dynl}, we find
\begin{equation}
    N=\pm\sqrt{1+g^{ij}N_iN_.}.
\end{equation}
Here "$+$" sign is for the phantom dust and the "$-$" sign is for the physical dust. In Cartesian coordinates, the asymptotic behaviors of $N_{j}$ \eqref{asympq} read
\begin{equation}
    N_j=O\lt(r^{-1}\rt).
\end{equation}
It turns out that\footnote{As a comparison, $N$ decays as $N=1+O\lt(r^{-2\epsilon}\rt)$ in \cite{giesel2010manifestly}.}
\begin{equation}\label{Nexp}
\begin{aligned}
N=\pm 1+O(r^{-2}).
\end{aligned}
\end{equation} 
Here $\gamma_{AB}$ is the 2-metric on the $S^2$ in the "$2+1$" decomposition \cite{hennADM}, which satisfies:
\begin{equation}
    \begin{aligned}
&\gamma_{AB}=g_{AB},\\
&\gamma^{AB}=g^{AB}-\frac{\lambda^A\lambda^B}{\lambda^2}.
    \end{aligned}
\end{equation}
$D_A$ is the covariant derivative compatible with $\gamma_{AB}$, and the angular indexes $A$ and $B$ are lowered and raised  by $\gamma_{AB}$ and its inverse $\gamma^{AB}$. 
By \eqref{asymsphere1}-\eqref{asymsphere6} and \eqref{Nexp}, the boundary term $\ck_N$ to be a differential
\begin{equation}
    \begin{aligned}
\mathcal{K}_N=&-\delta \mathbf{H}_{\text {bdy }}\\
=&\mp\oint_{S^2} \rmd^2 S\,
2\sqrt{\bar{\gamma}}\delta
(
    2\bar{\lambda}
    +\bar{D}_A\bar{\lambda}^A
)
\\
=&\mp4\oint_{S^2} \rmd^2 S\,
\sqrt{\bar{\gamma}}\delta\bar{\lambda}.
\\
    \end{aligned}
\end{equation}
The last step is because the integration region is closed. Here $\bar{D}_A$ is the covariant derivative compatible with $\bar{\gamma}_{AB}$, and the angular indices $A$ and $B$ are  lowered and raised by $\bar{\gamma}_{AB}$ and its inverse $\bar{\gamma}^{AB}$. We can add the boundary term $\mathbf{H}_{\text {bdy }}$
to the physical Hamiltonian to make its variation well-defined.
Where
\begin{equation}\label{eq: ADM_mass}
    \mathbf{H}_{\text {bdy }}=\pm4\oint_{S^2} \rmd^2 S\,
\sqrt{\bar{\gamma}}\bar{\lambda},
\end{equation}
and it is interpreted as the ADM mass. Here the $'+'$ is sign for the phantom dust, while the $'-'$ sign is for the physical dust. These signs are consistent with the convention that the physical time evolution corresponding to the phantom dust is future toward, whereas the physical time evolution corresponding to the physical dust is past toward, see \cite{giesel2010manifestly}.  

$\mathbf{H}_{\text {bdy }}$ is conserved under the physical time evolution, i.e., $\frac{\rmd \mathbf{H}_{\text {bdy }}}{\rmd\tau}=0$. Note that $\sqrt{\bar\gamma}$ is fiducial in \eqref{eq: ADM_mass}, we only need to compute the physical time derivative of $\bar \lambda$. We firstly compute the physical time derivative of $g_{rr}$. Since asymptotically flat cases necessitate the consideration of boundary terms, the physical time derivative formula \eqref{eq:phy_time_dire} needs to be modified as
\begin{equation}
    \frac{\rmd g_{rr}}{\rmd\tau}=\lt\{g_{rr},\mathbf{H}_{\rm {phys }}+\mathbf{H}_{\rm {bdy }}\rt\},
\end{equation}
and the result is
\begin{equation}\label{eq:time_deri_of_grr}
  \frac{\rmd g_{rr}}{\rmd\tau}= N g^{-\frac{1}{2}}\left(\pi_{rr}-\frac{1}{2} g_{rr} \pi\right)+ \mathcal{L}_{\vec{N}}g_{rr}=O\lt(r^{-2}\rt)
\end{equation}
with the fall-off conditions and boundary conditions introduced before. On the other hand, $\bar \lambda$ is proportional to the coefficient of the $O\lt(r^{-1}\rt)$ term in $g_{rr}$, and its physical time derivative is induced by expanding \eqref{eq:time_deri_of_grr} to the boundary. Therefore,
\begin{equation}
    \frac{\rmd\bar \lambda}{\rmd\tau}=0,
\end{equation}
and thus
\begin{equation}
    \frac{\rmd \mathbf{H}_{\text {bdy }}}{\rmd\tau}=0.
\end{equation}

The similar boundary Hamiltonian is obtained by the approach with the $\epsilon$-regulator in \cite{giesel2010manifestly}. However, as an advantage of our approach, the boundary conditions \eqref{asyc1}-\eqref{intial} result in more general symmetry charges, which turn out to relate to supertranslations on the boundary. We will show this in the next section.

\section{Boundary-preserving Symmetry Charges and Boundary Charges}\label{sec:bounchg}

In this section, we firstly define the boundary-preserving symmetry charges. Then these symmetry charges are explicitly constructed as phase space functions. The expressions of the symmetry charges contain both bulk and boundary terms. The boundary terms are constructed such that the variations of the functions are well-defined on the phase space.

\begin{definition}\label{defofchar}
 $G$ is a boundary-preserving symmetry charge if
\begin{enumerate}[label=(\roman*)]
  \item $G$ commutes with the physical Hamiltonian:
    \begin{equation}
    \lt\{G,\mathbf{H}_{\rm {phys }}+\mathbf{H}_{\rm {bdy }}\rt\}=0;
    \end{equation}
    \item The Hamiltonian flow of $G$ preserves the fall-off and boundary conditions in Section \ref{sec:phasym}.
\end{enumerate} 
\end{definition}

We first introduce the vector fields used to construct the boundary-preserving charges, which are denoted as $\xi^\mu:=\lt(\xi,\xi^j\rt)$. As $r\to\infty$, $\xi^\mu$ falls-off as:
\begin{equation}\label{asyvec}
\xi= f+O\left(r^{-1}\right), \quad \xi^{r}=W+O\left(r^{-1}\right), \quad \xi^{A}=Y^{A}+\frac{1}{r} I^{A}+O\left(r^{-2}\right),
\end{equation}
For convenience, we require that the expansions of $\lt(\xi,\vec{\xi}\rt)$ only contain integer order of $r^{-1}$.
$f$, $W$, and $I^A$ are functions and vector field on $S^2$, $Y^A$ is the killing vector field on the $S^2$.
Their geometric meaning will be discussed shortly.
The boundary-preserving symmetry $G\lt(\xi, \vec{\xi}\rt )$ is constructed as
\begin{equation}\label{expofchar}
G\lt(\xi,\vec{\xi}\rt):=J\lt(\xi,\vec{\xi}\rt)+\mathcal{B}\lt(\xi,\vec{\xi}\rt),
\end{equation}
where $J\lt(\xi,\vec{\xi}\rt)$ is the bulk term
\begin{equation}\label{eq:fin-J}
    J\lt(\xi,\vec{\xi}\rt)=\mathcal{T}\lt(\xi\rt)+\mathcal{P}\lt(\vec{\xi}\rt),
\end{equation}
with
\begin{equation}
\begin{aligned}
\mathcal{T}\lt(\xi\rt)=\int_{\cs} \rmd^3\mathscr{V}\,\xi H_{\text{phys}},\\
\mathcal{P}\lt(\vec{\xi}\rt)=\int_{\cs} \rmd^3\mathscr{V}\, \xi^i C_i.
\end{aligned}
\end{equation}
$\mathcal{B}\lt(\xi,\vec{\xi}\rt)$ is the boundary term, whose variation should cancel the boundary term from $\delta J\lt(\xi,\vec{\xi}\rt)$, where
\begin{equation}\label{varia-H}
\begin{aligned}
\delta J\lt(\xi,\vec{\xi}\rt)=&\int_{\cs} \rmd^3\mathscr{V}\,\left[\tilde{\xi}\delta C
+\left(\xi^{r}+\tilde{\xi}^{r}\right)\delta C_{r}
+\lt(\xi^{A}+\tilde{\xi}^{A}\rt)\delta C_{A}
+\frac{\xi}{2} H_{\text{phys}} N^iN^j\delta g_{ij}\right]\\
=&\int_{\cs} \rmd^3\mathscr{V}\,\lt[\mathscr{G}_{ij}\delta \pi^{ij}-\mathscr{F}^{ij}\delta g_{ij}\rt]
+\mathcal{K}\lt(\xi,\vec{\xi}\rt).
\end{aligned}
\end{equation}
In \eqref{varia-H}, $ \tilde{\xi}=\xi N$, $\tilde{\xi}^i=\xi N^i$, and $\mathcal{K}\lt(\xi,\vec{\xi}\rt)$ is the boundary term. 
The bulk terms are
\begin{equation}\label{eq:bul_term}
\begin{aligned}
    \mathscr{G}_{ij}=&2 \tilde{\xi} g^{-\frac{1}{2}}\left(\pi_{i j}-\frac{1}{2} g_{i j} \pi\right)+\mathcal{L}_{\vec{\xi}} g_{i j}+\mathcal{L}_{
\vec{\tilde\xi}} g_{i j},\\
    \mathscr{F}^{ij}=&-\tilde{\xi} g^{\frac{1}{2}}\left(R^{i j}-\frac{1}{2} g^{i j} R\right)+\frac{1}{2} \tilde{\xi} g^{-\frac{1}{2}}g^{ij}\left(\pi_{m n} \pi^{m n}-\frac{1}{2} \pi^2\right)\\
    &-2 \tilde{\xi} g^{-\frac{1}{2}}\left(\pi^{i m} \pi_m^j-\frac{1}{2} \pi^{i j} \pi\right)+g^{\frac{1}{2}}\left(D^iD^j\tilde{\xi}-g^{i j} D^mD_m\tilde{\xi}\right)\\
&+\mathcal{L}_{\vec{\xi}} \pi^{i j}+\mathcal{L}_{\vec{\tilde\xi}} \pi^{i j}+\frac{\xi}{2} H_{\text{phys}} N^iN^j.\\
    \end{aligned}
    \end{equation}
Unlike the typical general relativity, the $\tilde\xi$ and the $\tilde{\xi}^j$ in \eqref{eq:bul_term} are dynamical. 
The terms $\mathcal{L}_{\vec{\tilde\xi}} g_{i j}$ and $\mathcal{L}_{\vec{\tilde\xi}} \pi^{i j}$ are contributed by $\xi\delta H_{\text{phys}}$.
The term $\frac{\xi}{2} N^iN^j\delta g_{ij}$ is the correction due to the dust field. 

We discuss the geometric meaning of the parameters in \eqref{asyvec}. The boundary-preserving symmetry charges generate the diffeomorphisms in the bulk and the asymptotic symmetry transformations at the boundary. The asymptotic symmetry transformations at the boundary preserve the fall-off conditions and the boundary conditions. The charges are classified into the following four types:  
\begin{equation}
    \begin{aligned}
    &G\lt(f\rt):=\int_{\cs} \rmd^3\mathscr{V}\,f H_{\text{phys}}+\mathcal{B}\lt(f\rt),\\
    &G\lt(W\rt):=\int_{\cs} \rmd^3\mathscr{V}\,WC_r+\mathcal{B}\lt(W\rt),\\
    &G\lt(\vec{Y}\rt):=\int_{\cs} \rmd^3\mathscr{V}\,Y^AC_A+\mathcal{B}\lt(\vec{Y}\rt),\\
    &G\lt(\vec{I}\rt):=\int_{\cs} \rmd^3\mathscr{V}\,\frac{I^A}{r}C_A+\mathcal{B}\lt(\vec{I}\rt).
    \end{aligned}
\end{equation}
$\mathcal{B}\lt(f\rt), \mathcal{B}\lt(W\rt), \mathcal{B}\lt(\vec{Y}\rt), \mathcal{B}\lt(\vec{I}\rt)$ are the boundary charges. Physically, these four types of charges generate the following four classes of the asymptotic symmetry transformations at the boundary:
\begin{itemize}
    \item  $G\lt(\vec{Y}\rt)$ : spatial  rotations.

    \item $G\lt(f\rt)$ : supertranslations in the physical time direction.
    
    \item $G\lt(W\rt)$ : The radius component of the supertranslations in the spatial direction.
    
    \item $G\lt(\vec{I}\rt)$ : The angular components of the supertranslations in the spatial direction.
\end{itemize}
When $f=1$, $G\lt(f\rt)$ reduces to the physical Hamiltonian with the ADM mass as the boundary charge, and it generates the physical time translation. As is shown in \cite{hennADM}, when
    \begin{equation}
        \begin{aligned}
        &W=W_T=T_A\sin\sigma_\theta\cos\sigma_\varphi+T_Y\sin\sigma_\theta\sin\sigma_\varphi+T_Z\cos\sigma_\theta,\\
        &I^A=I^A_T=\bar{D}^AW_T,
        \end{aligned}
    \end{equation}
where $T_X, T_Y, T_Z$ are constants and other components of $\lt(\xi, \vec{\xi}\rt)$ are $0$,
$G\lt(\xi, \vec{\xi}\rt)$ generates spatial translation  since
    \begin{equation}
        W_T\frac{\partial}{\partial r}+\frac{I^A_T}{r}\frac{\partial}{\partial \sigma^A}
        =T_X\frac{\partial}{\partial \sigma_X}+T_Y\frac{\partial}{\partial \sigma_Y}+T_Z\frac{\partial}{\partial \sigma_Z}.
    \end{equation}

In order to obtain the boundary charge $\mathcal{B}\lt(\xi,\vec{\xi}\rt)$, we compute $\mathcal{K}\lt(\xi,\vec{\xi}\rt)$ (see appendix \ref{bunofc} for more details) 
\begin{equation}\label{Kexp}
\begin{aligned}
\mathcal{K}\lt(\xi,\vec{\xi}\rt)
=&\lim_{R\to \infty}\oint_{S^2}\rmd^2 S\,\lt(-2u^i
\delta \pi_{i}^{r}
+
u^r
\pi^{ij}\delta g_{ij}\rt)\\
&
+\lim_{R\to\infty}\oint_{S^2}\rmd^2 S\,\sqrt{\gamma}\lt(-\frac{1}{\lambda} \gamma^{AB}\lt(D_{r} \tilde{\xi}\rt) \delta \gamma_{AB}
+\frac{\lambda^A}{\lambda} \gamma^{BC}\lt(D_{A} \tilde{\xi}\rt) \delta  \gamma_{BC}
\rt)
\\
&
-\lim_{R\to\infty}\oint_{S^2}\rmd^2 S\,\sqrt{\gamma}\lt(
2\tilde{\xi}\delta K
+\tilde{\xi}\delta\lt(\gamma_{AC}\rt)\gamma^{AB}  K^{C}_B\rt)
\\
&+\lim_{R\to\infty}\oint_{S^2}\rmd^2 S\,\sqrt{\gamma}\frac{\tilde{\xi}}{\lambda}
\lt[
\delta
\lt(
    -\frac{\lambda^{A}}{\lambda}\left(\partial_{A} \lambda+K_{A C} \lambda^{C}\right)+D_{A} \lambda^{A}
\rt)
-D_{A} \delta\lambda^{A}
\rt]\\
&-\lim_{R\to\infty}\oint_{S^2}\rmd^2 S\,\sqrt{\gamma}\tilde{\xi}\frac{\lambda^A}{\lambda}\lt(
    \frac{1}{2}\gamma^{BC}D_A\delta \gamma_{BC}-\delta \lt(\frac{\lambda^{B}}{\lambda} K_{B A}\rt)
\rt)\\
&+\lim_{R\to\infty}\oint_{S^2}\rmd^2 S\,\sqrt{\gamma}\tilde{\xi}\frac{\lambda^A}{\lambda}
\delta
\lt(
    \frac{1}{\lambda}\left(\partial_{A} \lambda+K_{A B} \lambda^{B}\right)
\rt)\\
&+\lim_{R\to\infty}\oint_{S^2}\rmd^2 S\,\lambda\sqrt{\gamma}\lt(-\tilde{\xi}\frac{\lambda^A\lambda^B}{\lambda^5}\delta K_{AB}
+\tilde{\xi}K_{AB}\frac{\lambda^A\lambda^B}{\lambda^4}\frac{\delta\lambda}{\lambda^2}\rt)\\
&+\lim_{R\to\infty}\oint_{S^2}\rmd^2 S\,\sqrt{\gamma}\lt(
- \tilde{\xi}D^{A} \lt(\frac{1}{\lambda}\rt)\delta \lambda_A
-\frac{\lambda^A}{\lambda}  \gamma^{BC}\lt(D_{C} \tilde{\xi}\rt) \delta \gamma_{AB}\rt),\\
\end{aligned}
\end{equation}
with $u^i=\xi^i+\tilde \xi^i$.
From \eqref{asymsphere1}-\eqref{asymsphere6}, and \eqref{asydysh},
\begin{equation}
    \begin{aligned}
    \mathcal{K}\lt(\xi,\vec{\xi}\rt)
=&\lim_{R\to\infty}\oint_{S^2}\rmd^2 S\,
\lt(
    - 2RY^A\delta\bar \pi^{r}_A
-2Y^A\log (R)\delta\pi^{(\log)r}_A
-2Y^A\delta\lt(\bar\pi^{rr} \bar{\lambda}_A
+\pi^{r(2)}_A
+\bar \pi^{rB}\bar{h}_{AB}
\rt)
\right.\\
&\lt.
-2I^A\delta\bar \pi^{r}_A
-2W\delta\bar \pi^{rr}
\mp2\sqrt{\bar{\gamma}}f\delta
\lt(
     2\bar{\lambda}
    +\bar{D}_A\bar{\lambda}^A
\rt)
\rt).
    \end{aligned}
\end{equation}
Similar to the discussion in Section \ref{sec:varphyham}, we want to show $\mathcal{K}$ as a differential of certain boundary charge $\cb$. There are two terms in $\ck$ that would be divergent without the suitable boundary condition (see appendix B of \cite{beig1987poincare} for  similar results): As $R\to\infty$, the term $2R\oint_{S^2}\rmd^2 S\,Y^A\delta\bar \pi^{r}_A$ is generically divergent linearly but vanishes by imposing the boundary condition $C^{(0)}_A=0$ and the killing equation of $Y^A$. Note that $Y^A$ has odd parity and $\pi^{(\log)r}_A =\bar{\gamma}_{AB}\pi^{(\log)rB}$, the term $-2\oint_{S^2}\rmd^2 S\,Y^A\log (R)\delta\pi^{(\log)r}_A$ is generically divergent logarithmically but vanishes by \eqref{paroflogpi}. Finally,
\begin{equation}
    \begin{aligned}
    \mathcal{K}\lt(\xi,\vec{\xi}\rt)
=&\oint_{S^2}\rmd^2 S\,
\lt(
-2Y^A\delta\lt(\bar\pi^{rr} \bar{\lambda}_A
+\pi^{r(2)}_A
+\bar \pi^{rB}\bar{h}_{AB}
\rt)
-2I^A\delta\bar \pi^{r}_A\right.\\
&\lt.-2W\delta\bar \pi^{rr}
\mp2\sqrt{\bar{\gamma}}f\delta
\lt(
    2\bar{\lambda}
    +\bar{D}_A\bar{\lambda}^A
\rt)
\rt).
    \end{aligned}
\end{equation}
In order to fulfill $\mathcal{K}\lt(\xi,\vec{\xi}\rt)= -\delta\mathcal{B}\lt(\xi,\vec{\xi}\rt)$,
\begin{equation}\label{eq:fin-B}
    \begin{aligned}
   \mathcal{B} \lt(\xi,\vec{\xi}\rt)
=&\oint_{S^2}\rmd^2 S\,Y^A\left(4\lt(\tilde {\bar k}_{A B}-\bar\lambda\bar \gamma_{A B}\rt) \bar{\pi}^{r B}
+2\bar{\gamma}_{A B} \pi^{(2) r B}+\bar{\lambda}_{A} \lt(\bar p
+2\bar{\pi}^{B}_{B}\rt)\right)\\
&+\oint_{S^2}\rmd^2 S\,\left(2 I^{A} \bar{\gamma}_{A B} \bar{\pi}^{r B}+W  \lt(\bar p+2\bar{\pi}^{A}_{A}\rt)\pm2\sqrt{\bar{\gamma}} f \left(2 \bar{\lambda}+\bar{D}_{A} \bar{\lambda}^{A}\right)
\right),\\
=&\mathcal{B}\lt(\xi\rt)+\mathcal{B}\lt(\vec{\xi}\rt),
    \end{aligned}
\end{equation}
where
\begin{equation}\label{eq:se_of_boun}
    \begin{aligned}
        &\mathcal{B}\lt(\xi\rt)=\mathcal{B}\lt(f\rt),\\
        &\mathcal{B}\lt(\vec{\xi}\rt)=\mathcal{B}\lt(\vec{Y}\rt)+\mathcal{B}\lt(\vec{I}\rt)+\mathcal{B}\lt(W\rt).
    \end{aligned}
\end{equation}
Here $\mathcal{B}\lt(f\rt)$, $\mathcal{B}\lt(\vec{Y}\rt)$, $\mathcal{B}\lt(\vec{I}\rt)$, and $\mathcal{B}\lt(W\rt)$ are the four types of boundary charges:
\begin{equation}\label{eq:4tyboun}
\begin{aligned}
\mathcal{B}\lt(\vec{Y}\rt)=&\oint_{S^2}\rmd^2 S\,Y^A\left(4\lt(\tilde {\bar k}_{A B}-\bar\lambda\bar \gamma_{A B}\rt) \bar{\pi}^{r B}
+2\bar{\gamma}_{A B} \pi^{(2) r B}+\bar{\lambda}_{A} \lt(\bar p
+2\bar{\pi}^{B}_{B}\rt)\right),\\
\mathcal{B}\lt(\vec{I}\rt)=&\oint_{S^2}\rmd^2 S\, 2I^{A} \bar{\gamma}_{A B} \bar{\pi}^{rB},\\
\mathcal{B}\lt(W\rt)=&\oint_{S^2}\rmd^2 S\, W  \lt(\bar p+2\bar{\pi}^{A}_{A}\rt),\\
\mathcal{B}\lt(f\rt)=&\pm\oint_{S^2}\rmd^2 S\, 2\sqrt{\bar{\gamma}} f \left(2 \bar{\lambda}+\bar{D}_{A} \bar{\lambda}^{A}\right).
\end{aligned}
\end{equation}
The boundary charges \eqref{eq:fin-B} are finite. 
By \eqref{crexp}-\eqref{cexp}, \eqref{asyc1}, \eqref{intial}, and \eqref{asyvec},  $J(\xi,\vec{\xi})$ are also finite.
Thus, the boundary-preserving charges $G(\xi,\vec{\xi})=J(\xi,\vec{\xi})+\mathcal{B}(\xi,\vec{\xi})$ are finite on the phase space.
Additionally, by adding the boundary terms $\mathcal{B}(\xi,\vec{\xi})$, the variation of $ G(\xi,\vec{\xi})$ is well defined. {Furthermore, similar to the discussions at the end of Subsection \ref{sec:varphyham}, one can show the boundary charges in \eqref{eq:fin-B} are conserved, and the detailed discussion is in Appendix \ref{app:con_boun_char}.}

Before ending this section, we would like to compare our results with the results in \cite{hennADM}. Firstly, our work is conceptually different from \cite{hennADM}. The result in \cite{hennADM} bases on the traditional formulation of general relativity, where the bulk is a totally constrained system,
thus the symmetry charges vanish in the bulk and only have non-vanished terms at the boundary.
The result in our case, however, is based on the reduced phase space formulation. The gravity-dust system on the reduced phase space is not a constraint system any longer and has a physical Hamiltonian. Therefore, our symmetry charges have both non-vanishing bulk and boundary terms. The transformations generated by these charges are symmetries, but not gauge symmetries.
Secondly, although some boundary terms look similar to the ones in \cite{hennADM}, there are several additional boundary terms in our results, including $2\oint_{S^2}\rmd^2 S\, W\bar{\pi}^{A}_{A}$ and $4\oint_{S^2}\rmd^2 S\,Y^A\tilde {\bar k}_{A B}\bar{\pi}^{r B}$. These terms vanish when the parity condition in \cite{hennADM} is imposed. Finally, \cite{hennADM} considers the symmetry charge corresponding to the boundary boost, which requires $\bar{\lambda}_A=0$. However, we do not take into account the boundary boost in $G(\xi,\vec{\xi})$, so we do not need to introduce $\bar{\lambda}_A=0$ here.

\section{Algebra of the Boundary-preserving Symmetry Charges}\label{sec:ABSG}

In this section, we obtain the algebra of the boundary-preserving symmetry charges by computing their Poisson brackets. Subsection \ref{subsec:gfa} presents some detailed computations of the Poisson bracket between a pair of $G(\xi,\vec{\xi})$ for $\xi,\vec{\xi}$ satisfying the boundary condition \eqref{asyvec}. In Subsection \ref{subsec:efa}, we show that $G(\xi,\vec{\xi})$ form a closed Poisson algebra up to a central charge, and we also discuss the relation between this algebra and BMS algebra. Subsection \ref{subsec: pbd} checks that the boundary conditions \eqref{intial} are preserved by the transformations generated by $G(\xi,\vec{\xi})$.

\subsection{Computing Poisson Brackets between the Boundary-preserving Symmetry Charges}\label{subsec:gfa}

We first derive the following useful relations. In these derivations, the vector field $(\xi,\vec{\xi})$ satisfies \eqref{asyvec}. The boundary charge $\mathcal{B}(\vec{\xi})$ is given by the second equation of \eqref{eq:se_of_boun} and $\tilde{\mathcal{B}}\lt(\xi\rt)=\pm\mathcal{B}\lt(\xi\rt)=\oint_{S^2}\rmd^2 S\, 2\sqrt{\bar{\gamma}} f \left(2 \bar{\lambda}+\bar{D}_{A} \bar{\lambda}^{A}\right)$, where $\mathcal{B}\lt(\xi\rt)$ is given by the first equation of \eqref{eq:se_of_boun}. Firstly, 
\begin{equation}\label{poi1}
    \begin{aligned}
    &\lt\{\int_{\cs}\rmd^3\mathscr{V}\,\xi^i_1C_i+\mathcal{B}\lt({\vec{\xi}_1}\rt),\int_{\cs}\rmd^3\mathscr{V}\,\xi^i_2C_i+\mathcal{B}\lt({\vec{\xi}_2}\rt)\rt\}\\
=&\int_{\cs}\rmd^3\mathscr{V}
\lt[
-\mathcal{L}_{\vec{\xi_1}} \pi^{i j}
\mathcal{L}_{\vec{\xi_2}} g_{i j}+\mathcal{L}_{\vec{\xi_1}} g_{i j}
\mathcal{L}_{\vec{\xi_2}} \pi^{i j}
\rt]\\
=&\int_{\cs}\rmd^3\mathscr{V}\,\lt[\vec{\xi_1},\vec{\xi_2}\rt]^j C_j
 +2\lim_{R\to\infty}\int_{S^2}\rmd^2 S\,\lt[\vec{\xi_1},\vec{\xi_2}\rt]^j\pi^{r}_j,
    \end{aligned}
\end{equation}
Secondly,
\begin{equation}\label{poi2}
    \begin{aligned}
    &\lt\{\int_{\cs}\rmd^3\mathscr{V}\,\xi_1C+\tilde{\mathcal{B}}\lt({\xi_1}\rt),\int_{\cs}\rmd^3\mathscr{V}\,\xi_2C+\tilde{\mathcal{B}}\lt(\xi_2\rt)\rt\}\\
=&\int_{\cs}\rmd^3\mathscr{V}\,
\lt[
    \xi_1 g^{\frac{1}{2}}\left(R^{i j}-\frac{1}{2} g^{i j} R\right)
-\frac{1}{2} \xi_1 g^{ij} g^{-\frac{1}{2}}\left(\pi_{m n} \pi^{m n}-\frac{1}{2} \pi^{2}\right)\rt.\\
&\lt.+2 \xi_1 g^{-\frac{1}{2}}\left(\pi^{i m} \pi_{m}^{j}-\frac{1}{2} \pi^{i j} \pi\right)
-g^{\frac{1}{2}}\left(D^iD^j\xi_1-g^{i j}D_mD^m \xi_1\right)\rt]\\
&\times
   \xi_2 g^{-\frac{1}{2}}\left(\pi_{i j}-\frac{1}{2} g_{i j} \pi\right)-\lt(1\longleftrightarrow2\rt)\\
    =&\int_{\cs}\rmd^3\mathscr{V}\,
    \lt(\xi_1D^j\xi_2-\xi_2D^j\xi_1\rt)C_j
+2\lim_{R\to \infty}\oint_{S^2}\rmd^2 S\, \lt(\xi_1D^j\xi_2-\xi_2D^j\xi_1\rt)\pi^r_j.
    \end{aligned}
\end{equation}
Thirdly,
\begin{equation}\label{poi3}
    \begin{aligned}
       &\lt\{\int_{\cs}\rmd^3\mathscr{V}\,\xi^i_1C_i+\mathcal{B}\lt({\vec{\xi}_1}\rt),\int_{\cs}\rmd^3\mathscr{V}\,\xi_2C+\tilde{\mathcal{B}}\lt(\xi_2\rt)\rt\}\\
=& \int_{\cs}\rmd^3\mathscr{V}\,
\lt\{
    -2\mathcal{L}_{\vec{\xi}_1} \pi^{i j}
     \xi_2 g^{-\frac{1}{2}}\left(\pi_{i j}-\frac{1}{2} g_{i j} \pi\right)-
    \mathcal{L}_{\vec{\xi}_1} g_{i j}
    \lt[\xi_2 g^{\frac{1}{2}}\left(R^{i j}- g^{i j} R\right)
-\frac{1}{2} \xi_2 g^{ij}C\rt.\rt.\\
&\lt.\lt.+\,2 \xi_2 g^{-\frac{1}{2}}\left(\pi^{i m} \pi_{m}^{j}-\frac{1}{2} \pi^{i j} \pi\right)
-g^{\frac{1}{2}}\left(D^iD^j\xi_2-g^{i j}D_mD^m \xi_2\right)\rt]
\rt\}\\
=&\int_{\cs}\rmd^3\mathscr{V}\,
\xi^i_1\partial_i\lt(\xi_2\rt)C
-\delta_{\vec{\xi}_1}\tilde{\mathcal{B}}\lt(\xi_2\rt)
-\lim_{R\to \infty}\oint_{S^2}\rmd^2 S\,\xi_1^r\xi_2C.
    \end{aligned}
\end{equation}
Before continuing, we introduce the notation $\delta_{\vec{\xi}}$. For a vector field $\vec{\xi}$ in \eqref{asyvec}, $\delta_{\vec{\xi}}g_{ij}$ and $\delta_{\vec{\xi}}\pi^{ij}$ are given by
\begin{equation}
    \begin{aligned}
        \delta_{\vec{\xi}}g_{ij}:=&\left\{g_{ij},\int_{\cs}\rmd^3\mathscr{V}\,\xi^i_1C_i+\mathcal{B}\lt({\vec{\xi}_1}\rt)\rt\}=\mathcal{L}_{\vec{\xi}}g_{ij}
    \end{aligned}
\end{equation}
and
\begin{equation}
    \begin{aligned}
        \delta_{\vec{\xi}}\pi^{ij}:=&\left\{\pi^{ij},\int_{\cs}\rmd^3\mathscr{V}\,\xi^i_1C_i+\mathcal{B}\lt({\vec{\xi}_1}\rt)\rt\}
        =\mathcal{L}_{\vec{\xi}}\pi^{ij}
    \end{aligned}
\end{equation}
respectively.
We only consider $\delta_{\vec{\xi}}$ acting on phase space linear functions $f\lt[g_{ij},\pi^{ij}\rt]$,
\begin{equation}
    \delta_{\vec{\xi}}f\lt[g_{ij},\pi^{ij}\rt]:=f\lt[\delta_{\vec{\xi}}g_{ij},\delta_{\vec{\xi}}\pi^{ij}\rt].
\end{equation}


The boundary-preserving symmetry charges $G(\xi,\vec{\xi})$ can split into the temporal part $G(\xi)$ and the spatial part $G(\vec{\xi})$:
\begin{equation}
  G\lt(\xi,\vec{\xi}\rt)=G\lt(\xi\rt)+G\lt(\vec{\xi}\rt),
\end{equation}
where
\begin{equation}
    \begin{aligned}
      G\lt(\xi\rt) =\int_{\cs}\rmd^3\mathscr{V}\,\xi H_{\text{phys}}
+\mathcal{B}\lt(\xi\rt),\\
G\lt(\vec{\xi}\rt)=\int_{\cs}\rmd^3\mathscr{V}\,
\xi^i C_i
+\mathcal{B}\lt({\vec{\xi}}\rt).
    \end{aligned}
\end{equation}
The Poisson bracket is bi-linear,
\begin{equation}\label{eq:poitol}
    \begin{aligned}
      &\lt\{G\lt(\xi_1,\vec{\xi}_1\rt),G\lt(\xi_2,\vec{\xi}_2\rt)\rt\}\\
=&\lt\{G\lt(\xi_1\rt)+G\lt(\vec{\xi}_1\rt),G\lt(\xi_2\rt)+G\lt(\vec{\xi}_2\rt)\rt\}\\
=&
\lt\{G\lt(\vec{\xi}_1\rt),G\lt(\vec{\xi}_2\rt)\rt\}
+\lt\{G\lt(\vec{\xi}_1\rt),G\lt(\xi_2\rt)\rt\}
+\lt\{G\lt(\xi_1\rt),G\lt(\vec{\xi}_2\rt)\rt\}
+\lt\{G\lt(\xi_1\rt),G\lt(\xi_2\rt)\rt\}.
    \end{aligned}
\end{equation}
By \eqref{poi1}, the first term of \eqref{eq:poitol} becomes
\begin{equation}\label{poiCC}
    \begin{aligned}
      &\lt\{G\lt(\vec{\xi}_1\rt),G\lt(\vec{\xi}_2\rt)\rt\}
=\int_{\cs}\rmd^3\mathscr{V}\,\lt[\vec{\xi_1},\vec{\xi_2}\rt]^j C_j
 +2\lim_{R\to \infty}\oint_{S^2}\rmd^2 S\,\lt[\vec{\xi_1},\vec{\xi_2}\rt]^j\pi^{r}_j.
    \end{aligned}
\end{equation}
By \eqref{variaofph}, \eqref{poi1}, and \eqref{poi2}, 
\begin{equation}\label{pois2}
    \begin{aligned}
    &\lt\{G\lt(\vec{\xi}_1\rt),G\lt(\xi_2\rt)\rt\}\\
    =&\int_{\cs}\rmd^3\mathscr{V}\,
\xi^i_1\partial_i\lt(\tilde{\xi}_2\rt)C
-\frac{1}{2}\int_{\cs}\rmd^3\mathscr{V}\,\xi_2 H_{\text{phys}}N^jN^k
\mathcal{L}_{\vec{\xi}_1}g_{jk}
+\int_{\cs}\rmd^3\mathscr{V}\,\lt[\vec{\xi_1},\vec{\tilde\xi}_2\rt]^j C_j
\\
&
-\delta_{\vec{\xi}_1}\tilde{\mathcal{B}}\lt(\tilde\xi_2\rt)
-\lim_{R\to \infty}\oint_{S^2}\rmd^2 S\,\xi_1^r\tilde{\xi}_2C
 +2\lim_{R\to \infty}\oint_{S^2}\rmd^2 S\,\lt[\vec{\xi}_1,\vec{\tilde\xi}_2\rt]^j \pi^{r}_j\\
 =&
\lim_{R\to\infty}\int_{\cs}\rmd^3\mathscr{V}\,
\xi_1^iD_i\lt(\xi_2\rt)H_{\text{phys}}-\delta_{\vec{\xi}_1}\tilde{\mathcal{B}}\lt(\tilde\xi_2\rt)\\
&-\lim_{R\to \infty}\oint_{S^2}\rmd^2 S\,\xi_1^r\tilde{\xi}_2C
 +2\lim_{R\to \infty}\oint_{S^2}\rmd^2 S\,\lt[\vec{\xi}_1,\vec{\tilde\xi}_2\rt]^j \pi^{r}_j.
    \end{aligned}
\end{equation}
Similarly, 
\begin{equation}
    \begin{aligned}
\lt\{G\lt(\xi_1\rt),G\lt(\vec{\xi}_2\rt)\rt\}
=&-\lt\{G\lt(\vec{\xi}_2\rt),G\lt(\xi_1\rt)\rt\}\\
=&
-\int_{\cs}\rmd^3\mathscr{V}\,
\xi_2^iD_i\lt(\xi_1\rt)H_{\text{phys}}+\delta_{\vec{\xi}_2}\tilde{\mathcal{B}}\lt(\tilde\xi_1\rt)
+\lim_{R\to \infty}\oint_{S^2}\rmd^2 S\,\xi_2^r\tilde{\xi}_1C\\
 &-2\lim_{R\to \infty}\oint_{S^2}\rmd^2 S\,\lt[\vec{\xi}_2,\vec{\tilde\xi}_1\rt]^j \pi^{r}_j.
    \end{aligned}
\end{equation}
By \eqref{poi1}-\eqref{poi3},
\begin{equation}\label{poiHH}
    \begin{aligned}
    \lt\{G\lt(\xi_1\rt),G\lt(\xi_2\rt)\rt\}
=&\mathscr{B}+2\lim_{R\to \infty}\oint_{S^2}\rmd^2 S\, (\tilde\xi_1D^j\tilde\xi_2-\tilde\xi_2D^j\tilde\xi_1)\pi^r_j-\delta_{\vec{\tilde\xi}_1}\tilde{\mathcal{B}}\lt(\tilde\xi_2\rt)\\
&-\lim_{R\to \infty}\oint_{S^2}\rmd^2 S\,\tilde\xi_1^r\tilde\xi_2C
+\delta_{\vec{\tilde\xi}_2}\tilde{\mathcal{B}}\lt(\tilde\xi_1\rt)
+\lim_{R\to \infty}\oint_{S^2}\rmd^2 S\,\tilde\xi_2^r\tilde\xi_1C\\
&+2\lim_{R\to \infty}\oint_{S^2}\rmd^2 S\,[\vec{\tilde\xi}_1,\vec{\tilde\xi}_2]^j \pi^{r}_j,
    \end{aligned}
\end{equation}
where the bulk term $\mathscr{B}$ is
\begin{equation}
    \begin{aligned}
    \mathscr{B}
=&
\int_{\cs}\rmd^3\mathscr{V}\,   \lt(\tilde\xi_1D^j\tilde\xi_2-\tilde\xi_2D^j\tilde\xi_1\rt)C_j
    +
\int_{\cs}\rmd^3\mathscr{V}\,
\tilde\xi_1^jD_j\lt(\tilde\xi_2\rt)C\\
&-
\int_{\cs}\rmd^3\mathscr{V}\,
\tilde\xi_2^jD_j\lt(\tilde\xi_1\rt)C
+\int_{\cs}\rmd^3\mathscr{V}\,\lt[\vec{\tilde\xi}_1,\vec{\tilde\xi}_2\rt]^j C_j\\
&+\int_{\cs}\rmd^3\mathscr{V}\,H_{\rm phy}N^iN^j\xi_1\tilde\xi_2 g^{-\frac{1}{2}}\lt(\pi_{ij}-\frac{1}{2}\pi g_{ij}\rt)\\
&-\frac{1}{2}\int_{\cs}\rmd^3\mathscr{V}\,H_{\rm phy}N^iN^j\xi_2\tilde\xi_1 g^{-\frac{1}{2}}\lt(\pi_{ij}-\frac{1}{2}\pi g_{ij}\rt)\\
&+\int_{\cs}\rmd^3\mathscr{V}\,H_{\rm phy}N^iN^j\xi_1D_i\lt(\tilde\xi_{2}\rt)_j 
-\int_{\cs}\rmd^3\mathscr{V}\,H_{\rm phy}N^iN^j\xi_2D_i\lt(\tilde\xi_{1}\rt)_j. \\
    \end{aligned}
\end{equation}
This term turns out to vanish, in agree with the result of \cite{giesel2010algebraic}. The detailed computation is provided in appendix \ref{app:detaiofcal}.

In summary, with generic $\xi,\vec{\xi}$, the general form of the Poisson bracket between a pair of $G(\xi,\vec{\xi})$ reads 
\begin{equation}\label{whole}
   \begin{aligned}
   &\lt\{G\lt(\xi_1,\vec{\xi}_1\rt),G\lt(\xi_2,\vec{\xi}_2\rt)\rt\}\\
   =&\int_{\cs}\rmd^3\mathscr{V}\,\lt[\vec{\xi_1},\vec{\xi_2}\rt]^j C_j
   +\int_{\cs}\rmd^3\mathscr{V}\,
\lt[\xi_1^iD_i\lt(\xi_2\rt)-\xi_2^iD_i\lt(\xi_1\rt)\rt]H_{\text{phys}}
\\
 &-\delta_{\vec{\xi}_1}\tilde{\mathcal{B}}\lt(\tilde\xi_2\rt)
-\lim_{R\to \infty}\oint_{S^2}\rmd^2 S\,\xi_1^r\tilde{\xi}_2C
 +2\lim_{R\to \infty}\oint_{S^2}\rmd^2 S\,\lt[\vec{\xi}_1,\vec{\tilde\xi}_2\rt]^j \pi^{r}_j
 +\delta_{\vec{\xi}_2}\tilde{\mathcal{B}}\lt(\tilde\xi_1\rt)\\
&+\lim_{R\to \infty}\oint_{S^2}\rmd^2 S\,\xi_2^r\tilde{\xi}_1C
 -2\lim_{R\to \infty}\oint_{S^2}\rmd^2 S\,\lt[\vec{\xi}_2,\vec{\tilde\xi}_1\rt]^j \pi^{r}_j\\
 &+2\lim_{R\to \infty}\oint_{S^2}\rmd^2 S\, \lt(\tilde\xi_1D^j\tilde\xi_2-\tilde\xi_2D^j\tilde\xi_1\rt)\pi^r_j\\
 &-\delta_{\vec{\tilde\xi}_1}\tilde{\mathcal{B}}\lt(\tilde\xi_2\rt)
-\lim_{R\to \infty}\oint_{S^2}\rmd^2 S\,\tilde\xi_1^r\tilde\xi_2C
+\delta_{\vec{\tilde\xi}_2}\tilde{\mathcal{B}}\lt(\tilde\xi_1\rt)
+\lim_{R\to \infty}\oint_{S^2}\rmd^2 S\,\tilde\xi_2^r\tilde\xi_1C\\
&+2\lim_{R\to \infty}\oint_{S^2}\rmd^2 S\,[\vec{\tilde\xi}_1,\vec{\tilde\xi}_2]^j \pi^{r}_j
 +2\lim_{R\to \infty}\oint_{S^2}\rmd^2 S\,\lt[\vec{\xi_1},\vec{\xi_2}\rt]^j\pi^{r}_j,
   \end{aligned}
\end{equation}
where $ \tilde{\xi}=\xi N$, $\tilde{\xi}^i=\xi N^i$.

\subsection{The Poisson Algebra of Symmetry Charges }\label{subsec:efa}
We take into account the boundary condition \eqref{asyvec} for $\xi,\vec{\xi}$. The bulk terms of \eqref{whole} become
\begin{equation}
\int_{\cs}\rmd^3\mathscr{V}\,\lt(\hat{\xi}^j C_j
   +
\hat{\xi}H_{\text{phys}}\rt),
\end{equation}
where $\hat{\xi}, \hat{\xi}^j$ are 
\be\label{eq:def_of_hat_xi}
\hat{\xi}=\xi_1^iD_i\xi_2-\xi_2^iD_i\xi_1,\qquad \hat{\xi}^j=\lt[\vec{\xi}_1,\vec{\xi}_2\rt]^j.
\ee
The smearing fields $\xi_{1,2},\vec{\xi}_{1,2}$ satisfying \eqref{asyvec} imply that $\hat{\xi}, \hat{\xi}^j$ also satisfy \eqref{asyvec}, see Appendix \ref{app:dis_of_xi}. The boundary data of $\hat{\xi}, \hat{\xi}^j$ are denoted by $\widehat{Y}, \widehat{I}, \widehat{f}, \widehat{W}$, and they have the following relations with the boundary data of $\xi_{1,2},\vec{\xi}_{1,2}$, denoted by $\widehat{Y}_{1,2}, \widehat{I}_{1,2}, \widehat{f}_{1,2}, \widehat{W}_{1,2}$ 
\begin{equation}\label{algebofthefun}
    \begin{aligned}
\widehat{Y}^{A} &=Y_{1}^{B} \bar{D}_{B} Y_{2}^{A}-\lt(1 \leftrightarrow 2\rt), \\
\widehat{f} &=Y_{1}^{A} \partial_{A} f_{2}-\lt(1 \leftrightarrow 2\rt), \\
\widehat{W} &=Y_{1}^{A} \partial_{A} W_{2}-\lt(1 \leftrightarrow 2\rt),\\
\widehat{I}^{A}&=Y_{1}^{B} \bar{D}_{B} I_{2}^{A}+I_{1}^{B} \bar{D}_{B} Y_{2}^{A}-\lt(1 \leftrightarrow 2\rt).
\end{aligned}
\end{equation}

The boundary terms of \eqref{whole} include four parts: 
\begin{eqnarray}
    \mathcal{B}_{\xi_1,\xi_2}&=&-\delta_{\vec{\tilde\xi}_1}\tilde{\mathcal{B}}\lt(\tilde\xi_2\rt)-\lim_{R\to \infty}\oint_{S^2}\rmd^2 S\,\tilde\xi_1^r\tilde\xi_2C
+\delta_{\vec{\tilde\xi}_2}\tilde{\mathcal{B}}\lt(\tilde\xi_1\rt)
+\lim_{R\to \infty}\oint_{S^2}\rmd^2 S\,\tilde\xi_2^r\tilde\xi_1C\nonumber\\
&&+2\lim_{R\to \infty}\oint_{S^2}\rmd^2 S\,[\vec{\tilde\xi}_1,\vec{\tilde\xi}_2]^j \pi^{r}_j
+2\lim_{R\to \infty}\oint_{S^2}\rmd^2 S\, \lt(\tilde\xi_1D^j\tilde\xi_2-\tilde\xi_2D^j\tilde\xi_1\rt)\pi^r_j,\\
\mathcal{B}_{\vec{\xi}_1,\vec{\xi}_2}&=&2\lim_{R\to \infty}\oint_{S^2}\rmd^2 S\,\lt[\vec{\xi_1},\vec{\xi_2}\rt]^j\pi^{r}_j,\\
\mathcal{B}_{\vec{\xi}_1,\xi_2}
    &=&-\delta_{\vec{\xi}_1}\tilde{\mathcal{B}}\lt(\tilde\xi_2\rt)
-\lim_{R\to \infty}\oint_{S^2}\rmd^2 S\,\xi_1^r\tilde{\xi}_2C
 +2\lim_{R\to \infty}\oint_{S^2}\rmd^2 S\,\lt[\vec{\xi}_1,\vec{\tilde\xi}_2\rt]^j \pi^{r}_j,\\
    \mathcal{B}_{\xi_1,\vec{\xi}_2} &=&\delta_{\vec\xi_2}\tilde{\mathcal{B}}\lt(\tilde\xi_1\rt)
+\lim_{R\to \infty}\oint_{S^2}\rmd^2 S\,\xi_2^r\tilde{\xi}_1C
 -2\lim_{R\to \infty}\oint_{S^2}\rmd^2 S\,\lt[\vec{\xi}_2,\vec{\tilde\xi}_1\rt]^j \pi^{r}_j.
\end{eqnarray}
They come from
$\lt\{G\lt(\xi_1\rt),G\lt(\xi_2\rt)\rt\},$
$\{G(\vec{\xi}_1),G(\vec{\xi}_2)\},$
$\{G(\vec{\xi}_1),G(\xi_2)\}$, 
and $\{G(\xi_1),G(\vec{\xi}_2)\}$ respectively, and the boundary term of $\{G(\xi_1,\vec{\xi}_1),G(\xi_2,\vec{\xi}_2)\}$ is $\mathcal{B}_{\xi_1,\xi_2}+\mathcal{B}_{\vec{\xi}_1,\vec{\xi}_2}+\mathcal{B}_{\vec{\xi}_1,{\xi}_2}+\mathcal{B}_{{\xi}_1,\vec{\xi}_2}$. 
Next, we evaluate them individually. 
By \eqref{asymsphere1}-\eqref{asymsphere6}, \eqref{asydysh} and \eqref{asyvec}
\begin{equation}
    \begin{aligned}
        \mathcal{B}_{\xi_1,\xi_2}=&-\delta_{\vec{\tilde\xi}_1}\tilde{\mathcal{B}}\lt(\tilde\xi_2\rt)-\lim_{R\to \infty}\oint_{S^2}\rmd^2 S\,\tilde\xi_1^r\tilde\xi_2C
+\delta_{\vec{\tilde\xi}_2}\tilde{\mathcal{B}}\lt(\tilde\xi_1\rt)
+\lim_{R\to \infty}\oint_{S^2}\rmd^2 S\,\tilde\xi_2^r\tilde\xi_1C\\
&+2\lim_{R\to \infty}\oint_{S^2}\rmd^2 S\,[\vec{\tilde\xi}_1,\vec{\tilde\xi}_2]^j \pi^{r}_j
+2\lim_{R\to \infty}\oint_{S^2}\rmd^2 S\, \lt(\tilde\xi_1D^j\tilde\xi_2-\tilde\xi_2D^j\tilde\xi_1\rt)\pi^r_j\\
=&-\delta_{\vec{\tilde\xi}_1}\mathcal{B}\lt(\tilde\xi_2\rt)
+\delta_{\vec{\tilde\xi}_2}\tilde{\mathcal{B}}\lt(\tilde\xi_1\rt)
-\lim_{R\to \infty}\oint_{S^2}\rmd^2 S\,\xi_1\xi_2N^rNC+\lim_{R\to \infty}\oint_{S^2}\rmd^2 S\,\xi_1\xi_2N^rNC\\
&+2\lim_{R\to \infty}\oint_{S^2}\rmd^2 S\,\lt(\xi_1N^jN^k\partial_k\lt(\xi_2\rt)
-\xi_2N^jN^k\partial_k\lt(\xi_1\rt)\rt)\pi^r_j\\
&+2\lim_{R\to \infty}\oint_{S^2}\rmd^2 S\, \lt(\xi_1D^j\xi_2-\xi_2D^j\xi_1\rt)N^2\pi^r_j\\
=&-\delta_{\vec{\tilde\xi}_1}\tilde{\mathcal{B}}\lt(\tilde\xi_2\rt)
+\delta_{\vec{\tilde\xi}_2}\tilde{\mathcal{B}}\lt(\tilde\xi_1\rt).
    \end{aligned}
\end{equation}
Recall \eqref{eq:4tyboun}, we have 
\begin{equation}\label{eq: voila_b_xi}
    \delta_{\vec{\tilde\xi}_1}\tilde{\mathcal{B}}\lt(\tilde\xi_2\rt)
    =\mp\oint_{S^2}\rmd^2 S\, \sqrt{\bar{\gamma}} f_2\left(4\delta_{\vec{\tilde\xi}_1} \bar{\lambda}+2\bar{D}_{A} \lt(\delta_{\vec{\tilde\xi}_1}\bar{\lambda}^{A}\rt)\right).
\end{equation}
Here $\delta_{\vec{\tilde\xi}_1 }\bar{\lambda}$ is determined by the coefficient of $O\lt(r^{-1}\rt)$ of $\delta_{\vec{\tilde\xi}_1} g_{rr}$ and $\delta_{\vec{\tilde\xi}_1} \bar{\lambda}_A$ is determined by the coefficient of $O\lt(1\rt)$ of $\delta_{\vec{\tilde\xi}_1} g_{rA}$. With \eqref{asymsphere1}-\eqref{asymsphere3} and \eqref{asydysh}, we have
\begin{equation}
    \delta_{\vec{\tilde\xi}_1} g_{rr}=O\lt(r^{-2}\rt),\qquad  \delta_{\vec{\tilde\xi}_1} g_{rA}=O\lt(r^{-1}\rt).
\end{equation}
Henceforth,
\begin{equation}
  \delta_{\vec{\tilde\xi}_1} \bar{\lambda}=0,\qquad \delta_{\vec{\tilde\xi}_1} \bar{\lambda}_A=0.
\end{equation}
Consequently, we have
\begin{equation}
    \delta_{\vec{\tilde\xi}_1}\tilde{\mathcal{B}}\lt(\tilde\xi_2\rt)
    =\delta_{\vec{\tilde\xi}_2}\tilde{\mathcal{B}}\lt(\tilde\xi_1\rt)
    =0.
\end{equation}
Therefore, we obtain $\mathcal{B}_{\xi_1,\xi_2}=0$, so the symmetry charges $G\lt(\xi\rt)$ form an Abelian subalgebra. 
By \eqref{asymsphere1}-\eqref{asymsphere6}, \eqref{intial}, \eqref{asyvec}, \eqref{eq:4tyboun} and \eqref{poiCC},
\begin{equation}
    \begin{aligned}
    \mathcal{B}_{\vec{\xi}_1,\vec{\xi}_2}=&2\lim_{R\to \infty}\oint_{S^2}\rmd^2 S\,\lt[\vec{\xi_1},\vec{\xi_2}\rt]^j\pi^{r}_j\\
    =&2\lim_{R\to \infty}\oint_{S^2}\rmd^2 S\,\lt[\vec{\xi_1},\vec{\xi_2}\rt]^r\pi^{r}_r
    +2\lim_{R\to \infty}\oint_{S^2}\rmd^2 S\,\lt[\vec{\xi_1},\vec{\xi_2}\rt]^A\pi^{r}_A\\
    =&2\lim_{R\to \infty}\oint_{S^2}\rmd^2 S\,\hat{\xi}^r\pi^{r}_r
    +2\lim_{R\to \infty}\oint_{S^2}\rmd^2 S\,\hat{\xi}^A\pi^{r}_A\\
    =&\oint_{S^2}\rmd^2 S\,\widehat{Y}^A\left(4\lt(\tilde {\bar k}_{A B}-\bar\lambda\bar \gamma_{A B}\rt) \bar{\pi}^{r B}
+2\bar{\gamma}_{A B} \pi^{(2) r B}+\bar{\lambda}_{A} \lt(\bar p
+2\bar{\pi}^{B}_{B}\rt)\right)\\
&+\oint_{S^2}\rmd^2 S\, 2\widehat{I}^{A} \bar{\gamma}_{A B} \bar{\pi}^{rB}
\oint_{S^2}\rmd^2 S\, \widehat{W}  \lt(\bar p+2\bar{\pi}^{A}_{A}\rt)\\
&+2\lim_{R\to\infty}\oint_{S^2}\rmd^2 S\,
R\widehat{Y}^A\bar \pi^{r}_A
+2\lim_{R\to\infty}\oint_{S^2}\rmd^2 S\,
\log\lt(R\rt)\widehat{Y}^A\pi^{\lt(\log\rt)r}_A\\
    =&\mathcal{B}\lt(\widehat{W}\rt)+\mathcal{B}\lt(\widehat{\vec{Y}}\rt)+\mathcal{B}\lt(\widehat{\vec{I}}\rt)\\
    &+2\lim_{R\to\infty}\oint_{S^2}\rmd^2 S\,
R\widehat{Y}^A\bar \pi^{r}_A
+2\lim_{R\to\infty}\oint_{S^2}\rmd^2 S\,
\log\lt(R\rt)\widehat{Y}^A\pi^{\lt(\log\rt)r}_A.
    \end{aligned}
\end{equation}
The commutator of killing vector fields $\widehat{Y}^A=\lt[\vec{Y}_1,\vec{Y}_2\rt]^A$ is still a killing vector field on $S^2$, and it has odd parity. Similar to the discussions in section \ref{sec:bounchg}, the terms $\lim_{R\to\infty}\oint_{S^2}\rmd^2 S\,R\widehat{Y}^A\bar \pi^{r}_A$ and $\lim_{R\to\infty}\oint_{S^2}\rmd^2 S\,
\log\lt(R\rt)\widehat{Y}^A\pi^{\lt(\log\rt)r}_A$ are actually vanishing rather than divergent. As the result, we obtain
\begin{equation}
\mathcal{B}_{\vec{\xi}_1,\vec{\xi}_2}=\mathcal{B}\lt(\widehat{W}\rt)+\mathcal{B}\lt(\widehat{\vec{Y}}\rt)+\mathcal{B}\lt(\widehat{\vec{I}}\rt).
\end{equation}
By  \eqref{pois2}, we obtain
\begin{equation}
    \begin{aligned}
    \mathcal{B}_{\vec{\xi}_1,\xi_2}
    =&-\delta_{\vec{\xi}_1}\tilde{\mathcal{B}}\lt(\tilde\xi_2\rt)
-\lim_{R\to \infty}\oint_{S^2}\rmd^2 S\,\xi_1^r\tilde{\xi}_2C
 +2\lim_{R\to \infty}\oint_{S^2}\rmd^2 S\,\lt[\vec{\xi}_1,\vec{\tilde\xi}_2\rt]^j \pi^{r}_j\\
    =&\mp\oint_{S^2}\rmd^2 S\, \sqrt{\bar{\gamma}} f_2\left(4\delta_{\vec{\xi}_1} \bar{\lambda}+2\bar{D}_{A} \lt(\delta_{\vec{\xi}_1}\bar{\lambda}^{A}\rt)\right).\\
    &-\lim_{R\to \infty}\oint_{S^2}\rmd^2 S\,\xi_1^rN\xi_2C
 +2\lim_{R\to \infty}\oint_{S^2}\rmd^2 S\,\lt[\vec{\xi}_1,\xi_2\vec{N}\rt]^j \pi^{r}_j.
    \end{aligned}
\end{equation}
By \eqref{asymsphere1}-\eqref{asymsphere6}, \eqref{cexp}, \eqref{asyc1}, and \eqref{asydysh}, we find the last two terms vanish. Therefore,
\begin{equation}\label{bv1s2} 
    \mathcal{B}_{\vec{\xi}_1,\xi_2}=\mp\oint_{S^2}\rmd^2 S\, \sqrt{\bar{\gamma}} f_2\left(4\delta_{\vec{\xi}_1} \bar{\lambda}+2\bar{D}_{A} \lt(\delta_{\vec{\xi}_1}\bar{\lambda}^{A}\rt)\right).
\end{equation}
Here
\begin{equation}
\delta_{\vec{\xi}_1} \bar{\lambda}=Y_1^A\bar{D}_A\bar{\lambda},
\end{equation}
\begin{equation}
    \delta_{\vec{\xi}_1} \bar{\lambda}_A
    =\mathcal{L}_{\vec{Y}}\bar \lambda_{A}
    +\bar{D}_{A}W
    -\bar{\gamma}_{A B} I^B.
\end{equation}
In appendix \ref{app: proof}, we proof 
\begin{equation}\label{DlambA}
\bar{D}_A\lt(\mathcal{L}_{\vec{Y}_1}\bar\lambda_B\rt)=\mathcal{L}_{\vec{Y}_1}\lt(\bar{D}_A\bar\lambda_B\rt).
\end{equation}
By this result, \eqref{bv1s2} becomes
\begin{equation}
          \mathcal{B}_{\vec{\xi}_1,\xi_2}
=\pm\oint_{S^2}\rmd^2 S\,\lt( 2\sqrt{\bar{\gamma}}Y_1^B\bar{D}_B \lt(f_2\rt) \left(2 \bar{\lambda}+\bar{D}_{A} \bar{\lambda}^{A}\right)
-2  \sqrt{\bar{\gamma}} f_2 \lt(\bar{D}^{A}\bar{D}_{A}W_1
-
   \bar{D}_{A} I_1^A
\rt)\rt).
\end{equation}
Here we have used the Killing equations of $Y^A$. Similarly, we obtain
\begin{equation}
        \mathcal{B}_{\xi_1,\vec{\xi}_2}
=\mp\oint_{S^2}\rmd^2 S\, \lt(2\sqrt{\bar{\gamma}}Y_2^B\bar{D}_B \lt(f_1\rt) \left(2 \bar{\lambda}+\bar{D}_{A} \bar{\lambda}^{A}\right)
-2 \sqrt{\bar{\gamma}} f_1 \lt(\bar{D}^{A}\bar{D}_{A}W_2
-
   \bar{D}_{A} I_2^A
\rt)\rt).
\end{equation}
Then we have
\begin{equation}\label{eq:Boun_of_final}
    \begin{aligned}
        &\mathcal{B}_{\vec{\xi}_1,\xi_2}+
        \mathcal{B}_{\xi_1,\vec{\xi}_2}\\
        =&\pm\oint_{S^2}\rmd^2 S\,\lt( 2\sqrt{\bar{\gamma}}Y_1^B\bar{D}_B \lt(f_2\rt) \left(2 \bar{\lambda}+\bar{D}_{A} \bar{\lambda}^{A}\right)
-2  \sqrt{\bar{\gamma}} f_2 \lt(\bar{D}^{A}\bar{D}_{A}W_1
-
   \bar{D}_{A} I_1^A
\rt)\rt)\\
&\mp\oint_{S^2}\rmd^2 S\, \lt(2\sqrt{\bar{\gamma}}Y_2^B\bar{D}_B \lt(f_1\rt) \left(2 \bar{\lambda}+\bar{D}_{A} \bar{\lambda}^{A}\right)
-2 \sqrt{\bar{\gamma}} f_1 \lt(\bar{D}^{A}\bar{D}_{A}W_2
-
   \bar{D}_{A} I_2^A
\rt)\rt)\\
=&\cb\lt(\widehat{f}\rt)
+\mathscr{C}\lt(\hat{\xi}, \hat{\vec{\xi}}\rt),
    \end{aligned}
\end{equation}
where
\begin{equation}
    \cb\lt(\widehat{f}\rt)=
    \pm\oint_{S^2}\rmd^2 S\, 2\sqrt{\bar{\gamma}}\lt(Y_1^B\bar{D}_B f_2-Y_2^B\bar{D}_B f_1 \rt)\left(2 \bar{\lambda}+\bar{D}_{A} \bar{\lambda}^{A}\right)
\end{equation}
and
\begin{equation}\label{cenchar}
      \mathscr{C}\lt(\hat{\xi}, \hat{\vec{\xi}}\rt)=\mp2\oint_{S^2}\rmd^2 S\,  \lt(\sqrt{\bar{\gamma}} f_2 \lt(\bar{D}^{A}\bar{D}_{A}W_1
-
   \bar{D}_{A} I_1^A
\rt)
-\sqrt{\bar{\gamma}} f_1 \lt(\bar{D}^{A}\bar{D}_{A}W_2
-
   \bar{D}_{A} I_2^A
\rt)\rt).
\end{equation}
Therefore, we have the following Poisson algebra of $G\lt(\xi, \vec{\xi}\rt)$ 
\begin{equation}\label{algebra}
    \lt\{G\lt(\xi_1, \vec{\xi}_2\rt),G\lt(\xi_2, \vec{\xi}_2\rt)\rt\}
    =G\lt(\hat{\xi}, \hat{\vec{\xi}}\rt)
    +\mathscr{C}\lt(\hat{\xi}, \hat{\vec{\xi}}\rt).
\end{equation}
Although the Poisson algebra of $G(\xi, \vec{\xi})$ is not immediately close, the quantity $\mathscr{C}(\hat{\xi}, \hat{\vec{\xi}})$ is a phase-space independent  constant, and furthermore it is a central charge (see appendix \ref{prof-co} for detailed discussion). As a result, we find the symmetry charges $G(\xi, \vec{\xi})$ together with the central charge $\mathscr{C}(\hat{\xi}, \hat{\vec{\xi}})$ form an infinite-dimensional Lie algebra under the Poisson bracket. This algebra is denoted by $\sa_{\rm G}$.

In particular, if the $G\lt(\xi_2, \vec{\xi}_2\rt)$ in \eqref{algebra} is restricted to the physical Hamiltonian with boundary term, i.e., $\xi_2=1$, $\vec{\xi}_2=0$, the result of \eqref{algebra} vanishes. Indeed, it is obvious that the bulk terms of \eqref{algebra} vanish. For the boundary terms, \eqref{eq:Boun_of_final} reads
\begin{equation}
        \begin{aligned}
        \mathcal{B}_{\vec{\xi}_1,1}+
        \mathcal{B}_{\xi_1,0}
        =\mp2
\oint_{S^2}\rmd^2 S\, \sqrt{\bar{\gamma}}\lt(\bar{D}^{A}\bar{D}_{A}W_1
-
   \bar{D}_{A} I_1^A
\rt)
=0.
\end{aligned}
\end{equation}
The last step is because it is integrated on a closed surface. Therefore, the boundary-preserving symmetry charges commute with the physical Hamiltonian, which satisfies the first requirement of Definition \ref{defofchar}.

The central charge $\mathscr{C}(\hat{\xi}, \hat{\vec{\xi}})$ vanishes when we require the boundary data of $\xi,\vec{\xi}$ to satisfy certain conditions. For example, 
\begin{enumerate}

\item $I^A=\bar D^AW$ for the boundary data of $\vec{\xi}$.

\item The parity condition: we assign the odd parity to $W$ and the even parity to $f$ and $I^A$ \footnote{$\bar D_A\bar D^A$ preserves the parity of $W$: $\bar D_A\bar D^AW=\partial_A\partial^AW+{}^{\bar\gamma}\Gamma^A_{AB}\partial^BW$. The first term preserves the parity of $W$ obviously. For the second term, note that $\bar\gamma_{AB}$ has even parity, and ${}^{\bar\gamma}\Gamma^A_{AB}$ is first derivative of $\bar\gamma_{AB}$, thus ${}^{\bar\gamma}\Gamma^A_{AB}$ has odd parity. Therefore, the term ${}^{\bar\gamma}\Gamma^A_{AB}\partial^BW$ also preserves the parity conditions of $W$.}.

\end{enumerate}

We next compare $\sa_{\rm G}$ to the original BMS algebra, $\sa_{\rm BMS}$ \cite{bondi1962gravitational, sachs1962gravitational, barnich2010aspects,barnich2011bms}. Before discussing the similarities, it is important to note the distinctions between the two algebras. The original BMS charges are defined as boundary charges at the null infinity of an asymptotically flat spacetime. In contrast, the charges in $\sa_{\rm G}$ have both bulk and boundary terms, while being Dirac observables on the reduced phase space. Furthermore, the boundary in our study is defined at spatial infinity. Therefore, here we compare the algebraic structures of these two kinds of algebra, while their physical interpretations might be different.

The boundary-preserving generators defined in this work are significantly larger than the original BMS charges. In the bulk, they generate arbitrary physical transformations. Consequently, comparing $\sa_{\rm G}$ with the traditional BMS algebra  directly is meaningless. 
Nevertheless, we can construct a quotient Lie algebra of $\sa_{\rm G}$, which is denoted as $\hat\sa_{\rm G}$. $\hat\sa_{\rm G}$ is determined by boundary data of $\xi^\mu$, and we will compare it with the original BMS algebra. We consider the vector fields $v^\mu=\lt(v,\vec{v}\rt)$, which satisfy $f=W=0$, $Y^A=I^A=0$ in \eqref{asyvec}. The symmetry charges corresponding to $v^\mu$ constitute a subalgebra of $\sa_{\rm G}$. We denote these charges by $\tilde G\lt(v,\vec{v}\rt)$ and their algebra by $\tilde\sa_{\rm G}$.  $\tilde G\lt(v,\vec{v}\rt)$ is given by
\begin{equation}\label{eq:def_of_ompa_char}
  \tilde G\lt(v,\vec{v}\rt):=\int_{\cs} \rmd^3\mathscr{V}\,v H_{\text{phys}}+\int_{\cs} \rmd^3\mathscr{V}\,v^jC_j.
\end{equation}
Note that \eqref{eq:def_of_ompa_char} contains no boundary term. We next demonstrate that $\tilde\sa_{\rm G}$ is an ideal of $\sa_{\rm G}$. Firstly, it is obvious that $\tilde\sa_{\rm G}$ is a subalgebra of $\sa_{\rm G}$. The algebraic structure of $\sa_{\rm G}$ naturally induces an algebraic structure of $\tilde\sa_{\rm G}$, which is given by the Poisson bracket of the generators.  $\forall\,\tilde G\lt(v,\vec{v}\rt)\in \tilde\sa_{\rm G}$, $G\lt(\xi,\vec{\xi}\rt)\in\sa_{\rm G}$, we have
\begin{equation}
    \lt\{\tilde G\lt(v,\vec{v}\rt),G\lt(\xi,\vec{\xi}\rt)\rt\}=\tilde G\lt(\hat v,\hat{\vec{v}}\rt),
\end{equation}
with
\begin{equation}
    \hat{v}=v^j\partial_j\xi-\xi^j\partial_jv,
\end{equation}
and
\begin{equation}
    \hat{v}^j=\lt[\vec{v},\vec{\xi}\rt]^j.
\end{equation}
Therefore, $\tilde G\lt(\hat v,\hat{\vec{v}}\rt)\in \tilde\sa_{\rm G}$. Consequently, $\tilde\sa_{\rm g}$ is an ideal of $\sa_{\rm G}$. Next, we define a equivalent relation between the elements of $\sa_{\rm G}$:
\begin{definition}
    Given $G\lt(\xi_1,\vec{\xi}_1\rt)$, $G\lt(\xi_2,\vec{\xi}_2\rt)\in\sa_{\rm G}$. $G\lt(\xi_1,\vec{\xi}_1\rt)$ is equivalent to $G\lt(\xi_2,\vec{\xi}_2\rt)$ if
    \begin{equation}
        G\lt(\xi_1,\vec{\xi}_1\rt)-G\lt(\xi_2,\vec{\xi}_2\rt)\in\tilde\sa_{\rm G}.
    \end{equation}
\end{definition}
With the equivalent relation given above, we define a quotient Lie algebra of $\sa_{\rm G}$
\begin{equation}
    \hat\sa_{\rm G}=\sa_{\rm G}/\tilde\sa_{\rm G}.
\end{equation}
The algebraic structure of $\hat\sa_{\rm G}$ is naturally induced by the algebraic structure of $\sa_{\rm G}$.
Firstly, we define a projection $\Pi:\sa_{\rm G}\to\hat\sa_{\rm G}$ as following:
\begin{definition}
$\forall\,G\lt(\xi,\vec{\xi}\rt)\in \sa_{\rm G}$, $\tilde G\lt(v,\vec{v}\rt)\in\tilde\sa_{\rm G}$, the projection $\Pi$ is a linear map $\Pi:\sa_{\rm G}\to\hat\sa_{\rm G}$, s.t.,
\begin{equation}
    \Pi\lt[G\lt(\xi,\vec{\xi}\rt)\rt]=\Pi\lt[G\lt(\xi_1,\vec{\xi}_1\rt)+\tilde G\lt(v,\vec{v}\rt)\rt]
    =\hat G\lt(\xi,\vec{\xi}\rt),
\end{equation}
with $\hat G\lt(\xi,\vec{\xi}\rt)\in \hat\sa_{\rm G}$.
\end{definition}
The value of $\hat G\lt(\xi,\vec{\xi}\rt)$ only depends on the boundary data of $\xi^\mu$. The algebraic structure of $\sa_{G}$ naturally induces a algebraic structure of $\hat\sa_{\rm G}$ by $\Pi$:
\begin{enumerate}
    \item Addition: Given $G\lt(\xi_1,\vec{\xi}_1\rt),\,G\lt(\xi_2,\vec{\xi}_2\rt)\in \sa_{\rm G}$. Then we have
    \begin{equation}
        \hat G\lt(\xi_1,\vec{\xi}_1\rt)=\Pi\lt[G\lt(\xi_1,\vec{\xi}_1\rt)\rt],
        \quad \hat G\lt(\xi_2,\vec{\xi}_2\rt)=\Pi\lt[G\lt(\xi_2,\vec{\xi}_2\rt)\rt].
    \end{equation}
    Then the addition of $\hat\sa_{\rm G}$ is given by
    \begin{equation}
        \hat G\lt(\xi_1,\vec{\xi}_1\rt)+\hat G\lt(\xi_2,\vec{\xi}_2\rt):=\Pi\lt[G\lt(\xi_1,\vec{\xi}_1\rt)+G\lt(\xi_2,\vec{\xi}_2\rt)\rt].
    \end{equation} 
    The self-consistency of this definition is easy to check: Given $\tilde G\lt(v_1,\vec{v}_1\rt),\,\tilde G\lt(v_2,\vec{v}_2\rt)\in \tilde \sa_{\rm G}$, then
    \begin{equation}
        \hat G\lt(\xi_1,\vec{\xi}_1\rt)=\Pi\lt[G\lt(\xi_1,\vec{\xi}_1\rt)+\tilde G\lt(v_1,\vec{v}_1\rt)\rt],
        \quad \hat G\lt(\xi_2,\vec{\xi}_2\rt)=\Pi\lt[G\lt(\xi_2,\vec{\xi}_2\rt)+\tilde G\lt(v_2,\vec{v}_2\rt)\rt].
    \end{equation}
    As the result, we have 
    \begin{equation}
        \begin{aligned}
            \hat G\lt(\xi_1,\vec{\xi}_1\rt)+\hat G\lt(\xi_2,\vec{\xi}_2\rt)=&\Pi\lt[G\lt(\xi_1,\vec{\xi}_1\rt)+\tilde G\lt(v_1,\vec{v}_1\rt)
            +G\lt(\xi_2,\vec{\xi}_2\rt)+\tilde G\lt(v_2,\vec{v}_2\rt)\rt]\\
            =&\Pi\lt[G\lt(\xi_1,\vec{\xi}_1\rt)+G\lt(\xi_2,\vec{\xi}_2\rt)\rt].
        \end{aligned}
    \end{equation}
    \item Scalar Multiplication: Given $G\lt(\xi,\vec{\xi}\rt)\in \sa_{\rm G}$, with $\hat G\lt(\xi,\vec{\xi}\rt)=\Pi\lt[G\lt(\xi,\vec{\xi}\rt)\rt]$. $\forall\alpha\in R$, the scalar Multiplication is given by 
    \begin{equation}
        \alpha\hat G\lt(\xi,\vec{\xi}\rt)=\Pi\lt[\alpha G\lt(\xi,\vec{\xi}\rt)\rt].
    \end{equation}
    \item Multiplication: The multiplication is essential in the definition of a algebra. In our case, we denote it as $\{\cdot,\cdot\}_B$, which is a bilineal map $\{\cdot,\cdot\}_B: \hat\sa_{\rm G}\times\hat \sa_{\rm G}\to\hat \sa_{\rm G}$, and it is defined as following: Given $G\lt(\xi_1,\vec{\xi}_1\rt),\,G\lt(\xi_2,\vec{\xi}_2\rt)\in \sa_{\rm G}$. Then we have
    \begin{equation}
        \hat G\lt(\xi_1,\vec{\xi}_1\rt)=\Pi\lt[G\lt(\xi_1,\vec{\xi}_1\rt)\rt],
        \quad \hat G\lt(\xi_2,\vec{\xi}_2\rt)=\Pi\lt[G\lt(\xi_2,\vec{\xi}_2\rt)\rt].
    \end{equation}
    The $\{\cdot,\cdot\}_B$ is defined as
    \begin{equation}
        \lt\{\hat G\lt(\xi_1,\vec{\xi}_1\rt),\hat G\lt(\xi_2,\vec{\xi}_2\rt)\rt\}_B
        :=\Pi\lt[\lt\{ G\lt(\xi_1,\vec{\xi}_1\rt), G\lt(\xi_2,\vec{\xi}_2\rt)\rt\}\rt].
    \end{equation}
    Next, we check the self-consistency of the definition of $\{\cdot,\cdot\}_B$. Given $\tilde G\lt(v_1,\vec{v}_1\rt),\,\tilde G\lt(v_2,\vec{v}_2\rt)\in \tilde \sa_{\rm G}$, then
    \begin{equation}
        \hat G\lt(\xi_1,\vec{\xi}_1\rt)=\Pi\lt[G\lt(\xi_1,\vec{\xi}_1\rt)+\tilde G\lt(v_1,\vec{v}_1\rt)\rt],
        \quad \hat G\lt(\xi_2,\vec{\xi}_2\rt)=\Pi\lt[G\lt(\xi_2,\vec{\xi}_2\rt)+\tilde G\lt(v_2,\vec{v}_2\rt)\rt].
    \end{equation}
    Note that $\tilde\sa_{\rm G}$ is an ideal of $\sa_{\rm G}$, we have
    \begin{equation}
        \begin{aligned}
            &\lt\{ G\lt(\xi_1,\vec{\xi}_1\rt)+\tilde G\lt(v_1,\vec{v}_1\rt), G\lt(\xi_2,\vec{\xi}_2\rt)+\tilde G\lt(v_2,\vec{v}_2\rt)\rt\}\\
            =&\lt\{ G\lt(\xi_1,\vec{\xi}_1\rt), G\lt(\xi_2,\vec{\xi}_2\rt)\rt\}
            +\lt\{ G\lt(\xi_1,\vec{\xi}_1\rt), \tilde G\lt(v_2,\vec{v}_2\rt)\rt\}\\
            &+\lt\{ \tilde G\lt(v_1,\vec{v}_1\rt), G\lt(\xi_2,\vec{\xi}_2\rt)\rt\}
            +\lt\{ \tilde G\lt(v_1,\vec{v}_1\rt), \tilde G\lt(v_2,\vec{v}_2\rt)\rt\}\\
             =&\lt\{ G\lt(\xi_1,\vec{\xi}_1\rt), G\lt(\xi_2,\vec{\xi}_2\rt)\rt\}
             +\tilde G\lt(v,\vec{v}\rt),
        \end{aligned}
    \end{equation}
    with 
    \begin{equation}
        \tilde G\lt(v,\vec{v}\rt)=\lt\{ G\lt(\xi_1,\vec{\xi}_1\rt), \tilde G\lt(v_2,\vec{v}_2\rt)\rt\}
        +\lt\{ \tilde G\lt(v_1,\vec{v}_1\rt), G\lt(\xi_2,\vec{\xi}_2\rt)\rt\}
            +\lt\{ \tilde G\lt(v_1,\vec{v}_1\rt), \tilde G\lt(v_2,\vec{v}_2\rt)\rt\},
    \end{equation}
   and $\tilde G\lt(v,\vec{v}\rt)\in\tilde\sa_{\rm G}$. Consequently, we find 
    \begin{equation}
        \begin{aligned}
            \lt\{\hat G\lt(\xi_1,\vec{\xi}_1\rt),\hat G\lt(\xi_2,\vec{\xi}_2\rt)\rt\}_B
            =&\Pi\lt[\lt\{ G\lt(\xi_1,\vec{\xi}_1\rt)+\tilde G\lt(v_1,\vec{v}_1\rt), G\lt(\xi_2,\vec{\xi}_2\rt)+\tilde G\lt(v_2,\vec{v}_2\rt)\rt\}\rt]\\
        =&\Pi\lt[\lt\{ G\lt(\xi_1,\vec{\xi}_1\rt), G\lt(\xi_2,\vec{\xi}_2\rt)\rt\}\rt].
        \end{aligned}
    \end{equation}
\end{enumerate} 

In the original BMS algebra, the supertranslations form an Abelian subalgebra, which is denoted as $\sa_{\rm BMS}^{(ab)}$ in this paper. Similarly, in our work, as indicated by \eqref{algebra}, $G\lt(f\rt)$ and $G\lt(W\rt)$ also form an Abelian subalgebra when the central term $\mp2\oint_{S^2}\rmd^2 S\,  \lt(\sqrt{\bar{\gamma}} f_2 \bar{D}^{A}\bar{D}_{A}W_1-\sqrt{\bar{\gamma}} f_1 \bar{D}^{A}\bar{D}_{A}W_2\rt)$ vanishes. This central term arises from the contribution of $W$ in \eqref{cenchar}, and it vanishes if we assign even parity to $f$ and odd parity to $W$. Therefore, we have an Abelian subalgebra of $\hat\sa_{\rm G}$, which is denoted as $\hat\sa_{\rm G}^{(ab)}$. 
Furthermore, we can prove that $\hat\sa^{(ab)}_{\rm G}$ is isomorphic to $\sa_{\rm BMS}^{(ab)}$, and the proof is essentially the same as the one presented in \cite{hennADM}. We provide a brief review of the proof below.
    We firstly rewrite the asymptotically flat metric \eqref{asympflat} using a generalisation of the Beig-Schmidt ansatz \cite{beig1982einstein,compere2011relaxing,troessaert2018bms4} 
    \begin{equation}
\begin{aligned}
g_{\mu \nu} d x^\mu d x^\nu= & \left(1+\frac{2 \sigma}{\eta}+\frac{\sigma^2}{\eta^2}+o\left(\eta^{-2}\right)\right) d \eta^2+o\left(\eta^{-1}\right) d \eta d x^b \\
& +\left(\eta^2 h_{a b}^{(0)}+\eta\left(k_{a b}-2 \sigma h_{a b}^{(0)}\right)+\log \eta i_{a b}+h_{a b}^{(2)}+o\left(\eta^0\right)\right) d x^a d x^b.
\end{aligned}
\end{equation}
Here
\begin{equation}
    \eta^2=-t^2+r^2,
\end{equation} The coordinates $x^a$ are
\begin{equation}
    x^a=\lt(s,x^A\rt)(\,A=1,2),
\end{equation}
where $X^A(\,A=1,2)$ are the angular coordinates, and
\begin{equation}
    s=\pm t/r.
\end{equation}
$\sigma$, $k_{ab}$, $i_{ab}$ and $h_{ab}^{(n)}$ depend on $x^a$ only. The spatial infinity is given by the limitation $\eta\to\infty$, and the boundary metric $h_{ab}^{(0)}$ is the metric of a unit hyperboloid, which reads
\begin{equation}
h_{a b}^{(0)} d x^a d x^b=\frac{-1}{\left(1-s^2\right)^2} d s^2+\frac{1}{1-s^2} \gamma_{A B} d x^A d x^B.
\end{equation}
The covariant derivative compatible with $h_{a b}^{(0)}$ is denoted as $\mathcal{D}_a$.

Next, we aim to establish a connection between the null infinity $\mathscr{I}^\pm$ and the spatial infinity $i^0$ of the asymptotically flat spacetime. In the geometrical definition of the asymptotically flat spacetime $\lt(M,g_{\mu\nu}\rt)$, there exists a conformal spacetime $\lt(\tilde M,\tilde g_{\mu\nu}\rt)$, such that 
\begin{equation}
    \tilde g_{\mu\nu}=\Omega^2g_{\mu\nu}.
\end{equation}
Where $\O$ is a scalar field satisfying certain conditions \cite{ashtekar1978unified}.
To construct $\O$ explicitly, we follow the scheme described in \cite{troessaert2018bms4}.  
Assuming some smoothness conditions for the metric around the spatial infinity (for more details, refer to \cite{friedrich1998gravitational, friedrich2000calculating, friedrich2000bondi} ), we can introduce a new coordinate system in a neighbourhood of $i^0$, which is $\lt(\rho,s,x^A\rt)$. $\rho$ is a new coordinate we introduce, which is given by
\begin{equation}
\eta=\frac{1}{\rho \sqrt{1-s^2}}.
\end{equation}
Note that the spatial infinity is given by $\eta\to\infty$. Therefore, it is also given by $\rho\to0$.
Then $\O$ reads
\begin{equation}
\Omega=\widetilde{\Omega}\left(1-\left(\frac{s}{\tilde\omega}\right)^2\right),
\end{equation}
where $\tilde\O$ and $\tilde\o$ are some scalar functions, satisfying 
\begin{equation}
\lim _{\rho \rightarrow 0} \rho^{-1} \widetilde{\Omega}=1, \quad \lim _{\rho \rightarrow 0} \tilde\omega=1.
\end{equation}
The null infinity $\mathscr{I}^\pm$ locates at 
\begin{equation}
    s=\pm\tilde\o.
\end{equation}
At the neighborhood of $i^o$, the null infinity $\mathscr{I}^\pm$ locates at 
\begin{equation}
    s=\pm1.
\end{equation}
Then we can connect spatial infinity and the null infinity by taking the following limitation:
\begin{equation}
    \rho\to0,\quad s\to \pm1.
\end{equation}
The parameter that characterizes the supertranslation is govern by the following equation
\begin{equation}\label{eq:for_om}
    \mathcal{D}_a\mathcal{D}^a\o=0,
\end{equation}
where $\o$ is a function of $\lt(s,x^A\rt)$. Ref. \cite{troessaert2018bms4} states that the solutions with odd parity\footnote{here odd parity means $\o\lt(s,x^A\rt)=-\o\lt(-s,-x^A\rt)$.} of \eqref{eq:for_om} correspond to the supertanslations, which are isomorphic to the usual BMS supertanslations. These solutions are
\begin{equation}
    \o=\lt(1-s^2\rt)^{-1/2}\hat \o.
\end{equation}
Here $\hat \o$ can be expanded by the spherical harmonics
\begin{equation}
\widehat{\omega}=\sum_{l m} \omega_{l, m} \Psi_l(s) Y_{l, m}\left(x^A\right), \quad \Psi_l=\frac{1}{2}\left(1-s^2\right)^2 \partial_s^2 Q_l.
\end{equation}
Here $Q_l\lt(s\rt)$ are Legendre functions of the second kind and can be written in terms of Legendre Polynomials $P_l\lt(s\rt)$ as
\begin{equation}
Q_l(s)=P_l(s) \frac{1}{2} \log \left(\frac{1+s}{1-s}\right)+\widetilde{Q}_l(s),
\end{equation}
with $\widetilde{Q}_l(s)$ are polynomials. Furthermore, we have the following limitation
\begin{equation}
    \lim_{s\to1}\Psi_l=1,
\end{equation}
which can be found in \cite{troessaert2018bms4}.
We then introduce 
\begin{equation}
\mathcal{T}=\lim _{s \rightarrow 1}\left(\sqrt{1-s^2} \omega\right).
\end{equation}
Here $\mathcal{T}$ can be viewed as the parameter of usual BMS supertranslations, and it is expanded as
\begin{equation}\label{eq:exp_T}
    \mathcal{T}=\sum_{l m} \omega_{l, m} Y_{l, m}\left(x^A\right).
\end{equation}
We now turn back to the subalgebra formed by $f$ and $W$ in our work. We show that in the special cases of $f$ is assigned with even parity and $W$ is assigned with odd parity, the subalgebra formed by $f$ and $W$ is isomorphic to  the subalgebra formed by $\mathcal{T}$. We just need to show the coefficients $\o_{ml}$ in \eqref{eq:exp_T} can be determined by $f$ and $W$ completely, and vice versa. Note that $\hat{\o}$ is odd, we can set its initial conditions at $s=0$:
\begin{equation}
\left.\omega\right|_{s=0}=\left.\widehat{\omega}\right|_{s=0}=W\left(x^A\right),\left.\quad \partial_s \omega\right|_{s=0}=\left.\partial_s \widehat{\omega}\right|_{s=0}=f\left(x^A\right) .
\end{equation}
Note that $W$ is odd and $f$ is even, they are expanded by the spherical haomonics as 
\begin{equation}
W=\sum_k \sum_{m=-2 k-1}^{2 k+1} W_{2 k+1, m} Y_{2 k+1, m}, \quad f=\sum_k \sum_{m=-2 k}^{2 k} f_{2 k, m} Y_{2 k, m}.
\end{equation}
As the result, the coefficients $\o_{ml}$ are determined by 
\begin{equation}
\left.\omega_{2 k+1, m} \Psi_{2 k+1}\right|_{s=0}=W_{2 k+1, m},\left.\quad \omega_{2 k, m} \partial_s \Psi_{2 k}\right|_{s=0}=f_{2 k, m}.
\end{equation}
Therefore, $\hat\sa^{(ab)}_{\rm G}$ and $\sa_{\rm BMS}^{(ab)}$ are isomorphic.

In general, $f$ and $W$ are not required to have the parity conditions. Moreover, we have additionally the supertranslations contributed by $I^A$. Therefore, the subalgebra in $\hat \sa_{\rm G}$ of supertranslations represents the generalization of the original BMS subalgebra of supertranslations.

The asymptotic spatial rotations in $\hat \sa_{\rm G}$ form an SO(3) algebra thus is isomorphic to the corresponding subalgebra in BMS. The commutator between the supertranslations and spatial rotations also involves the central charge of the algebra\footnote{As a comparison, the central charges obtained in \cite{barnich2011bms} are field dependent, whereas the central charge here is not.}. 

It is also interesting to compare $\hat \sa_{\rm G}$ to the results in \cite{hennADM}, which constructs the BMS group at the spatial infinity. The analysis there relies on the usual fall-off \eqref{asympq000} and \eqref{asympp000} and the parity conditions on $\bar{h}_{ij}$ and $\bar{\pi}_{ij}$. The resulting BMS algebra in \cite{hennADM} does not have the central charge due to the additional boundary condition $\bar{\l}_A=0$. Preserving $\bar{\l}_A=0$ needs to require $I^A=\frac{2 b}{\sqrt{\bar{\gamma}}} \bar{\pi}^{r A}+\bar{D}^A W$, where $b$ is the boost parameter. $\hat \sa_{\rm G}$ does not take into account the boost, and $b=0$ precisely reduces this requirement to $I^A=\bar{D}^A W$ for vanishing $\mathscr{C}(\xi,\vec{\xi})$ mentioned above. Another way to make $\mathscr{C}(\xi,\vec{\xi})$ vanishing is to assign even parity to $f$ and odd parity to $W$. If we impose these parity conditions to $f$ and $W$ and require $I^A=\bar{D}^A W$, $\hat \sa_{\rm G}$ recovers the BMS algebra at spatial infinity given by \cite{hennADM} with vanishing $b$ (vanishing boost generators).

\subsection{Preservation of the Fall-off Conditions and the Boundary Conditions}\label{subsec: pbd}
This subsection shows that the boundary-preserving symmetry transformations preserve the fall-off conditions \eqref{asymsphere1}-\eqref{asymsphere6} and the boundary conditions \eqref{intial}. This means that the transformations generated by the boundary-preserving charges are restricted in the phase space defined in section \ref{sec:ABC}.

As stated in Section \ref{sec:bounchg}, 
$G(\xi,\vec{\xi})$ is finite and its variation is well-defined on the phase space.
Therefore, the Hamiltonian flow of $G(\xi,\vec{\xi})$ is well-defined and we can compute the transformations of the canonical variables with Poisson bracket:
\be
\delta_G g_{i j}&=&\lt\{g_{ij},G\lt(\xi,\vec{\xi}\rt)\rt\}\nonumber\\
&=&2\tilde{\xi} g^{-\frac{1}{2}}\left(\pi_{i j}-\frac{1}{2} g_{i j} \pi\right)+2D_{(i}\tilde{\xi}_{j)}
+2D_{(i}\xi_{j)},\\
\delta_G\pi^{i j}&=&\lt\{\pi^{ij},G\lt(\xi,\vec{\xi}\rt)\rt\}\nonumber\\
&=&-\tilde{\xi} g^{\frac{1}{2}}\left(g^{ik}g^{jl}{}^{\lt(3\rt)}R_{lk}-\frac{1}{2} g^{i j} {}^{\lt(3\rt)}R\right)+\frac{1}{2} \tilde{\xi} g^{-\frac{1}{2}}g^{ij}\left(\pi_{m n} \pi^{m n}-\frac{1}{2} \pi^{2}\right)\nonumber \\
&&-2 \tilde{\xi} g^{-\frac{1}{2}}\left(\pi^{i m} \pi_{m}^{j}-\frac{1}{2} \pi^{i j} \pi\right)+g^{\frac{1}{2}}\left(D^iD^j\tilde{\xi}-g^{i j} D_mD^m\tilde{\xi}\right)\nonumber \\
&&+\mathcal{L}_{\vec{\tilde{\xi}}} \pi^{i j}+\frac{\xi}{2}H_{\text{phys}}N^iN^j+\mathcal{L}_{\vec{\xi}} \pi^{i j}.
\ee
Here $\lt((\xi,\vec{\xi}\rt)$ is given by \eqref{asyvec} and $\tilde\xi^j=N\xi^j$. ${}^{\lt(3\rt)}R_{lk}$ is the Ricci tensor of $g_{ij}$ and ${}^{\lt(3\rt)}R$ is the corresponding Ricci scalar. With \eqref{asympq}, we have
\begin{equation}
    {}^{\lt(3\rt)}R_{lk}=
\frac{\bar{R}_{lk}}{r^{3}}+\frac{\log\lt(r\rt)}{r^{4}}R^{\lt(\log\rt)}_{lk}+O\lt(r^{-4}\rt)
\end{equation}
in the asymptotically Cartesian coordinates. Then we find
\be
\delta_G g_{i j}&=&\frac{\bar{\mathscr{H}}_{ij}}{r}
+\frac{\log\lt(r\rt)}{r^2}\mathscr{H}^{\lt(\log\rt)}_{ij}
+O\lt(r^{-2}\rt),\label{eq:transg}\\
\delta_G \pi^{i j}&=&\frac{\bar{\mathscr{P}}^{ij}}{r^2}
+\frac{\log\lt(r\rt)}{r^3}\mathscr{P}^{\lt(\log\rt)ij}
+\frac{\mathscr{P}^{\lt(2\rt)ij}}{r^3}
+\frac{\log\lt(r\rt)}{r^4}\mathscr{P}^{\lt(ll\rt)ij}
+O\lt(r^{-4}\rt).
\label{eq:transpi}
\ee
We have used $\bar{\mathscr{H}}_{ij},\bar{\mathscr{P}}^{ij}, {\mathscr{H}}_{ij}^{\rm (log)},{\mathscr{P}}^{{\rm (log)}ij}, {\mathscr{P}}^{{(2)}ij}, {\mathscr{P}}^{{(ll)}ij}$ to label the expansions coefficients, whose explicit expression may not be useful for the discussion. Eqs. \eqref{eq:transg}-\eqref{eq:transpi} satisfy the fall-off conditions \eqref{asympq}-\eqref{asympp}. 
Additionally, \eqref{eq:transpi} gives
\begin{equation}
    \delta_G \pi^{rA}=\frac{\bar{\mathscr{P}}^{rA}}{r}
+\frac{\log\lt(r\rt)}{r^2}\mathscr{P}^{\lt(\log\rt)rA}
+\frac{\mathscr{P}^{\lt(2\rt)rA}}{r^3}
+\frac{\log\lt(r\rt)}{r^3}\mathscr{P}^{\lt(ll\rt)rA}
+O\lt(r^{-3}\rt),
\end{equation}
Especially, $\mathscr{P}^{\lt(\log\rt)rA} =\mathcal{L}_{\vec{Y}}\pi^{\lt(\log\rt)rA}$. Therefore, the parity of  $\pi^{\lt(\log\rt)rA}$ is preserved. 

The charge $G(\xi,\vec{\xi})$ infinitesimally transforms $C_j$ as
\footnote{We apply $\left(D_i D_j-D_j D_i\right) v^k=-R_{ijl}{ }^k v^l$ in the calculation.}
\begin{equation}\label{trancj}
    \begin{aligned}
      \lt\{C_j,G\lt(\xi,\vec{\xi}\rt)\rt\}
      =&\lt\{C_j,G\lt(\xi\rt)\rt\}
      +\lt\{C_j,G\lt(\vec{\xi}\rt)\rt\}\\
      =&\partial_j\lt(\xi\rt)H_{\text{phys}}+\mathcal{L}_{\vec{\xi}}C_j\\
      =&\partial_j\lt(\xi\rt)H_{\text{phys}}+\xi^iD_iC_j+C_iD_j\xi^i+C_jD_i\xi^i
     .
    \end{aligned}
\end{equation}
To check the boundary conditions of $C_j$ are preserved by \eqref{trancj}, the asymptotic Cartesian coordinates are convenient. With \eqref{crexp}-\eqref{cexp}, \eqref{asyc1}, and \eqref{intial}, we can rewrite the asymptotic behaviors of $C_j$ and $H_{\text{phys}}$ in the asymptotic Cartesian coordinates
\be
C_j&=&\frac{\log r}{r^5}C_j^{\lt(ll\rt)}+O\lt(r^{-5}\rt),\label{eq:car_asy_ci}\\
H_{\text{phys}}&=&\frac{\log r}{r^4}H^{\lt(\log\rt)}+O\lt(r^{-4}\rt)\label{eq:car_asy_H}.
\ee
Additionally, with \eqref{asyvec}, $\xi=O\lt(1\rt)$ and $\xi^i=O\lt(r\rt)$ in the asymptotic Cartesian coordinates. Thus \eqref{trancj} gives
\begin{equation}
    \begin{aligned}
      &\lt\{C_j,G\lt(\xi,\vec{\xi}\rt)\rt\}=O\lt(\frac{\log r}{r^5}\rt),
    \end{aligned}
\end{equation}
which agrees with \eqref{eq:car_asy_ci}. Since $H_{\text{phys}}$ commutes with itself \cite{gieselmanifestly}, $G(\xi,\vec{\xi})$ infinitesimally transforms $H_{\text{phys}}$ as
\begin{equation}
    \begin{aligned}
    \lt\{H_{\text{phys}},G\lt(\xi,\vec{\xi}\rt)\rt\}
     =&\lt\{H_{\text{phys}},G\lt(\xi\rt)\rt\}
     +\lt\{H_{\text{phys}},G\lt(\xi,\vec{\xi}\rt)\rt\}\\
    =&\mathcal{L}_{\vec{\xi}}H_{\text{phys}}
      =D_i\lt(\xi^iH_{\text{phys}}\rt)
    =\partial_i\lt(\xi^iH_{\text{phys}}\rt).
    \end{aligned}
\end{equation}
Note that $H_{\text{phys}}$ is a density of weight one, $D_i$ can be replaced by $\partial_i$ in the last step.  Similar to the discussions above, we have
\begin{equation}
    \lt\{H_{\text{phys}},G\lt(\xi,\vec{\xi}\rt)\rt\}=O\lt(\frac{\log r}{r^4}\rt),
\end{equation}
which agrees with \eqref{eq:car_asy_H}.

\section{On the Boost Transformation}\label{sec: boost}
In this section we discuss the boost component of the boundary-preserving generators.
We will explore some challenges that arise when considering the boost component, which is why we have chosen not to include it in this work.
 
 The vector field of the boost component is \cite{hennADM, beig1987poincare}
\begin{equation}
    \xi_b=rb,
\end{equation}
where $b(\sig^A)$ is the boost parameter, satisfying 
\begin{equation}\label{eq:req_of_boo}
\bar{D}_A \bar{D}_B b+\bar{\gamma}_{A B} b=0.
\end{equation}
In the traditional spherical coordinates $\{\theta,\varphi\}$, $b$ takes the following formula
\begin{equation}
b=b_1 \sin \theta \cos \varphi+b_2 \sin \theta \sin \varphi+b_3 \cos \theta.
\end{equation}
Here $b_1$, $b_2$ and $b_3$ are some constants. It is obvious that $b$ has odd perity:
\begin{equation}
    b\lt(\pi-\theta,\varphi+\pi\rt)=-b\lt(\pi,\varphi\rt).
\end{equation}
We formally construct the  boost component of the  boundary-preserving symmetry generators as
\begin{equation}\label{bosgen}
    G_b=\int_{\cs}\rmd^3\mathscr{V}\, \xi_bH_{\text{phys}}+\mathcal{B}_b.
\end{equation}
To ensure the finiteness of the bulk term of $G_b$, it is necessary for $H_{\text{phys}}$  to decay at a faster rate than $O\lt(r^{-2}\rt)$ in the spherical coordinates. We first introduce the higher orders fall-off conditions of the canonical variables:
\be
\label{booboun1} g_{r r}&=&1+\frac{1}{r} \bar{h}_{r r}+\frac{\log  r}{r^2} h^{(\log)}_{r r}+\frac{1}{r^{2}} h_{r r}^{(2)}+\frac{\log  r}{r^3} h^{(ll)}_{r r}+\frac{1}{r^{3}} h_{r r}^{(3)}+\frac{\log  r}{r^4} h^{(lll)}_{r r}+o\left(\frac{\log  r}{r^4}\right),\\
g_{r A}&=&\bar{h}_{rA}+\frac{\log  r}{r} h_{r A}^{(\log)}+\frac{1}{r} h_{r A}^{(2)}+\frac{\log  r}{r^2} h_{r A}^{(ll)}+\frac{1}{r^2} h_{r A}^{(3)}+\frac{\log  r}{r^3} h_{r A}^{(lll)}+o\left(\frac{\log  r}{r^3}\right),\\
\label{booboun3} g_{A B}&=&r^{2} \bar{\gamma}_{A B}+r \bar{h}_{A B}+\log (r) h_{A B}^{(\log)}+h_{A B}^{(2)}+\frac{\log r}{r}h_{A B}^{(ll)}\nonumber\\
&&+\frac{1}{r}h_{A B}^{(3)}+\frac{\log r}{r^2}h_{A B}^{(lll)}+o\lt(\frac{\log r}{r^2}\rt),\\
\pi^{r r}&=&\bar{\pi}^{r r}+\frac{\log r}{r} \pi^{(\log)r r}+\frac{1}{r} \pi^{(2) r r}
+\frac{\log r}{r^2}\pi^{(ll)rr}
+\frac{\pi^{(3)rr}}{r^2}\nonumber\\
&&+\frac{\log r}{r^3}\pi^{(lll)rr}
+\frac{\pi^{(4)rr}}{r^3}
+\frac{\log r}{r^4}\pi^{(llll)rr}
+{o}\left(\frac{\log r}{r^4}\right),\\
\pi^{r A}&=&\frac{1}{r} \bar{\pi}^{r A}+\frac{\log  r}{r^2} \pi^{(\log)r A}
+\frac{1}{r^{2}} {\pi^{(2)rA}}
+\frac{\log r}{r^3}\pi^{(ll)rA}
+\frac{\pi^{(3)rA}}{r^3}\nonumber\\
&&+\frac{\log r}{r^4}\pi^{(lll)rA}
+\frac{\pi^{(4)rA}}{r^4}
+\frac{\log r}{r^5}\pi^{(llll)rA}
+{o}\left(\frac{\log r}{r^5}\right),\\
\pi^{A B}&=&\frac{1}{r^{2}} \bar{\pi}^{A B}+\frac{\log  r}{r^3} \pi^{(\log)A B}
+\frac{1}{r^{3}} \pi^{(2) A B}
+\frac{\log r}{r^4}\pi^{(ll)AB}
+\frac{\pi^{(3)AB}}{r^4}\nonumber\\
&&+\frac{\log r}{r^5}\pi^{(lll)AB}
+\frac{\pi^{(4)AB}}{r^5}
+\frac{\log r}{r^6}\pi^{(llll)AB}
+{o}\left(\frac{\log r}{r^6}\right).\label{booboun6}
\ee
By \eqref{booboun1}-\eqref{booboun6},
\begin{equation}\label{eq:Boun_with_boo}
    \begin{aligned}
     C=&\frac{C^{(1)}}{r}
+\frac{\log r}{r^{2}}C^{\lt(\log\rt)}
+\frac{C^{(2)}}{r^{2}}
+\frac{\log r}{r^{3}}C^{\lt(ll\rt)}
+\frac{C^{(3)}}{r^{3}}\\
&+\frac{(\log r)^2}{r^{4}}\tilde{C}^{\lt(lll\rt)}
+\frac{\log r}{r^{4}}C^{\lt(lll\rt)}
+o\lt(\frac{\log r}{r^{4}}\rt),
\\
C_r=&\frac{C^{(1)}_r}{r}
+\frac{\log r}{r^2}C^{\lt(\log\rt)}_r
+\frac{ C^{(2)}_r}{r^2}
+\frac{\log r}{r^3}C^{(ll)}_r
+\frac{ C^{(3)}_r}{r^3}
+\frac{(\log r)^2}{r^4}\tilde{C}^{(lll)}_r
+\frac{\log r}{r^4}C^{(lll)}_r\\
&+\frac{ C^{(4)}_r}{r^4}
+\frac{(\log r)^2}{r^5}\tilde{C}^{(llll)}_r
+\frac{\log r}{r^5}C^{(llll)}_r
+o\lt(\frac{\log r}{r^{5}}\rt),
\\
C_A=&C^{(1)}_A
+\frac{\log r}{r}C^{(\log)}_A
+\frac{ C^{(2)}_A}{r}
+\frac{\log r}{r^2}C^{(ll)}_A
+\frac{ C^{(3)}_A}{r^2}
+\frac{(\log r)^2}{r^3}\tilde{C}^{(lll)}_A
+\frac{\log r}{r^3}C^{(lll)}_A\\
&+\frac{ C^{(4)}_A}{r^3}
+\frac{(\log r)^2}{r^4}\tilde{C}^{(llll)}_A
+\frac{\log r}{r^4}C^{(llll)}_A
+o\lt(\frac{\log r}{r^{4}}\rt).
    \end{aligned}
\end{equation}
In this case, the boundary conditions for $C$, $C_j$ are
\begin{equation}\label{eq:boo_boun}
\begin{aligned}
&C_{A}^{(1)}=C^{(\log)}_A=C_{A}^{(2)}=C^{(ll)}_A=C_{A}^{(3)}=\tilde{C}^{(lll)}_A=C^{(lll)}_A=C_A^{(4)}\\
&=C_{r}^{(1)}=C_{r}^{(2)}=C^{(ll)}_r=C_{r}^{(3)}=C^{(\log)}_{r}=\tilde{C}^{(lll)}_r=C^{(lll)}_r=C_r^{(4)}\\
&=C^{(1)}=C^{(\log)}=C^{(2)}=C^{(ll)} =0,\\
&C^{(3)} \neq 0.
\end{aligned}
\end{equation}
These boundary conditions ensure
\begin{equation}
    N^r=O\lt(\frac{\lt(\log r\rt)^2}{r^2}\rt),\quad
    N^A=O\lt(\frac{\lt(\log r\rt)^2}{r^3}\rt).
\end{equation}
Similar to the discussions in Subsection \ref{sec:varphyham}, the non-differentiable terms of $\delta\int_{\cs}\rmd^3\mathscr{V}\, \xi_bH_{\text{phys}}$ contributed by $\delta\int_{\cs}\rmd^3\mathscr{V}\, \xi_b N^jC_j$ is then vanished.
Furthermore, similar to Section \ref{sec:bounchg}, the boundary terms $\mathcal{B}_b$ are given by variating $\int_{\cs}\rmd^3\mathscr{V}\,\xi_bH_{\text{phys}}$
and integrating by part:
\begin{equation}\label{eq:boost_varia}
\begin{aligned}
\delta\int_{\cs}\rmd^3\mathscr{V}\, \xi_bH_{\text{phys}}
=\int_{\cs} \rmd^3\mathscr{V}\,\lt[\mathscr{A}_{ij}\delta \pi^{ij}-\mathscr{B}^{ij}\delta g_{ij}\rt]
+\mathcal{K}_b,
\end{aligned}
\end{equation}
where
\begin{equation}
\begin{aligned}
    \mathscr{A}_{ij}=&2 rb g^{-\frac{1}{2}}\left(\pi_{i j}-\frac{1}{2} g_{i j} \pi\right)+\mathcal{L}_{\vec{b}} g_{i j},\\
    \mathscr{B}^{ij}=&-rb g^{\frac{1}{2}}\left(R^{i j}-\frac{1}{2} g^{i j} R\right)+\frac{1}{2} rb g^{-\frac{1}{2}}g^{ij}\left(\pi_{m n} \pi^{m n}-\frac{1}{2} \pi^2\right)\\
    &-2 rb g^{-\frac{1}{2}}\left(\pi^{i m} \pi_m^j-\frac{1}{2} \pi^{i j} \pi\right)+g^{\frac{1}{2}}\left(D^iD^j\lt(rb\rt)-g^{i j} D^mD_m\lt(rb\rt)\right)\\
&+\mathcal{L}_{\vec{b}} \pi^{i j}+\frac{rb}{2} H_{\text{phys}} N^iN^j,\\
    \end{aligned}
    \end{equation}
with $\vec{b}:=rb\vec{N}$. Before computing the boundary terms of \eqref{eq:boost_varia}, we firstly introduce some useful notations. Following \cite{hennADM}, the extrinsic curvature of a 2-sphere reads
\begin{equation}
K_{A B}=\frac{1}{2 \lambda}\left(-\partial_r g_{A B}+D_A \lambda_B+D_B \lambda_A\right).
\end{equation}
Here $\lambda_A=g_{rA}$. With \eqref{asymsphere1}-\eqref{asymsphere3}, its trace $K:=\gamma^{AB}K_{AB}$ asymptotically behaviors as
\begin{equation} \label{eq:exp_k}
    K=-\frac{2}{r}+\frac{\bar{k}}{r^2}+\frac{\log\lt(r\rt)}{r^3}k^{\lt(\log\rt)}+\frac{k^{\lt(2\rt)}}{r^3}+o\lt(r^{-3}\rt).
\end{equation}
Then the boundary terms of \eqref{eq:boost_varia} read
\begin{equation}
    \begin{aligned}
        \mathcal{K}_b=\pm\lim_{R\to\infty}\oint_{S^2}\rmd^2 S\,\sqrt{\bar{\gamma}}&\lt(-2Rb\delta\bar{k}-2\log\lt(R\rt)b\delta k^{\lt(\log\rt)} -\frac{1}{4}b\delta\lt(\bar{h}^{2}+\bar{h}^{AB}\bar{h}_{AB}\rt)-2b\delta k^{\lt(2\rt)}\rt.\\
        &\lt.+\left(\bar{\lambda}^C \partial_C b \bar{\gamma}^{A B}-b \bar{D}^A \bar{\lambda}^B\right) \delta \bar{h}_{A B}-\bar{h} b \delta\left(2 \bar{\lambda}+\bar{D}_A \bar{\lambda}^A\right)\rt).
    \end{aligned}
\end{equation}
The terms $\left(\bar{\lambda}^C \partial_C b \bar{\gamma}^{A B}-b \bar{D}^A \bar{\lambda}^B\right) \delta \bar{h}_{A B}$ are obviously not to be differentials, so we introduce an additional boundary $\bar{\lambda}_A=0$ to make them vanish. To deal with $-\bar{h} b \delta\left(2 \bar{\lambda}+\bar{D}_A \bar{\lambda}^A\right)$, we require the function $f$ in \eqref{asyvec} to be phase-space dependent:
\begin{equation}
    f=-b\bar{\lambda}-b\bar{k}+T,
\end{equation}
where $T$ is an arbitrary function on $S^2$. The detailed derivation can be found in \cite{hennADM}. 
There are a linear divergent term $\mp2R\oint_{S^2}\rmd^2 S\,\sqrt{\bar{\gamma}}b\delta\bar{k}$ and a logarithmic divergent term $\mp2\log\lt(R\rt)\oint_{S^2}\rmd^2 S\,\sqrt{\bar{\gamma}}b\delta k^{\lt(\log\rt)}$ in \eqref{eq:boost_varia}. By imposing the boundary condition $C^{(1)}=0$, we find that $\mp2R\oint_{S^2}\rmd^2 S\,\sqrt{\bar{\gamma}}b\delta\bar{k}$ vanishes with
\eqref{eq:boost_varia}, see \cite{hennADM}. One way to get rid of $\mp2\log\lt(R\rt)\oint_{S^2}\rmd^2 S\,\sqrt{\bar{\gamma}}b\delta k^{\lt(\log\rt)}$ is to assign even parity to $k^{\lt(\log\rt)}$. The parity condition of $k^{\lt(\log\rt)}$ is not preserved by $G_b$ in general but we can assign additional parity conditions to $\pi^{\lt(\log\rt) ij}$ to ensure the parity condition of $k^{\lt(\log\rt)}$ be preserved, which will be shown shortly. Finally we get
\begin{equation}\label{eq:boost_boun}
    \begin{aligned}
    \mathcal{K}_b=\mp\lim_{R\to\infty}\oint_{S^2}\rmd^2 S\,\sqrt{\bar{\gamma}}b\delta\left(2 k^{(2)}+\bar{k}^2+\bar{k}_B^A \bar{k}_A^B-6 \overline{\lambda k}\right).
    \end{aligned}
\end{equation} Then we can compute the transformations of the canonical variables:
\begin{equation}\label{boostran}
    \begin{aligned}
      \delta_b g_{i j}=&2rbN g^{-\frac{1}{2}}\left(\pi_{i j}-\frac{1}{2} g_{i j} \pi\right)+2D_{(i}\lt(rbN_{j)}\rt),\\
      \delta_ b\pi^{i j}=&-rbN g^{\frac{1}{2}}\left(R^{i j}-\frac{1}{2} g^{i j} R\right)+\frac{1}{2} rbN g^{-\frac{1}{2}}g^{ij}\left(\pi_{m n} \pi^{m n}-\frac{1}{2} \pi^{2}\right) \\
&-2 rbN g^{-\frac{1}{2}}\left(\pi^{i m} \pi_{m}^{j}-\frac{1}{2} \pi^{i j} \pi\right)+g^{\frac{1}{2}}\left(D^iD^j\lt(rbN\rt)-g^{i j} D_mD^m\lt(rbN\rt)\right) \\
&+\mathcal{L}_{\vec{b}} \pi^{i j}+\frac{rb}{2}HN^iN^j.\end{aligned}
\end{equation}
Here $b^j:=rbN^j$. With \eqref{eq:exp_k}, we have    
\begin{equation}
    k^{\lt(\log\rt)}=h^{\lt(\log\rt)A}_A+2\lambda^{\lt(\log\rt)}+\bar{D}^A\lambda^{\lt(\log\rt)}_A,
\end{equation}
where $\lambda^{\lt(\log\rt)}_A=h_{rA}^{\lt(\log\rt)}$. We further introduce $C^{(llll)}_A=C^{(llll)}_r=0$, then $\delta_b g_{i j}$ gives
\begin{equation}
    \begin{aligned}
        \delta_b h_{AB}^{\lt(\log\rt)}=&2b\lt(\pi_{AB}^{\lt(\log\rt)}-\frac{1}{2}\bar{\gamma}_{AB}\pi^{\lt(\log\rt)}\rt),\\
    \delta_b \lambda^{\lt(\log\rt)}=&2b\lt(\pi_{rr}^{\lt(\log\rt)}-\frac{1}{2}\pi^{\lt(\log\rt)}\rt),\\
    \delta_b \lambda_{A}^{\lt(\log\rt)}=&2b\pi_{Ar}^{\lt(\log\rt)},
    \end{aligned}
\end{equation}
with
\begin{equation}
    \pi_{AB}^{\lt(\log\rt)}=\bar{\gamma}_{AC}\bar{\gamma}_{BD}\pi^{\lt(\log\rt)CD},\,
    \pi_{rr}^{\lt(\log\rt)}=\pi^{\lt(\log\rt)rr},\,
    \pi_{rA}^{\lt(\log\rt)}=\bar{\gamma}_{AB}\pi^{\lt(\log\rt)rB},
\end{equation}
and
\begin{equation}
    \pi^{\lt(\log\rt)}=\pi^{\lt(\log\rt)rr}+\bar{\gamma}_{AB}\pi^{\lt(\log\rt)AB}.
\end{equation}
Therefore,
\begin{equation}
    \begin{aligned}
\delta_bk^{\lt(\log\rt)}=&\delta_bh^{\lt(\log\rt)A}_A+2\delta_b\lambda^{\lt(\log\rt)}+\bar{D}_A\delta_b\lambda^{\lt(\log\rt)A}\\
        =&2b\lt(\pi_{A}^{\lt(\log\rt)A}+2\pi_{rr}^{\lt(\log\rt)}-2\pi^{\lt(\log\rt)}+\bar{D}_A\pi^{\lt(\log\rt)rA}
        \rt). 
    \end{aligned}
\end{equation}
As mentioned before, $\pi^{\lt(\log\rt)rA}$ has even parity. We assign odd parity to $\pi^{\lt(\log\rt)rr}$ and $\pi^{\lt(\log\rt)AB}$, then the parity condition of $\delta_bk^{\lt(\log\rt)}$ is preserved by $G_b$. 

Although $G_b$ preserves the parity condition of $k^{\lt(\log\rt)}$, it breaks the boundary conditions conditions of $C_j$.  To see this, we firstly transform \eqref{eq:Boun_with_boo} and \eqref{eq:boo_boun} to the asymptotic Cartesian Coordinates
\begin{equation}
    \begin{aligned}
     C=&\frac{C^{(1)}}{r^3}
+\frac{\log r}{r^{4}}C^{\lt(\log\rt)}
+\frac{C^{(2)}}{r^{4}}
+\frac{\log r}{r^{5}}C^{\lt(ll\rt)}
+\frac{C^{(3)}}{r^{5}}\\
&+\frac{(\log r)^2}{r^{6}}\tilde{C}^{\lt(lll\rt)}
+\frac{\log r}{r^{6}}C^{\lt(lll\rt)}
+o\lt(\frac{\log r}{r^{6}}\rt),
\\
C_j=&\frac{C^{(1)}_j}{r^3}
+\frac{\log r}{r^4}C^{\lt(\log\rt)}_j
+\frac{ C^{(2)}_j}{r^4}
+\frac{\log r}{r^5}C^{(ll)}_j
+\frac{ C^{(3)}_j}{r^5}
+\frac{(\log r)^2}{r^6}\tilde{C}^{(lll)}_j
+\frac{\log r}{r^6}C^{(lll)}_j\\
&+\frac{ C^{(4)}_j}{r^6}
+\frac{(\log r)^2}{r^7}\tilde{C}^{(llll)}_j
+\frac{\log r}{r^7}C^{(llll)}_j
+o\lt(\frac{\log r}{r^{7}}\rt),
\end{aligned}
\end{equation}
\begin{equation}\label{eq:boo_boun_eu}
\begin{aligned}
&C_{j}^{(1)}=C_{j}^{(2)}=C^{(ll)}_j=C_{j}^{(3)}=C^{(\log)}_{j}=\tilde{C}^{(lll)}_j=C^{(lll)}_j=C_j^{(4)}\\
&=C^{(1)}=C^{(\log)}=C^{(2)}=C^{(ll)} =0,\\
&C^{(3)} \neq 0.
\end{aligned}
\end{equation}
By \eqref{eq:boo_boun_eu}, $C_j=O\lt(\frac{\lt(\log r\rt)^2}{r^7}\rt)$ and $H_{\text{phys}}=O\lt(r^{-5}\rt)$. From \eqref{trancj}
\begin{equation}
    \begin{aligned}
        \lt\{C_j,G_b\rt\}=&\partial_j\lt(rb\rt)H_{\text{phys}}\\
        =&O\lt(r^{-5}\rt),
    \end{aligned}
\end{equation}
Then $C_j^{(3)}$ becomes non-vanishing, so the boundary conditions of $C_j$ in \eqref{eq:boo_boun_eu} are broken.
These boundary conditions are important for ensuring that both $\delta \int_{\cs} \rmd^3\mathscr{V}\,  H_{\text{phys}}$ and $\delta\int_{\cs} \rmd^3\mathscr{V}\,rb  H_{\text{phys}}$ are well-defined by adding appropriate boundary terms. Indeed, similar to the discussions in Subsection \ref{sec:varphyham}, there are boundary terms $-2\oint_{S^2} \rmd^2 S\, N^j\delta 
\lt(\pi^{rk}g_{jk}\rt)$ and $-2R\oint_{S^2} \rmd^2 S\, bN^j\delta 
\lt(\pi^{rk}g_{jk}\rt)$ from $\delta\int_{\cs} \rmd^3\mathscr{V}\, H_{\text{phys}}$ and $\delta \int_{\cs} \rmd^3\mathscr{V}\,rb H_{\text{phys}}$ respectively. $b$ has odd parity, so the boundary terms cannot vanish at the same time, no matter what parity condition is assigned to $N^j\delta \lt(\pi^{rk}g_{jk}\rt)$. Imposing the boundary conditions in \eqref{eq:Boun_with_boo} can resolve this problem. Since the boundary conditions in \eqref{eq:Boun_with_boo} are broken, $G_b$ is not a boundary-preserving symmetry charge.

\section{Conclusion and outlook}\label{conclude}
In this paper, we introduce asymptotic boundary conditions for the asymptotic flatness in BK formalism and investigate the symmetries preserving these conditions. A counter term is added to the symplectic form to make it finite. We add the boundary term to the physical Hamiltonian ${\bf H}_{{\rm phys}}$ to make $\delta \mathbf{H}_{\text {phys }}$ well-defined. The boundary term coincides with the ADM mass.

We define the boundary-preserving symmetry charges on the reduced phase space consisting contributions from both bulk and boundary of the dust space. Unlike the usual formulation in general relativity (on the non-reduced phase space), the bulk terms of these charges do not generally vanish on-shell. The Poisson brackets of the charges form a closed Lie algebra up to a central term. The resulting Lie algebra of symmetry charges relates to the BMS algebra by a suitable quotient.

Our work is the first step of investigating the boundary terms in relational formalism. There are several projects we plan to do in future:
\begin{itemize}
    \item We can apply our work to the connection dynamics formalism \cite{ashtekar1986new}. In this formalism, the canonical variables are converted to the densitized tetrads and Ashtekar-Barbero connections $\lt\{E^a,A^a\rt\}$. 
    In the canonical formalism of loop quantum gravity, $\lt\{E^a,A^a\rt\}$ are the basic variables in the quantization (e.g., see \cite{han2007fundamental, thiemann2008modern,ashtekar2004background}). 
    Applying our work in connection dynamics may benefit our understanding of asymptotically flat quantum gravity.
    
    \item It may be interesting to investigate the BK formalism in asymptotically anti-de Sitter (AdS) spacetimes.
    Unlike asymptotically flat spacetimes, the boundary of an asymptotically AdS spacetime is a $3$-dimensional timelike hypersurface \cite{ashtekar2000asymptotically,marolf2014conserved}. 
    This may relate to the bulk/boundary duality.
    
    \item  In this work, we require the asymptotic symmetry preserve the metric on $S^2$. We may relax this requirement and allow generic diffeomorphisms on $S^2$ as the asymptotic symmetries. Some existing results along this direction are given in e.g. \cite{barnich2010aspects, freidel2021weyl, freidel2021extended}.

    \item In general relativity with asymptotically flat spacetime, asymptotic symmetry charges have close relation with Weinberg’s soft theorem \cite{strominger2018lectures}. 
    For example, Refs.\cite{strominger2014bms, he2015bms} show the equivalence between the BMS charges and soft graviton theorem \cite{weinberg1965infrared,weinberg1995quantum} at the null infinity by scattering a massless scalar field.
    There is a subtlety to applying this analysis to BK formalism, since BK formalism can only analyze the structure at the spatial infinity, whereas there is no dynamics at the spatial infinity. Therefore, we expect that need the analysis of the scattering process should be carried out in the regime near spatial infinity, but still with finite distance. We plan to do this analysis in our future work.
\end{itemize}

\section*{Acknowledgments}
This work receives support from the National Science Foundation through grants PHY-2207763 and the Blaumann foundation. M.H. acknowledges IQG at FAU Erlangen-N\"urnberg, Perimeter Institute for Theoretical Physics, and University of Western Ontario for the hospitality during his visits.

\appendix

\section{The Boundary terms from $N\delta C$}\label{bunofc}
Use  (1.5.10) In \cite{thiemann2008modern}, the boundary terms from $N\delta C$ are:
\begin{equation}\label{bounofndc}
    \begin{aligned}
       \lt[N\delta C\rt]_{{\rm boundary}}= &-\lim_{R\to \infty}\oint_{S^2}\, \sqrt{g} g^{c d} g^{e f}\left[\left(D_{c} N\right)\left(d S_{d} \delta g_{e f}\right)-\left(D_{e} N\right)\left(d S_{c} \delta g_{d f}\right)\right] \\
&-\lim_{R\to \infty}\oint_{S^2} \sqrt{g} g^{c d} N\left[-d S_{c} \delta \Gamma_{e d}^{e}+d S_{e} \delta \Gamma_{c d}^{e}\right],
    \end{aligned}
\end{equation}
with $\sqrt{g}=\lambda\sqrt{\gamma}$. With the results in appendix A of \cite{hennADM}, eq.\eqref{bounofndc} can be expressed in the spherical coordinates:

\begin{equation}
    \begin{aligned}
         \lt[N\delta C\rt]_{{\rm boundary}}
=&\lim_{R\to\infty}\oint_{S^2}\rmd^2 S\,\sqrt{\gamma}\lt(-\frac{1}{\lambda} \gamma^{AB}\lt(D_{r} N\rt) \delta \gamma_{AB}
+\frac{\lambda^A}{\lambda} \gamma^{BC}\lt(D_{A} N\rt) \delta  \gamma_{BC}\rt)\\
&-2\lim_{R\to\infty}\oint_{S^2}\rmd^2 S\,\lt(N\delta K
+N\delta\lt(\gamma_{AC}\rt)\gamma^{AB}  K^{C}_B\rt)
\\
&+\lim_{R\to\infty}\oint_{S^2} \rmd^2 S\,\sqrt{\gamma}\frac{N}{\lambda}
\delta
\lt[
\lt(
    -\frac{\lambda^{A}}{\lambda}\left(\partial_{A} \lambda+K_{A C} \lambda^{C}\right)+D_{A} \lambda^{A}
\rt)
-D_{A} \delta\lambda^{A}
\rt]\\
&-\lim_{R\to\infty}\oint_{S^2} \rmd^2 S\,\sqrt{\gamma}N\frac{\lambda^A}{\lambda}\lt(
    \frac{1}{2}\gamma^{BC}D_A\delta \gamma_{BC}-\delta \lt(\frac{\lambda^{B}}{\lambda} K_{B A}\rt)
\rt)\\
&+\lim_{R\to\infty}\oint_{S^2}\rmd^2 S\,\sqrt{\gamma}N\frac{\lambda^A}{\lambda}
\delta
\lt(
    \frac{1}{\lambda}\left(\partial_{A} \lambda+K_{A B} \lambda^{B}\right)
\rt)\\
&+\lim_{R\to\infty}\oint_{S^2}\rmd^2 S\,\lambda\sqrt{\gamma}\lt(-N\frac{\lambda^A\lambda^B}{\lambda^5}\delta K_{AB}
+NK_{AB}\frac{\lambda^A\lambda^B}{\lambda^4}\frac{\delta\lambda}{\lambda^2}\rt)\\
&+\lim_{R\to\infty}\oint_{S^2}\rmd^2 S\,\sqrt{\gamma}\lt(
- ND^{A} \lt(\frac{1}{\lambda}\rt)\delta \lambda_A
-\frac{\lambda^A}{\lambda}  \gamma^{BC}\lt(D_{C} N\rt) \delta \gamma_{AB}\rt).
    \end{aligned}
\end{equation}

\section{The conservation of the  Boundary Charges}\label{app:con_boun_char}
This appendix demonstrates the boundary charges defined in  \eqref{eq:fin-B} are conserved. It is convenient to demonstrate this statement in the asymptotic Cartesian coordinates. We firstly compute the physical time derivative of $g_{ij}$:
\begin{equation}\label{eq:time_deri_of_gij}
  \frac{\rmd g_{ij}}{\rmd\tau}= N g^{-\frac{1}{2}}\left(\pi_{ij}-\frac{1}{2} g_{ij} \pi\right)+ \mathcal{L}_{\vec{N}}g_{ij}=O\lt(r^{-2}\rt).
\end{equation}
The physical time derivative of the $\bar h_{ij}$ in \eqref{asympq} is given by expanding \eqref{eq:time_deri_of_gij} to the boundary, which reads 
\begin{equation}
    \frac{\rmd\bar h_{ij}}{\rmd\tau}=0.
\end{equation}
Note that all of the metric variables in \eqref{eq:fin-B} are the components of $\bar h_{ij}$, so their physical time derivatives vanish. Next, we compute the physical time derivative of $\pi^{ij}$:
\begin{equation}\label{eq:time_deri_of_pij}
\begin{aligned}
      \frac{\rmd \pi^{ij}}{\rmd\tau}
  =&-N g^{\frac{1}{2}}\left(g^{ik}g^{jl}{}^{\lt(3\rt)}R_{lk}-\frac{1}{2} g^{i j} {}^{\lt(3\rt)}R\right)+\frac{1}{2} N g^{-\frac{1}{2}}g^{ij}\left(\pi_{m n} \pi^{m n}-\frac{1}{2} \pi^{2}\right)\nonumber \\
&-2 N g^{-\frac{1}{2}}\left(\pi^{i m} \pi_{m}^{j}-\frac{1}{2} \pi^{i j} \pi\right)+g^{\frac{1}{2}}\left(D^iD^jN-g^{i j} D_mD^mN\right)\nonumber \\
&+\frac{1}{2}H_{\text{phys}}N^iN^j+\mathcal{L}_{\vec{N}} \pi^{i j}\\
=&O(r^{-3}).
\end{aligned}
\end{equation}
It turns out that
\begin{equation}
    \frac{\rmd\bar \pi^{ij}}{\rmd\tau}=0.
\end{equation}
Thus, the physical time derivatives of the  conjugate momenta with a bar on the top in \eqref{eq:fin-B} vanish, due to they are the components of $\bar{\pi}^{ij}$. Finally, we deal with the term with $\pi^{(2)rA}$. By \eqref{eq:time_deri_of_pij},
\begin{equation}
 \frac{\rmd \pi^{(2)rA}}{\rmd\tau}=\mp\sqrt{\bar{\gamma}}\bar{\gamma}^{AB}{}^{(3)}R^{(1)}_{rB}.
\end{equation}
Here ${}^{(3)}R^{(1)}_{rA}$ is the coefficient of the leading order of ${}^{(3)}R_{rA}$, which reads
\begin{equation}
    \begin{aligned}
        {}^{(3)}R^{(1)}_{rA}=&\bar{D}_B\bar{D}_A\bar{\lambda}^B
    -\bar{D}_B\bar{k}^B_A
+\frac{1}{2}
    \lt(
        \bar\gamma^{BC}\partial_A\bar h_{BC}
        -\bar h^{BC}\partial_A\bar\gamma_{BC}
    \rt)
    -\bar{\lambda}_A
    +2\partial_A\bar{\lambda}\\
    =&\bar{D}_A\bar{D}_B\bar{\lambda}^B
    +{}^{\bar\gamma}R_{AB}\bar{\lambda}^B
    -\bar{D}_B\bar{k}^B_A
+\frac{1}{2}\bar\gamma^{BC}\bar D_A\bar h_{BC}
-\bar{\lambda}_A
    +2\partial_A\bar{\lambda}\\
    =&\bar{D}_A\bar{D}_B\bar{\lambda}^B
    -\bar{D}_B\bar{k}^B_A
+\frac{1}{2}\bar\gamma^{BC}\bar D_A\bar h_{BC}
    +2\partial_A\bar{\lambda}.
    \end{aligned}
\end{equation}
In the last step, we use the fact that for a unit sphere, we have
\begin{equation}
{}^{\bar\gamma}R_{AB}=\bar{\gamma}_{AB}. 
\end{equation}
Then we find
\begin{equation}
    \begin{aligned}
        2\oint_{S^2}\rmd^2 S\, Y^A\bar{\gamma}_{AB}\frac{\rmd \pi^{(2)rB}}{\rmd\tau}
    =&\mp2\oint_{S^2}\rmd^2 S\, Y^A
    \lt(
    \bar{D}_A\bar{D}_B\bar{\lambda}^B
    -\bar{D}_B\bar{k}^B_A
+\frac{1}{2}\bar\gamma^{BC}\bar D_A\bar h_{BC}
    +2\partial_A\bar{\lambda}
    \rt)\\=&0.
    \end{aligned}
\end{equation}
In the last step, we integrate by part and use the Killing equation of $Y^A$. As the result, we demonstrate that the boundary charges defined in  \eqref{eq:fin-B} are conserved.

\section{Properties of $\lt(\hat{\xi},\hat{\xi}^j\rt)$}\label{app:dis_of_xi}
From \eqref{eq:def_of_hat_xi}, we have
\begin{equation}
\hat{\xi}= \widehat{f}+O\left(r^{-1}\right), \quad \hat{\xi}^{r}=\widehat{W}+O\left(r^{-1}\right), \quad \hat{\xi}^{A}=\widehat{Y}^{A}+\frac{1}{r} \widehat{I}^{A}+O\left(r^{-2}\right),
\end{equation}
\begin{equation}
    \begin{aligned}
\widehat{Y}^{A} &=Y_{1}^{B} \bar{D}_{B} Y_{2}^{A}-\lt(1 \leftrightarrow 2\rt), \\
\widehat{f} &=Y_{1}^{A} \partial_{A} f_{2}-\lt(1 \leftrightarrow 2\rt), \\
\widehat{W} &=Y_{1}^{A} \partial_{A} W_{2}-\lt(1 \leftrightarrow 2\rt),\\
\widehat{I}^{A}&=Y_{1}^{B} \bar{D}_{B} I_{2}^{A}+I_{1}^{B} \bar{D}_{B} Y_{2}^{A}-\lt(1 \leftrightarrow 2\rt).
\end{aligned}
\end{equation}
Since $f$, $W$ are functions of $S^2$ and $Y^A$ is the Killing vector of $S^2$, $\widehat{f}$, $\widehat{W}$ are also functions of $S^2$. similarly,  $\widehat{I}^{A}$ is the vector field of $S^2$.
Note that $Y_1^A$, $Y_2^A$ are the Killing vector fields on $S^2$, their commutator $\widehat{Y}^{A}:=\lt[Y_1,Y_2\rt]^A$ is also the Killing vector field on $S^2$. Therefore, $\lt(\hat{\xi},\hat{\xi}^j\rt)$ satisfies  \eqref{asyvec}.

\section{Details of Subsection \ref{subsec:gfa}}\label{app:detaiofcal}
This subsection gives some detail calculations of Subsection \ref{subsec:gfa}.
We firstly give the details of \eqref{pois2}.
By \eqref{dynl} and \eqref{dyns}, \eqref{pois2} yields 
\begin{equation}
    \begin{aligned}
     &\lt\{G\lt(\vec{\xi}_1\rt),G\lt(\xi_2\rt)\rt\}\\
      =&
\int_{\cs}\rmd^3\mathscr{V}\,
\xi_1^iD_i\lt(N\xi_2\rt)C
+\int_{\cs}\rmd^3\mathscr{V}\,\lt[\xi_1^iD_i\lt(N^j\xi_2\rt)
-N^i\xi_2D_i\xi_1^j \rt] C_j\\
&-\int_{\cs}\rmd^3\mathscr{V}\,\xi_2 H_{\rm phy}N^jN^k
D_j\lt(\xi_{1}\rt)_k
 -\delta_{\vec{\xi}_1}\tilde{\mathcal{B}}\lt(\tilde\xi_2\rt)\\
&-\lim_{r\to \infty}\oint_{S^2}\rmd^2 S\,\xi_1^r\tilde{\xi}_2C
 +2\lim_{r\to \infty}\oint_{S^2}\rmd^2 S\,\lt[\vec{\xi}_1,\vec{\tilde\xi}_2\rt]^j \pi^{r}_j\\
=&
\int_{\cs}\rmd^3\mathscr{V}\,\lt[
\pm\xi_1^i\frac{g^{jk}N_jD_iN_k}{\sqrt{1+g^{jk}N_jN_k}}\xi_2C
+\xi_1^iD_i\lt(\xi_2\rt)\frac{C^2}{H_{\text{phys}}}\rt]\\
&+\int_{\cs}\rmd^3\mathscr{V}\,
\lt[
    \xi_1^iD_i\lt(N^j\rt)\xi_2
    -\frac{C^j}{H_{\text{phys}}}\xi_1^iD_i\lt(\xi_2\rt)
    +\frac{C^i}{H_{\text{phys}}}\xi_2D_i\xi_1^j 
\rt] C_j\\
 &-\int_{\cs}\rmd^3\mathscr{V}\,\xi_2 \frac{C^jC^k}{H_{\text{phys}}}
D_j\lt(\xi_{1}\rt)_k-\delta_{\vec{\xi}_1}\tilde{\mathcal{B}}\lt(\tilde\xi_2\rt)\\
&-\lim_{r\to \infty}\oint_{S^2}\rmd^2 S\,\xi_1^r\tilde{\xi}_2C
 +2\lim_{r\to \infty}\oint_{S^2}\rmd^2 S\,\lt[\vec{\xi}_1,\vec{\tilde\xi}_2\rt]^j \pi^{r}_j\\
=&
\int_{\cs}\rmd^3\mathscr{V}\,
\lt[
\xi_1^iN_jD_i\lt(N^j\rt)H_{\rm phy}\xi_2
+\xi_1^iD_i\lt(N^j\rt)\xi_2C_j\rt]
+\int_{\cs}\rmd^3\mathscr{V}\,
    \xi_1^iD_i\lt(\xi_2\rt)\frac{C^2-g^{jk}C_jC_k}{H_{\text{phys}}}
\\
&+\int_{\cs}\rmd^3\mathscr{V}\,\xi_2\lt[\frac{C^jC^k}{H_{\text{phys}}}\xi_2D_j\lt(\xi_{1}\rt)_k- \frac{C^jC^k}{H_{\text{phys}}}
D_j\lt(\xi_{1}\rt)_k\rt]\\
&-\delta_{\vec{\xi}_1}\tilde{\mathcal{B}}\lt(\xi_2\rt)
-\oint_{S^2}\rmd^2 S\,\xi_1^r\tilde{\xi}_2C
 +2\oint_{S^2}\rmd^2 S\,\lt[\vec{\xi}_1,\vec{\tilde\xi}_2\rt]^j \pi^{r}_j\\
=&
\int_{\cs}\rmd^3\mathscr{V}\,
\xi_1^iD_i\lt(\xi_2\rt)H_{\text{phys}}-\delta_{\vec{\xi}_1}\tilde{\mathcal{B}}\lt(\tilde\xi_2\rt)
-\lim_{r\to \infty}\oint_{S^2}\rmd^2 S\,\xi_1^r\tilde{\xi}_2C
 +2\lim_{r\to \infty}\oint_{S^2}\rmd^2 S\,\lt[\vec{\xi}_1,\vec{\tilde\xi}_2\rt]^j \pi^{r}_j.
    \end{aligned}
\end{equation}
Next, we show that $ \mathscr{B}$ in \eqref{poiHH} is vanished. 
By \eqref{dynl} and \eqref{dyns}, 
\begin{equation}
    \begin{aligned}
    \mathscr{B}
=&
\int_{\cs}\rmd^3\mathscr{V}\,
    \lt[N\xi_1D^j\lt(N\xi_2\rt)-N\xi_2D^j\lt(N\xi_1\rt)\rt]C_j\\
   & +
\int_{\cs}\rmd^3\mathscr{V}\,
\lt[N^j\xi_1D_j\lt(N\xi_2\rt)
-
N^j\xi_2D_j\lt(N\xi_1\rt)\rt]C
\\
&+\int_{\cs}\rmd^3\mathscr{V}\,
\lt[
    N^i\xi_1D_i\lt(N^j\xi_2\rt)
    -N^i\xi_2D_i\lt(N^j\xi_1\rt)
\rt] C_j\\
&+\frac{1}{2}\int_{\cs}\rmd^3\mathscr{V}\,H_{\text{phys}}N^iN^j\xi_1N\xi_2 g^{-\frac{1}{2}}\lt(\pi_{ij}-\frac{1}{2}\pi g_{ij}\rt)\\
&-\frac{1}{2}\int_{\cs}\rmd^3\mathscr{V}\,H_{\text{phys}}N^iN^j\xi_2N\xi_1 g^{-\frac{1}{2}}\lt(\pi_{ij}-\frac{1}{2}\pi g_{ij}\rt)\\
&+\int_{\cs}\rmd^3\mathscr{V}\,H_{\text{phys}}N^iN^j\xi_1D_i\lt(N_j\xi_{2} \rt)\\
&-\int_{\cs}\rmd^3\mathscr{V}\,H_{\text{phys}}N^iN^j\xi_2D_i\lt(N_j\xi_{1} \rt)\\              
    =&
\int_{\cs}\rmd^3\mathscr{V}\,
    \lt(N^2\xi_1D^j\xi_2
    -N^2\xi_2D^j\xi_1
    \rt)
    C_j\\
    &+
   \int_{\cs}\rmd^3\mathscr{V}\,
   \lt(
    NN^j\xi_1D_j\xi_2
    -NN^j\xi_2D_j\xi_1
   \rt)C
\\
&+\int_{\cs}\rmd^3\mathscr{V}\,
\lt(
    \xi_1N^iN^jD_i\xi_2
    -\xi_2N^iN^jD_i\xi_1
\rt) C_j\\
&+\int_{\cs}\rmd^3\mathscr{V}\,H_{\text{phys}}N^iN^j\xi_1N_jD_i\xi_{2}\\
&-\int_{\cs}\rmd^3\mathscr{V}\, H_{\text{phys}}N^iN^j\xi_2N_jD_i\xi_{1} \\
=&
\int_{\cs}\rmd^3\mathscr{V}\,
    \lt(\frac{C^2}{H_{\text{phys}}^2}\xi_1C_jD^j\xi_2
    -\frac{C^2}{H_{\text{phys}}^2}\xi_2C_jD^j\xi_1
    \rt)\\
    &-\int_{\cs}\rmd^3\mathscr{V}\,
      \lt(\frac{C^2}{H_{\text{phys}}^2}\xi_1C_jD^j\xi_2
    -\frac{C^2}{H_{\text{phys}}^2}\xi_2C_jD^j\xi_1
    \rt)
\\
&+\int_{\cs}\rmd^3\mathscr{V}\,
\lt(
    \xi_1N^iN^jC_jD_i\xi_2
    -\xi_2N^iN^jC_jD_i\xi_1
\rt) \\
&-\int_{\cs}\rmd^3\mathscr{V}\,\lt(N^iN^j\xi_1C_jD_i\xi_{2}
- N^iN^j\xi_2C_jD_i\xi_{1}\rt)\\
=&0.
    \end{aligned}
\end{equation}

\section{Proof of \eqref{DlambA}}\label{app: proof}
\begin{equation}
    \begin{aligned}
      &\bar{D}_A\lt(\mathcal{L}_{\vec{Y}}\bar\lambda_B\rt)\\
=&\bar{D}_A
\lt(
    Y^C\bar D_{C}\bar \lambda_B
    +\bar\lambda_C\bar{D}_{B}Y^C
\rt)\\
=&
     Y^C\bar{D}_A\bar D_{C} \bar\lambda_B
    +\bar{D}_A\lt(Y^C\rt)\bar D_{C} \bar\lambda_B
         +\bar{D}_A\lt(\bar\lambda_C\rt)\bar{D}_{B}Y^C
    +\bar\lambda_C\bar{D}_A\bar{D}_{B}Y^C
\\
=&
     Y^C\bar{D}_C\bar D_{A} \bar\lambda_B
     +Y^C{}^{\bar\gamma} R_{ACB}{}^{ D} \bar\lambda_D
    +\bar{D}_A\lt(Y^C\rt)\bar D_{C} \bar\lambda_B
         +\bar{D}_A\lt(\bar\lambda_C\rt)\bar{D}_{B}Y^C
    +\bar\lambda_C\bar{D}_A\bar{D}_{B}Y^C.
    \end{aligned}
\end{equation}
On the other hand
\begin{equation}
    \begin{aligned}
      &\mathcal{L}_{\vec{Y}}\lt(\bar{D}_A\bar\lambda_B\rt)\\
=&
     Y^C\bar{D}_C\bar D_{A} \bar\lambda_B
    +\bar D_{C} \lt(\bar\lambda_B\rt)\bar{D}_AY^C
         +\bar{D}_A(\bar\lambda_C)\bar{D}_{B}Y^C
    \end{aligned}
\end{equation}
To prove \eqref{DlambA}, we need to show
\begin{equation}
    Y^C{}^{\bar\gamma} R_{ACB}{}^{ D} \bar\lambda_D+\bar\lambda_C\bar{D}_A\bar{D}_{B}Y^C=0.
\end{equation}
For a Killing vector field on the spacetime $\xi^\mu$, we have 
\begin{equation}
    \nabla_{\mu} \nabla_{\nu} \xi_{\sigma}=-R_{\nu \sigma \mu}{}^{ \rho} \xi_\rho,
\end{equation}
then
\begin{equation}
    \begin{aligned}
      &\bar\lambda_C\bar{D}_A\bar{D}_{B}Y^C\\
=&\bar\lambda^{C}\bar{D}_A\bar{D}_{B}Y_C\\
=&-\bar\lambda^{C}{}^{\bar\gamma} R_{BCA}{}^{ D}Y_D,\\
    \end{aligned}
\end{equation}
We have
\begin{equation}
    \begin{aligned}
      &Y^C{}^{\bar\gamma} R_{ACB}{}^{D} \bar\lambda_D+\bar\lambda_C\bar{D}_A\bar{D}_{B}Y^C\\
=&Y^C{}^{\bar\gamma} R_{ACB}{}^{ D} \bar\lambda_D-\bar\lambda^{C}{}^{\bar\gamma} R_{BCA}{}^{ D}Y_D\\
=&Y^C\bar\lambda^{D}{}^{\bar\gamma} R_{ACBD}
-Y^D\bar\lambda^{C}{}^{\bar\gamma} R_{BCAD}\\
=&Y^C\bar\lambda^{D}{}^{\bar\gamma} R_{ACBD}
-Y^C\bar\lambda^{D}{}^{\bar\gamma} R_{ACBD}\\
=&0.
    \end{aligned}
\end{equation}
\section{Central Charges}\label{prof-co}
We first give the definition of the central extension of a Lie algebra $L_g$, this definition can be find in e.g. \cite{kohno2002conformal}. For a Lie algebra $L \mathfrak{g}$, its central extension $\widehat{g}$ is the direct sum of itself and a one dimensional complex vector space with the basis $c$
\begin{equation}
L \mathfrak{g} \oplus \mathbf{C} c,
\end{equation}
satisfying the following requirements:
\begin{equation}
    \begin{aligned}
        &[\xi+\alpha c, \eta+\beta c]=[\xi, \eta]+\omega(\xi, \eta) c, \quad \xi, \eta \in L \mathfrak{g}, \quad \alpha, \beta \in \mathbf{C},\\
        &[c, \xi]=0.
    \end{aligned}
\end{equation}
Here the square bracket denotes the Lie bracket of the algebra, and the $\omega$ is a bilinear form $\omega:L \mathfrak{g} \times L \mathfrak{g} \rightarrow \mathbf{C}$ satisfying the 2-cocycle condition:
\begin{equation}
\begin{gathered}
\omega(x, y)=-\omega(y, x), \\
\omega([x, y], z)+\omega([y, z], x)+\omega([z, x], y)=0,
\end{gathered}
\end{equation}
with $x, y, z \in L \mathfrak{g}$.

With the definition above, we can check that the $\mathscr{C}\lt(\hat{\xi}, \hat{\vec{\xi}}\rt)$ in \eqref{algebra} provides a central extension of the algebra.  Introduce $\xi^\mu_1:=\lt(\xi_1,\vec{\xi}_1\rt)$, $\xi^\mu_2:=\lt(\xi_2,\vec{\xi}_2\rt)$, and $\xi^\mu_3:=\lt(\xi_3,\vec{\xi}_3\rt)$, which satisfy \eqref{asyvec}. Define their commutators with Poisson bracket:
\begin{equation}
    \lt\{G\lt(\xi^\mu_1\rt),G\lt(\xi^\mu_2\rt)\rt\}=G\lt(\lt[\xi^\mu_1,\xi^\mu_2\rt]\rt)+\mathscr{C}\lt(\lt[\xi^\mu_1,\xi^\mu_2\rt]\rt).
\end{equation}
Here $\lt[\xi^\mu_1,\xi^\mu_2\rt]$ is the commutator of $\xi^\mu_1$ and $\xi^\mu_2$ defined by the Poisson bracket. In our case, $\xi^\mu \in L \mathfrak{g}$. the complex vector space $\mathbf{C}$ is restricted to be the real vector $\mathbf{R}$. Especially, $c=1$.  The $\omega$ is given by
$$
\omega\left(\xi^\mu_1,\xi^\mu_1\rt)\equiv\mathscr{C}\lt(\lt[\xi^\mu_1,\xi^\mu_2\rt]\rt).
$$
By \eqref{cenchar}, 
\begin{equation}
    \omega\lt(\lt[\xi^\mu_1,\xi^\mu_2\rt]\rt)=-\omega\lt(\lt[\xi^\mu_2,\xi^\mu_1\rt]\rt).
\end{equation}
By \eqref{algebofthefun} and \eqref{cenchar},
\begin{equation}\label{centcom}
    \begin{aligned}
      &\omega\lt(\lt[\xi^\mu_1,\xi^\mu_2\rt],\xi^\mu_3\rt)\\
=&\pm2\oint_{S^2}\rmd^2 S\, \sqrt{\bar{\gamma}} \lt[
\lt(
    Y_{1}^{A} \bar{D}_{A} f_{2}
    -Y_{2}^{A} \bar{D}_{A} f_{1}
\rt)
\lt(
    \bar{D}^{B}\bar{D}_{B}W_3
    -\bar{D}_{B} 
    I_3^{B}
\rt)\rt.\\
&-f_3 
    \bar{D}^{A}\bar{D}_{A}
\lt(
        Y_{1}^{B} \bar{D}_{B} W_{2}
        -Y_{2}^{B} \bar{D}_{B} W_{1}
    \rt)\\
    &+\lt.f_3\bar{D}_{A}
    \lt(
        Y_{1}^{B} \bar{D}_{B} I_{2}^{A}
        +I_{1}^{B} \bar{D}_{B} Y_{2}^{A}
        -Y_{2}^{B} \bar{D}_{B} I_{1}^{A}
        -I_{2}^{B} \bar{D}_{B} Y_{1}^{A}
    \rt)
\rt].
    \end{aligned}
\end{equation}
Since
\begin{equation}
    \begin{aligned}
      &\bar{D}_{A}
    \lt(
        Y_{1}^{B} \bar{D}_{B} I_{2}^{A}
        +I_{1}^{B} \bar{D}_{B} Y_{2}^{A}
        -Y_{2}^{B} \bar{D}_{B} I_{1}^{A}
        -I_{2}^{B} \bar{D}_{B} Y_{1}^{A}
    \rt)
\\
=&
\bar{D}_{A}\lt(Y_{1}^{B}\rt) \bar{D}_{B} I_{2}^{A}
+Y_{1}^{B} \bar{D}_{A}\bar{D}_{B} I_{2}^{A}
+\bar{D}_{A}\lt(I_{1}^{B}\rt) \bar{D}_{B} Y_{2}^{A}
+I_{1}^{B} \bar{D}_{A}\bar{D}_{B} Y_{2}^{A}\\
&-\bar{D}_{A}\lt(Y_{2}^{B}\rt) \bar{D}_{B} I_{1}^{A}
-Y_{2}^{B} \bar{D}_{A}\bar{D}_{B} I_{1}^{A}
-\bar{D}_{A}\lt(I_{2}^{B}\rt) \bar{D}_{B} Y_{1}^{A}
-I_{2}^{B} \bar{D}_{A}\bar{D}_{B} Y_{1}^{A}
\\
=&
\bar{D}_{A}\lt(Y_{1}^{B}\rt) \bar{D}_{B} I_{2}^{A}
+Y_{1}^{B} \bar{D}_{B}\bar{D}_{A} I_{2}^{A}
+Y_{1}^{B} R^{\quad\quad A}_{BAC} I_{2}^{C}\\
&+\bar{D}_{A}\lt(I_{1}^{B}\rt) \bar{D}_{B} Y_{2}^{A}
+I_{1}^{B} \bar{D}_{B}\bar{D}_{A} Y_{2}^{A}
+I_{1}^{B} R^{\quad\quad A}_{BAC} Y_{2}^{C}\\
&-\bar{D}_{A}\lt(Y_{2}^{B}\rt) \bar{D}_{B} I_{1}^{A}
-Y_{2}^{B} \bar{D}_{B}\bar{D}_{A} I_{1}^{A}
-Y_{2}^{B} R^{\quad\quad A}_{BAC} I_{1}^{C}\\
&-\bar{D}_{A}\lt(I_{2}^{B}\rt) \bar{D}_{B} Y_{1}^{A}
-I_{2}^{B} \bar{D}_{B}\bar{D}_{A} Y_{1}^{A}
-I_{2}^{B} R^{\quad\quad A}_{BAC} Y_{1}^{C}
\\
=&
\bar{D}_{A}\lt(Y_{1}^{B}\rt) \bar{D}_{B} I_{2}^{A}
+Y_{1}^{B} \bar{D}_{B}\bar{D}_{A} I_{2}^{A}
+Y_{1}^{B} R_{BC} I_{2}^{C}\\
&+\bar{D}_{A}\lt(I_{1}^{B}\rt) \bar{D}_{B} Y_{2}^{A}
+I_{1}^{B} \bar{D}_{B}\bar{D}_{A} Y_{2}^{A}
+I_{1}^{B} R_{BC} Y_{2}^{C}\\
&-\bar{D}_{A}\lt(Y_{2}^{B}\rt) \bar{D}_{B} I_{1}^{A}
-Y_{2}^{B} \bar{D}_{B}\bar{D}_{A} I_{1}^{A}
-Y_{2}^{B} R_{BC} I_{1}^{C}\\
&-\bar{D}_{A}\lt(I_{2}^{B}\rt) \bar{D}_{B} Y_{1}^{A}
-I_{2}^{B} \bar{D}_{B}\bar{D}_{A} Y_{1}^{A}
-I_{2}^{B} R_{BC} Y_{1}^{C}
\\
=&
\bar{D}_{A}\lt(Y_{1}^{B}\rt) \bar{D}_{B} I_{2}^{A}
+Y_{1}^{B} \bar{D}_{B}\bar{D}_{A} I_{2}^{A}
+\bar{D}_{A}(I_{1}^{B}) \bar{D}_{B} Y_{2}^{A}
\\
&-\bar{D}_{A}\lt(Y_{2}^{B}\rt) \bar{D}_{B} I_{1}^{A}
-Y_{2}^{B} \bar{D}_{B}\bar{D}_{A} I_{1}^{A}
-\bar{D}_{A}(I_{2}^{B}) \bar{D}_{B} Y_{1}^{A}
\\
=&
Y_{1}^{B} \bar{D}_{B}\bar{D}_{A} I_{2}^{A}
-Y_{2}^{B} \bar{D}_{B}\bar{D}_{A} I_{1}^{A},
    \end{aligned}
\end{equation}
Eq. \eqref{centcom} becomes
\begin{equation}
    \begin{aligned}      &\omega\lt(\lt[\xi^\mu_1,\xi^\mu_2\rt],\xi^\mu_3\rt)\\
=&\pm2\oint_{S^2}\rmd^2 S\, \sqrt{\bar{\gamma}} \lt[
\lt(
    Y_{1}^{A} \bar{D}_{A} f_{2}
    -Y_{2}^{A} \bar{D}_{A} f_{1}
\rt)
\lt(
    \bar{D}^{B}\bar{D}_{B}W_3
    -\bar{D}_{B} 
    I_3^{B}
\rt)\rt.\\
&-f_3 
    \bar{D}^{A}\bar{D}_{A}
\lt(
        Y_{1}^{B} \bar{D}_{B} W_{2}
        -Y_{2}^{B} \bar{D}_{B} W_{1}
    \rt)\\
    &+\lt.f_3\bar{D}_{A}
    \lt(
       Y_{1}^{B} \bar{D}_{B}\bar{D}_{A} I_{2}^{A}
-Y_{2}^{B} \bar{D}_{B}\bar{D}_{A} I_{1}^{A}
    \rt)
\rt].
    \end{aligned}
\end{equation}
Similarly 
\begin{equation}
    \begin{aligned}
&\omega\lt(\lt[\xi^\mu_3,\xi^\mu_1\rt],\xi^\mu_2\rt)\\
=&\pm2\oint_{S^2}\rmd^2 S\, \sqrt{\bar{\gamma}} \lt[
\lt(
    Y_{3}^{A} \bar{D}_{A} f_{1}
    -Y_{1}^{A} \bar{D}_{A} f_{3}
\rt)
\lt(
    \bar{D}^{B}\bar{D}_{B}W_2
    -\bar{D}_{B} 
    I_2^{B}
\rt)\rt.\\
&-f_2 
    \bar{D}^{A}\bar{D}_{A}
\lt(
        Y_{3}^{B} \bar{D}_{B} W_{1}
        -Y_{1}^{B} \bar{D}_{B} W_{3}
    \rt)\\
    &+\lt.f_2\bar{D}_{A}
    \lt(
       Y_{3}^{B} \bar{D}_{B}\bar{D}_{A} I_{1}^{A}
-Y_{1}^{B} \bar{D}_{B}\bar{D}_{A} I_{3}^{A}
    \rt)
\rt],
    \end{aligned}
\end{equation}
and
\begin{equation}
    \begin{aligned}      &\omega\lt(\lt[\xi^\mu_2,\xi^\mu_3\rt],\xi^\mu_1\rt)\\
=&\pm2\oint_{S^2}\rmd^2 S\, \sqrt{\bar{\gamma}} \lt[
\lt(
    Y_{2}^{A} \bar{D}_{A} f_{3}
    -Y_{3}^{A} \bar{D}_{A} f_{2}
\rt)
\lt(
    \bar{D}^{B}\bar{D}_{B}W_1
    -\bar{D}_{B} 
    I_1^{B}
\rt)\rt.\\
&-f_1 
    \bar{D}^{A}\bar{D}_{A}
\lt(
        Y_{2}^{B} \bar{D}_{B} W_{3}
        -Y_{3}^{B} \bar{D}_{B} W_{2}
    \rt)\\
    &+\lt.f_1\bar{D}_{A}
    \lt(
       Y_{2}^{B} \bar{D}_{B}\bar{D}_{A} I_{3}^{A}
-Y_{3}^{B} \bar{D}_{B}\bar{D}_{A} I_{2}^{A}
    \rt)
\rt].
    \end{aligned}
\end{equation}
With the results above, we can show that $\omega\lt(\xi^\mu\rt)$ satisfies the Jacobi identity:
\begin{equation}\label{jocoofcen}
    \omega\lt(\lt[\xi^\mu_1,\xi^\mu_2\rt],\xi^\mu_3\rt)
    +\omega\lt(\lt[\xi^\mu_3,\xi^\mu_1\rt],\xi^\mu_2\rt)
    +\omega\lt(\lt[\xi^\mu_2,\xi^\mu_3\rt],\xi^\mu_1\rt)
    =0.
\end{equation}
Firstly
\begin{equation}
    \begin{aligned}
&\omega\lt(\lt[\xi^\mu_1,\xi^\mu_2\rt],\xi^\mu_3\rt)
    +\omega\lt(\lt[\xi^\mu_3,\xi^\mu_1\rt],\xi^\mu_2\rt)
    +\omega\lt(\lt[\xi^\mu_2,\xi^\mu_3\rt],\xi^\mu_1\rt)\\
    =&\pm 2\oint_{S^2}\rmd^2 S\, \sqrt{\bar{\gamma}}
      \lt[
             \lt(
           Y_{1}^{A} \bar{D}_{A} f_{2}
    -Y_{2}^{A} \bar{D}_{A} f_{1}
\rt)
\lt(
    \bar{D}^{B}\bar{D}_{B}W_3
    -\bar{D}_{B} 
    I_3^{B}
\rt)\rt.\\
&-f_3 
    \bar{D}^{A}\bar{D}_{A}
\lt(
        Y_{1}^{B} \bar{D}_{B} W_{2}
        -Y_{2}^{B} \bar{D}_{B} W_{1}
    \rt)-\lt.f_3\bar{D}_{A}
    \lt(
       Y_{1}^{B} \bar{D}_{B}\bar{D}_{A} I_{2}^{A}
-Y_{2}^{B} \bar{D}_{B}\bar{D}_{A} I_{1}^{A}
    \rt)\rt.\\
    &+\lt(
    Y_{3}^{A} \bar{D}_{A} f_{1}
    -Y_{1}^{A} \bar{D}_{A} f_{3}
\rt)
\lt(
    \bar{D}^{B}\bar{D}_{B}W_2
    -\bar{D}_{B} 
    I_2^{B}
\rt)\\
&-f_2 
    \bar{D}^{A}\bar{D}_{A}
\lt(
        Y_{3}^{B} \bar{D}_{B} W_{1}
        -Y_{1}^{B} \bar{D}_{B} W_{3}
    \rt)-f_2\bar{D}_{A}
    \lt(
       Y_{3}^{B} \bar{D}_{B}\bar{D}_{A} I_{1}^{A}
-Y_{1}^{B} \bar{D}_{B}\bar{D}_{A} I_{3}^{A}
    \rt)\\
    &+\lt(
    Y_{2}^{A} \bar{D}_{A} f_{3}
    -Y_{3}^{A} \bar{D}_{A} f_{2}
\rt)
\lt(
    \bar{D}^{B}\bar{D}_{B}W_1
    -\bar{D}_{B} 
    I_1^{B}
\rt)\\
&-f_1 
    \bar{D}^{A}\bar{D}_{A}
\lt(
        Y_{2}^{B} \bar{D}_{B} W_{3}
        -Y_{3}^{B} \bar{D}_{B} W_{2}
    \rt) -\lt.f_1\bar{D}_{A}
    \lt(
       Y_{2}^{B} \bar{D}_{B}\bar{D}_{A} I_{3}^{A}
-Y_{3}^{B} \bar{D}_{B}\bar{D}_{A} I_{2}^{A}
    \rt)
      \rt].
    \end{aligned}
\end{equation}
Next since
\begin{equation}
    \begin{aligned}
      &\bar{D}^{A}\bar{D}_{A}
    \lt(
        Y_{B} \bar{D}^{B} W
    \rt)\\
=&\bar{D}^{A}
\lt(
    \bar{D}_{A}(Y_{B}) \bar{D}^{B} W
    +Y_{B} \bar{D}_{A}\bar{D}^{B} W
\rt)\\
=&\bar{D}^{A}
\lt(
    -\bar{D}_{B}(Y_{A}) \bar{D}^{B} W
    +Y_{B} \bar{D}_{A}\bar{D}^{B} W
\rt)\\
=&
-\bar{D}^{A}(\bar{D}_{B}Y_{A}) \bar{D}^{B} W
-\bar{D}_{[B}\lt(Y_{A]}\rt) \bar{D}^{(A}\bar{D}^{B)} W  
+\bar{D}_{[A}\lt(Y_{B]} \rt)\bar{D}^{(A}\bar{D}^{B)} W
+Y_{B} \bar{D}^{A}\bar{D}_{A}\bar{D}^{B} W
\\
=&
-\bar{D}_{B}\lt(\bar{D}_{A}Y^{A}\rt) \bar{D}^{B} W
-Y^{C}R_{BC}\bar{D}^{B} W
+Y^{A} \bar{D}_{A}\bar{D}_{B}\bar{D}^{B} W
+Y^{A}R_{AC}\bar{D}^{C} W
\\
=&Y^{A} \bar{D}_{A}\bar{D}_{B}\bar{D}^{B} W,
    \end{aligned}
\end{equation}   
we have
\begin{equation}
    \begin{aligned}
      &\pm2\oint_{S^2}\rmd^2 S\, \sqrt{\bar{\gamma}}
      \lt[\rt. 
 \lt(
    Y_{1}^{A} \bar{D}_{A} f_{2}
    -Y_{2}^{A} \bar{D}_{A} f_{1}
\rt)\bar{D}^{B}\bar{D}_{B}W_3\\
&- 
 f_3\bar{D}^{A}\bar{D}_{A}
    \lt(
        Y_{1}^{B} \bar{D}_{B} W_{2}
        -Y_{2}^{B} \bar{D}_{B} W_{1}
    \rt)\\
&+
\lt(
    Y_{3}^{A} \bar{D}_{A} f_{1}
    -Y_{1}^{A} \bar{D}_{A} f_{3}
\rt)\bar{D}^{B}\bar{D}_{B}W_2\\
&- 
    f_2\bar{D}^{A}\bar{D}_{A}
    \lt(
        Y_{3}^{B} \bar{D}_{B} W_{1}
        -Y_{1}^{B} \bar{D}_{B} W_{3}
    \rt)\\
&+
\lt(
    Y_{2}^{A} \bar{D}_{A} f_{3}
    -Y_{3}^{A} \bar{D}_{A} f_{2}
\rt)\bar{D}^{B}\bar{D}_{B}W_1\\
&- 
    f_1\bar{D}^{A}\bar{D}_{A}
    \lt(
        Y_{2}^{B} \bar{D}_{B} W_{3}
        -Y_{3}^{B} \bar{D}_{B} W_{2}
    \rt)\lt.\rt]\\
=&\pm2\oint_{S^2}\rmd^2 S\, \sqrt{\bar{\gamma}}
      \lt[\rt.
 (
    Y_{1}^{A} \bar{D}_{A} f_{2}
    -Y_{2}^{A} \bar{D}_{A} f_{1}
)\bar{D}^{B}\bar{D}_{B}W_3\\
&- 
 f_3
    \lt(
        Y_1^{A} \bar{D}_{A}\bar{D}_{B}\bar{D}^{B} W_2
        -Y_2^{A} \bar{D}_{A}\bar{D}_{B}\bar{D}^{B} W_1
    \rt)\\
&+
\lt(
    Y_{3}^{A} \bar{D}_{A} f_{1}
    -Y_{1}^{A} \bar{D}_{A} f_{3}
\rt)\bar{D}^{B}\bar{D}_{B}W_2\\
&-
    f_2
    \lt(
        Y^{A}_3 \bar{D}_{A}\bar{D}_{B}\bar{D}^{B} W_1
        -Y^{A}_1 \bar{D}_{A}\bar{D}_{B}\bar{D}^{B} W_3
    \rt)\\
&+
\lt(
    Y_{2}^{A} \bar{D}_{A} f_{3}
    -Y_{3}^{A} \bar{D}_{A} f_{2}
\rt)\bar{D}^{B}\bar{D}_{B}W_1\\
&-f_1
    \lt(
        Y^{A}_2 \bar{D}_{A}\bar{D}_{B}\bar{D}^{B} W_3
        -Y^{A}_3 \bar{D}_{A}\bar{D}_{B}\bar{D}^{B} W_2
    \rt)\lt.\rt]\\
=&\pm2\oint_{S^2}\rmd^2 S\, \sqrt{\bar{\gamma}}
      \lt[\rt.-
 \lt(
    Y_{1}^{A} f_{2}\bar{D}_{A}\bar{D}^{B}\bar{D}_{B} W_3
    -Y_{2}^{A}  f_{1}\bar{D}_{A}\bar{D}^{B}\bar{D}_{B} W_3
\rt)\\
&- 
    \lt(
        Y_1^{A}  f_3\bar{D}_{A}\bar{D}_{B}\bar{D}^{B} W_2
        -Y_2^{A} f_3 \bar{D}_{A}\bar{D}_{B}\bar{D}^{B} W_1
    \rt)\\
&-
\lt(
    Y_{3}^{A}  f_{1}\bar{D}_{A}\bar{D}^{B}\bar{D}_{B}W_2
    -Y_{1}^{A}  f_{3}\bar{D}_{A}\bar{D}^{B}\bar{D}_{B}W_2
\rt)\\
&-\lt(
        Y^{A}_3f_2 \bar{D}_{A}\bar{D}_{B}\bar{D}^{B} W_1
        -Y^{A}_1f_2\bar{D}_{A}\bar{D}_{B}\bar{D}^{B} W_3
    \rt)\\
&-
\lt(
    Y_{2}^{A} f_{3}\bar{D}_{A}\bar{D}^{B}\bar{D}_{B}W_1
    -Y_{3}^{A} f_{2}\bar{D}_{A}\bar{D}^{B}\bar{D}_{B}W_1
\rt)\\
&-  
    \lt(
        Y^{A}_2f_1 \bar{D}_{A}\bar{D}_{B}\bar{D}^{B} W_3
        -Y^{A}_3f_1 \bar{D}_{A}\bar{D}_{B}\bar{D}^{B} W_2
    \rt)\lt.\rt]\\
=&0.
    \end{aligned}
\end{equation}
Furthermore,
\begin{equation}
    \begin{aligned}
      &\pm2\oint_{S^2}\rmd^2 S\, \sqrt{\bar{\gamma}}\lt[\rt.- 
\lt(
    Y_{3}^{A} \bar{D}_{A} f_{1}
    -Y_{1}^{A} \bar{D}_{A} f_{3}
\rt)\bar{D}_{B} 
    I_2^{B}\\
&-  
\lt(
-f_2
\lt(
    Y_{3}^{B} \bar{D}_{B}\bar{D}_{A} I_{1}^{A}
    -Y_{1}^{B} \bar{D}_{B}\bar{D}_{A} I_{3}^{A}
\rt)\rt)\\
&- 
\lt(
    Y_{1}^{A} \bar{D}_{A} f_{2}
    -Y_{2}^{A} \bar{D}_{A} f_{1}
\rt)\bar{D}_{B} 
    I_3^{B}
\\
&-  
\lt(
-f_3
\lt(
    Y_{1}^{B} \bar{D}_{B}\bar{D}_{A} I_{2}^{A}
    -Y_{2}^{B} \bar{D}_{B}\bar{D}_{A} I_{1}^{A}
\rt)\rt)\\
&- 
\lt(
    Y_{2}^{A} \bar{D}_{A} f_{3}
    -Y_{3}^{A} \bar{D}_{A} f_{2}
\rt)\bar{D}_{B} 
    I_1^{B}
\\
&-  
\lt(
-f_1
\lt(
    Y_{2}^{B} \bar{D}_{B}\bar{D}_{A} I_{3}^{A}
    -Y_{3}^{B} \bar{D}_{B}\bar{D}_{A} I_{2}^{A}
\rt)\rt)\lt.\rt]\\
=&\pm2\oint_{S^2}\rmd^2 S\, \sqrt{\bar{\gamma}}\lt[\rt.
\lt(
    Y_{3}^{A}  f_{1}\bar{D}_{A}\bar{D}_{B} 
    I_2^{B}
    -Y_{1}^{A}  f_{3}\bar{D}_{A}\bar{D}_{B} 
    I_2^{B}
\rt)
\\
&+  
\lt(
    f_2Y_{3}^{B} \bar{D}_{B}\bar{D}_{A} I_{1}^{A}
    -f_2Y_{1}^{B} \bar{D}_{B}\bar{D}_{A} I_{3}^{A}
\rt)\\
&+ 
\lt(
    Y_{1}^{A} f_{2}\bar{D}_{A} \bar{D}_{B} 
    I_3^{B}
    -Y_{2}^{A}  f_{1}\bar{D}_{A}\bar{D}_{B} 
    I_3^{B}
\rt)
\\
&+  
\lt(
   f_3 Y_{1}^{B} \bar{D}_{B}\bar{D}_{A} I_{2}^{A}
    -f_3Y_{2}^{B} \bar{D}_{B}\bar{D}_{A} I_{1}^{A}
\rt)\\
&+ 
\lt(
    Y_{2}^{A}  f_{3}\bar{D}_{A}\bar{D}_{B} 
    I_1^{B}
    -Y_{3}^{A}  f_{2}\bar{D}_{A}\bar{D}_{B} 
    I_1^{B}
\rt)
\\
&+  
\lt(
    f_1Y_{2}^{B} \bar{D}_{B}\bar{D}_{A} I_{3}^{A}
    -f_1Y_{3}^{B} \bar{D}_{B}\bar{D}_{A} I_{2}^{A}
\rt)\lt.\rt]\\
=&0.
    \end{aligned}
\end{equation}
As the result, we verify \eqref{jocoofcen}.

\bibliographystyle{jhep}
\bibliography{myref}
\end{document}